\documentclass[acmtog,final]{acmart}
\acmSubmissionID{420}

\usepackage{booktabs} 
\usepackage[utf8]{inputenc}
\usepackage[T1]{fontenc}
\usepackage{graphicx}
\usepackage{overpic}
\usepackage{trimclip}
\usepackage{contour}\contourlength{1pt}
\usepackage[pscoord]{eso-pic}

\theoremstyle{acmplain}

\newtheorem{defn}{Definition}[section]
\newtheorem{lem}[defn]{Lemma}
\newtheorem{prop}[defn]{Proposition}
\theoremstyle{remark}
\newtheorem{remark}[defn]{Remark}

\def\lput(#1,#2)#3{\put(#1,#2){\hbox to 0pt{\hss{#3}}}}
\def\rput(#1,#2)#3{\put(#1,#2){\hbox to 0pt{{#3}\hss}}}
\def\cput(#1,#2)#3{\put(#1,#2){\hbox to 0pt{\hss{#3}\hss}}}
\def\ctput(#1,#2)#3{\cput(#1,#2){\vbox to 0pt{\hbox{#3}\vss}}}
\def\ccput(#1,#2)#3{\cput(#1,#2){\vbox to 0pt{\vss\hbox{#3}\vss}}}
\def\ltput(#1,#2)#3{\lput(#1,#2){\vbox to 0pt{\hbox{#3}\vss}}}
\def\lcput(#1,#2)#3{\lput(#1,#2){\vbox to 0pt{\vss\hbox{#3}\vss}}}
\def\rtput(#1,#2)#3{\put(#1,#2){\vbox to 0pt{\hbox{#3}\vss}}}
\def\rcput(#1,#2)#3{\put(#1,#2){\vbox to 0pt{\vss\hbox{#3}\vss}}}

\def\R{{\mathbb R}\!}

\def\Bffont{\sffamily\bfseries}
\def\xw{{\text{\Bffont x}}}
\def\yw{{\text{\Bffont y}}}
\def\cw{{\text{\Bffont c}}}
\def\nw{{\text{\Bffont n}}}
\def\vw{{\text{\Bffont v}}}
\def\ww{{\text{\Bffont w}}}

\def\ew{{\text{\Bffont e}}}

\def\rw{{\text{\Bffont r}}}

\def\Fnfont{\sffamily}
\def\iso{\text{\Fnfont iso}}
\def\pos{\text{\Fnfont pos}}
\def\prox{\text{\Fnfont prox}}
\def\dev{\text{\Fnfont dev}}
\def\rul{\text{\Fnfont rul}}
\def\fair{{\text{\Fnfont fair}}}
\def\ev{{\text{\Fnfont ev}}}
\def\vert{{\text{\Fnfont vert}}}
\def\norm{{\text{\Fnfont norm}}}

\def\wt{\widetilde}
\def\wh{\widehat}

\def\II{{\mathord{\rm II}}}

\def\Bullet{\smallskip\noindent$\bullet$\hspace*{1.5ex}}

\def\Reg{\textsuperscript{\textregistered}}

\definecolor{rot}{rgb}{0.7,0.0,0.1}
\definecolor{gelb}{rgb}{.7,.5,.2}
\definecolor{blau}{rgb}{0,0,0.5}

\def\<{\langle}
\def\>{\rangle}

\def\Inc{\includegraphics}

\newenvironment{pp}
{\unitlength.01\columnwidth\begin{picture}(.01,.01)}
	{\end{picture}}

\outer\def\proclaim #1. #2\par{\medbreak
  \noindent{\sc#1.\enspace}{\it#2\par}%
  \ifdim\lastskip<\medskipamount \removelastskip\penalty55\medskip\fi}

\usepackage[ruled]{algorithm2e} 

\SetAlFnt{\small}
\SetAlCapFnt{\small}
\SetAlCapNameFnt{\small}
\SetAlCapHSkip{0pt}

\acmJournal{TOG}
\acmVolume{42}
\acmNumber{6}
\acmArticle{183}
\acmYear{2023}
\acmMonth{12}

\acmDOI{10.1145/3618355}

\setcopyright{acmlicensed}

\begin{document}

\title{Developable Quad Meshes and Contact Element Nets}

\author{Victor Ceballos Inza}
\affiliation{%
  \institution{KAUST}
  \country{Saudi Arabia}
}
\author{Florian Rist}
\affiliation{%
  \institution{KAUST}
  \country{Saudi Arabia}
}
\author{Johannes Wallner}
\affiliation{%
  \institution{TU Graz}
  \country{Austria}
}
\author{Helmut Pottmann}
\affiliation{%
  \institution{KAUST}
  \country{Saudi Arabia}
}


\renewcommand\shortauthors{Ceballos Inza, V. \ et al.}

\begin{abstract}
The property of a surface being developable can be expressed in different equivalent ways, by vanishing Gauss curvature, or by the existence of isometric mappings to planar domains. Computational contributions to this topic range from special parametrizations to discrete-isometric mappings. However, so far a local criterion expressing developability of general quad meshes has been lacking. In this paper, we propose a new and efficient discrete developability criterion that is applied to quad meshes equipped with vertex weights, and which is motivated by a well-known characterization in differential geometry, namely a rank-deficient second fundamental form. We assign contact elements to the faces of meshes and ruling vectors to the edges, which in combination yield a developability condition per face. Using standard optimization procedures, we are able to perform interactive design and developable lofting. The meshes we employ are combinatorially regular quad meshes with isolated singularities but are otherwise not required to follow any special curves on a developable surface. They are thus easily embedded into a design workflow involving standard operations like remeshing, trimming, and merging operations. An important feature is that we can directly derive a watertight, rational bi-quadratic spline surface from our meshes. Remarkably, it occurs as the limit of weighted Doo-Sabin subdivision, which acts in an interpolatory manner on contact elements.
\end{abstract}

\ccsdesc[500]{Computing methodologies~Shape modeling}
\ccsdesc[300]{Computing methodologies~Optimization algorithms}

\keywords{developable surface, discrete differential geometry,
computer-aided design, shape optimization, checkerboard pattern,
contact elements}

\begin{teaserfigure}
\clipbox{6mm 5mm 5mm 2mm}{\Inc[height=52mm]{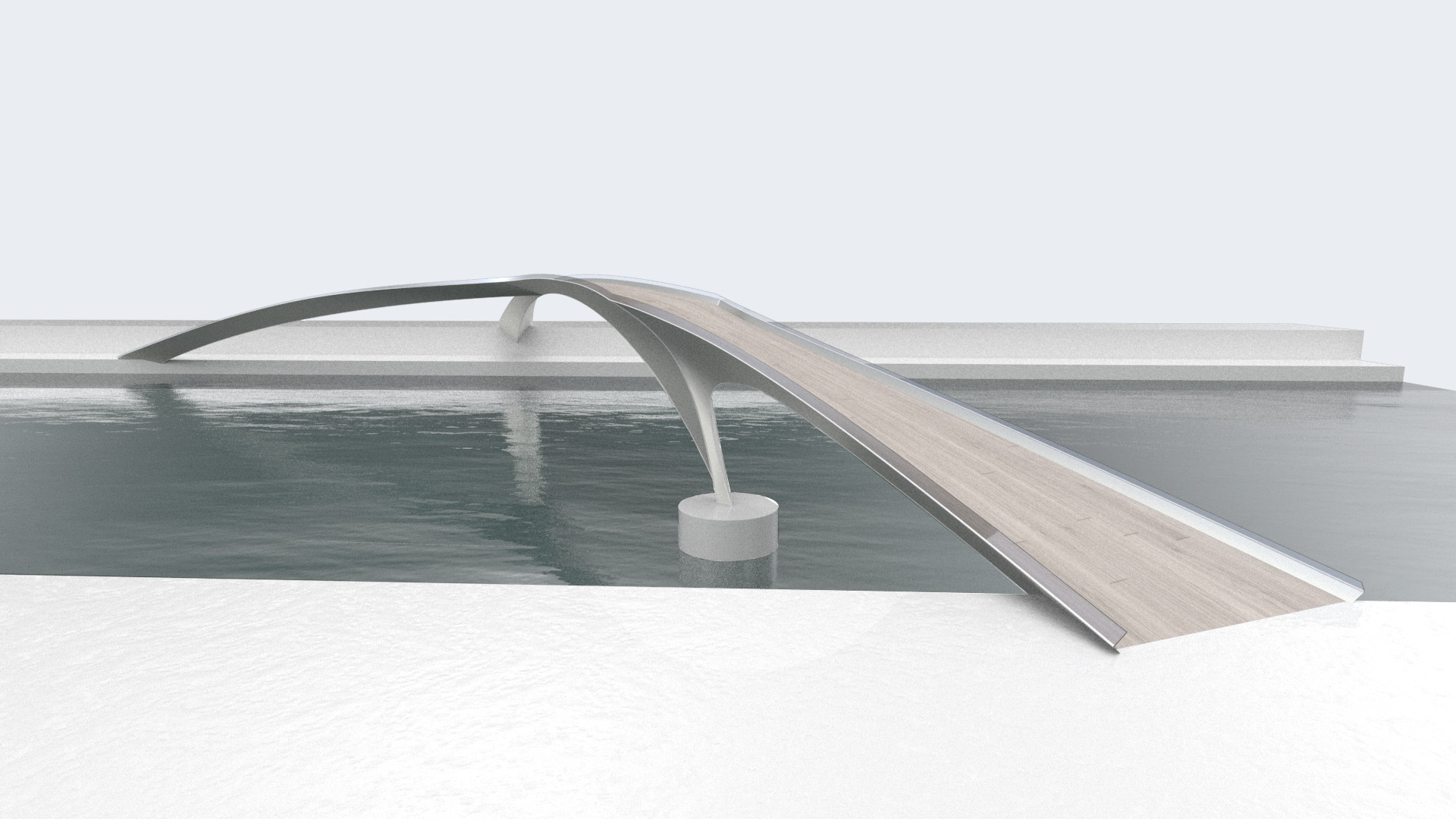}}\hfill
\clipbox{70mm 12mm 9mm 3mm}{\Inc[height=60mm]{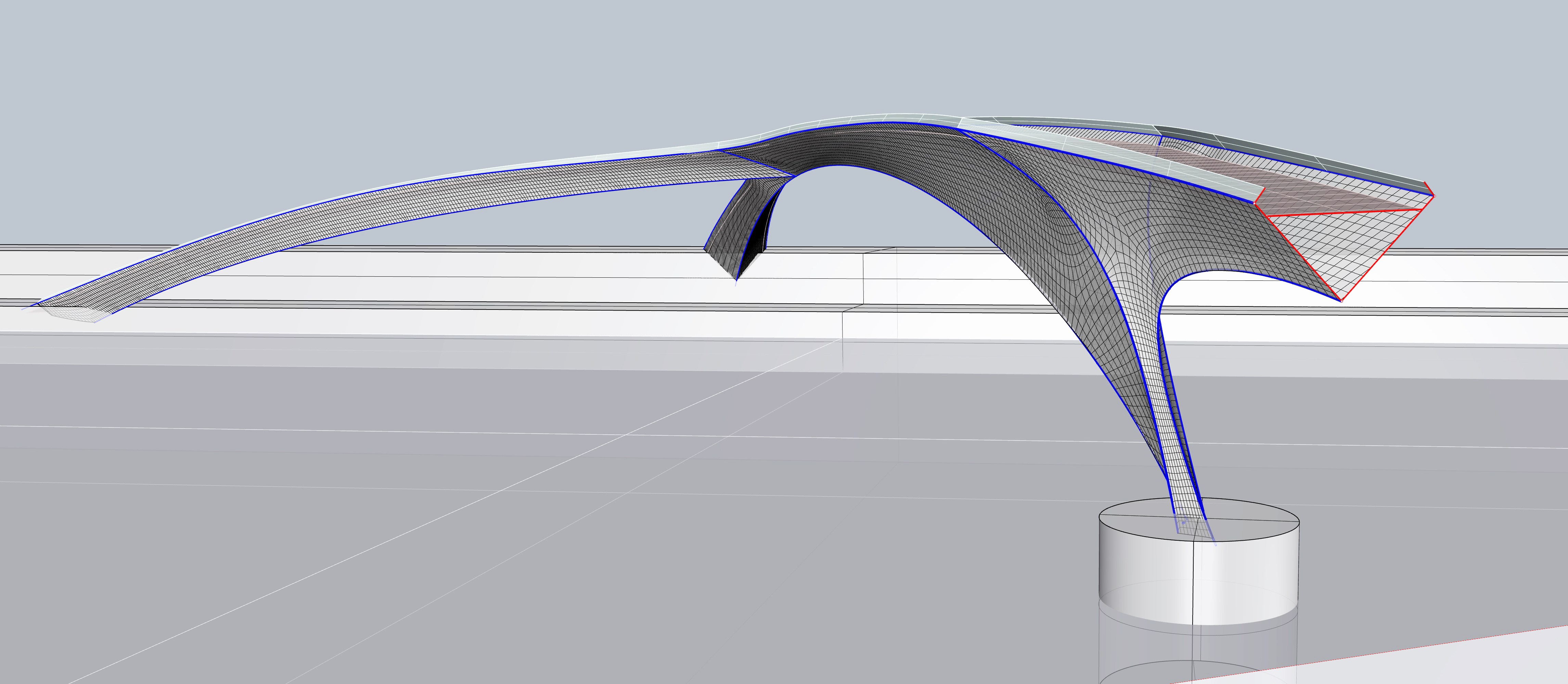}}\hfill
\clipbox{26mm 10mm 20mm 7mm}{\Inc[height=63mm]{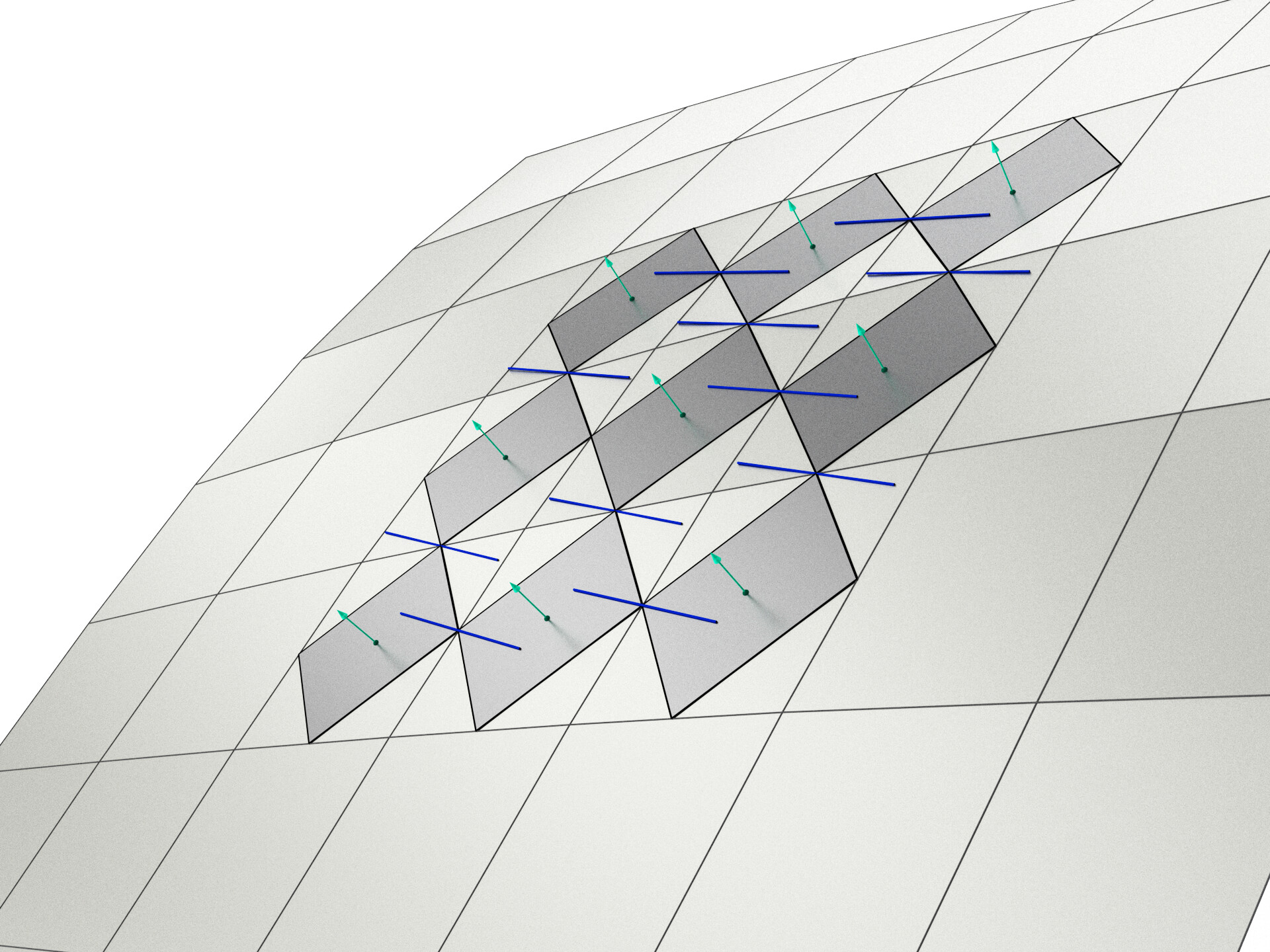}}
\caption{An architectural design based on challenging developable lofting
tasks. CAD software has been used to connect given boundaries by
a NURBS loft, followed by quad remeshing. Subsequently, our
optimization towards developability has been applied.
Observe that patches have several boundaries,
and meshes exhibit combinatorial singularities. From left
to right, we show the resulting design, meshes representing developables, and
a zoomed-in detail revealing a discrete-developable quad mesh, endowed with inscribed
contact elements, and the ruling line field we get from intersecting
neighboring contact elements.}
\label{fig:bridge}
\end{teaserfigure}

\maketitle


\section{Introduction}
\label{sec:introduction}

The numerical and geometric modeling of developable surfaces has attracted
attention for many years, starting with the first mesh representations
proposed by R.\ Sauer \shortcite{sauer:1970}.
One reason for that is the great practical
importance of developables, which represent shapes made by bending
flat pieces of inextensible sheet material. New algorithms continue
to emerge, as do new applications --- we only point to the construction
of metamaterials that can be speedily laser-cut and fill
volumes in the manner of ruffles by Signer et al.\ \shortcite{signer:2021},
and the use of developables in interactive physical book simulation by
Wolf et al.\ \shortcite{wolf-2021}.

A developable surface is
unanimously defined by the existence of local isometric mappings to planar
domains. As it turns out, the physical reality of bending thin sheets
is modeled by surfaces exhibiting piecewise $C^2$ smoothness, which means
surfaces exhibiting curvature continuity except for creases along curves.
Geometric modeling of developables has been confined to this case.
The continuous new proposals for the computational treatment of
developables are a sign that the problem has not yet been satisfactorily
solved. Another reason is the mathematical complexity of the
subject itself, as well as the fact that $C^2$ developables enjoy many different
but equivalent characterizations.  These include
vanishing Gauss curvature or other equivalent infinitesimal properties;
line contact with tangent planes; or the existence of a local
planar development.
Each of these properties
has been the basis of a computational approach, and each serves as
motivation for defining a certain kind of discrete developable surface.
This is also true for the present paper: We use the fact that
developability is characterized by a rank deficient second fundamental
form, and we employ the checkerboard patterns proposed by
Peng et al.\ \shortcite{checkerboard-2019} to express this in a discrete way.
The basic entity we work with is a {\em contact element}, which is a weighted
point plus a normal vector.

\subsection{Overview and Contributions}

\Bullet We present a new quad-mesh-based
discrete model of developable surfaces
which does not require the use of a development. Neither do edges
have to be aligned with special curves on the surface under consideration.
It is based on so-called contact elements inscribed in the faces
of the mesh which are defined via vertex weights (\S\,\ref{ss:contactelements}).

\Bullet The contact element net derived from a quad mesh is used
to express  discrete developability (\S\,\ref{ss:discretedevelopables}),
and also to derive a watertight
spline surface interpolating contact elements. Incidentally that spline
surface is the limit of weighted Doo-Sabin subdivision which acts
in an interpolatory manner on contact elements (\S\,\ref{sss:subdivision}).

\Bullet The discrete developability property is achieved by optimization,
essentially performing a projection onto the constraint manifold,
guided by soft constraints like fairness and proximity to a reference
surface (\S\,\ref{ss:computation}).
When needed, we can combine our method with the isometric mappings
proposed by Jiang et al.\ \shortcite{jiang-2020}.

\Bullet Interactive design of developables can mean the isometric
deformation of a given flat piece as well as generally modifying
a design surface such that developability is maintained as a constraint.
We are able to do both (\S\,\ref{sec:desig}).

\Bullet Modeling tools treated in this paper include developable lofting,
which is an old problem not easily accessible with
previous methods (\S\,\ref{ss:lofting}).
A user can interactively modify developables by pulling on handles and letting
developables glide through guiding curves;
attaching a developable patch to a surface; and positioning
singular curves. The refinement property of contact element nets allows for
a multiresolution approach to modeling developables (\S\,\ref{ss:multiresolution}).

\subsection{Previous Work}

There is a large body of literature about geometric modeling
of and with developable surfaces. Our brief overview is subdivided according
to the way developables are represented, either as spline surfaces or as
discrete surfaces.

\subsubsection{Previous work based on splines}
It is known that developables consist of ruled surfaces enjoying
{\em torsality}, i.e., the tangent plane along a ruling is constant.
A ruled surface can be modeled as a degree $(1,n)$ B-spline
surface --- one family of parameter lines then will be the surface's
rulings. Torsality is a nonlinear constraint \cite{langroeschel92}
that can be achieved by optimization
\cite{tang-dev-2015,ciang-2019-cf}. One limitation of such a method is
the necessity to decompose developables into ruled pieces. The method thus
may not be suitable for modeling deformations of developables where
that decomposition may change.

A torsal ruled surface is the envelope of its tangent planes, and
thus essentially is a curve in the dual space of planes. This fact
has been exploited by Bodduluri and Ravani\ \shortcite{bodduluri-93}
and follow-up publications. It
reduces the design of ruled developables to the design of curves.
The drawback of this method is that, in addition to the one mentioned in the previous paragraph, working in plane space is not intuitive and does not naturally avoid singularities.

Finally, we point to Jiang et al.\ \shortcite{jiang-2020} who
impose approximate developability on spline surfaces   via conversion
to a quad mesh. Here rulings do not have to coincide with parameter lines, cf.\
\S\,\ref{sss:previous:quadmeshes}.

\subsubsection{Previous work based on quad meshes}
\label{sss:previous:quadmeshes}

The textbook \cite{sauer:1970} proposes discrete developables based on the
fact that developables are the envelopes of their tangent planes: A
discrete ruled developable is simply a sequence of flat quads, and the
edges between them play the role of rulings.
This property lies on the basis of quad-meshing of
developables, see recent work by Verhoeven at al.\
\shortcite{Verhoeven-2022}. Such ruling-based
developables are the basis of
work by Liu et al.\ \shortcite{liu+2006} and
Solomon et al.\ \shortcite{solomon-2012}. Their disadvantages
are the same as for other
ruling-based methods: it is difficult to model situations where the ruling pattern of a developable changes.

A different characterization of developables, the existence of a network of orthogonal
geodesics, is the basis of work by
Rabinovich et al.\
\shortcite{rabinovich+2018a,rabinovich+2018,Rabinovich:CurvedFolds:2019}
and Ion et al.\ \shortcite{ion-2020}. Here developables are represented by
quad meshes whose edges are not necessarily aligned with the rulings, but
nevertheless are in a special position --- they discretize a network of
orthogonally intersecting geodesics.

Jiang et al.\ \shortcite{jiang-2020} use
discrete-isometric mappings to handle developables: A surface is
represented by a quad mesh whose edges do not have any special relation to
the surface. Developability is imposed by maintaining a second mesh
in $\R^2$ isometric to the first mesh.
Similarly, the work by Chern et al.\ \shortcite{chern-2018} is also
capable of handling developables via isometric mappings to planar domains.

\subsubsection{Previous work based on other discretizations}
A triangle mesh is intrinsically flat except at the vertices, and it is
in fact locally flat if and only if the angle sum in its vertices equals
$2\pi$. This natural discretization of the concept of a developable surface
has not been in much use in geometric modeling. An exception is provided
by meshes consisting of equilateral triangles as used by Jiang
et at.\ \shortcite{jiang:2014}. Better suited for geometric design of developables
is the `local hinge' condition imposed by Stein et al.\
\shortcite{Stein:2018:DSF}. It ensures the existence of rulings and also implies
that the position of edges is not arbitrary; every face has an edge that represents the direction of a ruling.
Binninger et al.\ \shortcite{binninger-2021} approximate surfaces
with piecewise-\hskip0pt developable ones by thinning the Gauss image; their method
operates on a triangle mesh.

Sell\'an et al.\ \shortcite{sellan-2020} use an entirely different way
of imposing developability. A height field $z=f(x,y)$ represents a
developable surface if and only if the Hessian of $f$ has a vanishing
determinant. A certain convex optimization converts a given height
field to one where the Hessian is nonsingular only along 1D curves. This
leads to a piecewise-developable surface. Our method is also based on
the characterization of developables as surfaces with low-rank second
fundamental forms. Our discretization, however, works for any quad mesh,
and is not limited to height fields.

\subsubsection{Previous work on Contact Elements}
In the continuous setting, contact elements have been used in differential
geometry and analysis since the 19th century. In the discrete setting,
contact elements are implicitly present every time vertices are treated
together with normal vectors. There are only a few contributions explicitly
based on contact elements, such as the discrete principal meshes proposed
by Bobenko and Suris \shortcite{bobenko:org:2007} and subsequent treatment
of a discrete curvature theory and related topics
\cite{schroecker:2010,bobenko-2010-ct,roerig2021}.

\section{Discrete Developables}
\label{sec:basic}

\subsection{Differential Geometry of Smooth Developables}

We consider surfaces that are piecewise curvature
continuous and which enjoy the property of being locally
isometric to a planar domain. It is well known (see e.g.\
\cite{guggenheimer:1963}) that this intrinsic
flatness is characterized by the vanishing of Gauss curvature,
$K=0$. It is also well known that one family of principal curvature lines of
such surfaces is composed of straight lines, and that the tangent plane along these
straight lines (\emph{rulings}) is constant.  The rulings extend all the way
to the boundary of the surface. A developable thus decomposes into
ruled surface pieces and planar parts. In geometric modeling, the number
of pieces is assumed to be finite. We also consider surfaces
consisting of individual developable pieces.

\paragraph{Gauss Image and 2nd Fundamental Form}
The {\em Gauss image} of a developable is the set of its unit normal vectors.
It is contained in the unit sphere $S^2$, and it decomposes into
curves, one for each ruled piece.

In each point $p$ of the surface, the second fundamental form $\II_p(\vw,\ww)$
governs curvatures.  It takes as arguments vectors $\vw,\ww$ tangent to
the surface. If $\xw(u,v)$ is a parametrization, and $\nw(u,v)$ is
the corresponding unit normal vector field, then the second fundamental
form relates their derivatives via
\begin{align*}
	\II_p(\xw_u,\xw_u)&=\<\xw_u,-\nw_u\>, \quad
	\II_p(\xw_u,\xw_v)=\<\xw_u,-\nw_v\>,
	\\
	\II_p(\xw_v,\xw_v)&=\<\xw_v,-\nw_v\>.
\end{align*}
By linearity, $\II$ is now defined for all tangent vectors.

\begin{figure}[b]
{\centerline{\begin{minipage}[b]{.37\columnwidth}
\caption{A developable touching another surface along a curve $\cw(t)$. Its
rulings $r(t)$ are conjugate to the derivative vectors $\dot\cw(t)$. \newline}
\label{fig:beruehrtorse}
\end{minipage}
\begin{overpic}[width=.6\columnwidth]{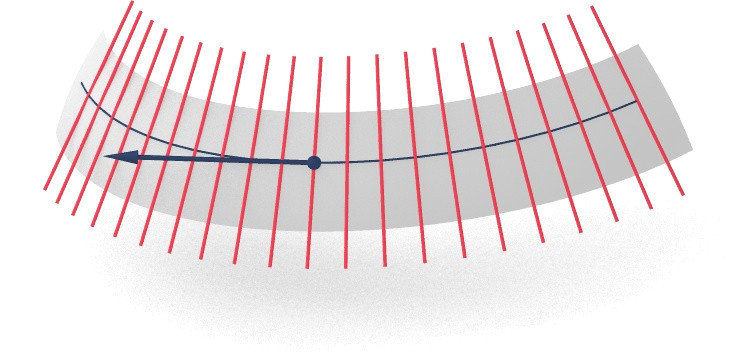}
	\color{rot}\cput(40,40){\contour{white}{$r(t)$}}
	\color{blau}\rtput(37,23){{$\cw(t)$}}
	\ctput(20,24){{$\dot\cw(t)$}}
\end{overpic}}}
\end{figure}

\paragraph{The Conjugacy Relation and Developability}
Tangent vectors
$\vw,\ww$ attached to the same point $p$ are called conjugate, if
\begin{align*}
	\II_p(\vw,\ww)=0.
\end{align*}
Gauss curvature vanishes in $p$ if and only if $\II_p$ has rank less than $2$ and
exhibits a kernel (the {\em ruling})
which is conjugate to {\em all} tangent vectors. This
can equivalently be expressed by the existence of a tangent
vector which obeys
\begin{align}
	\II(\xw_u,\rw)=\II(\xw_v,\rw)=0 \quad (\rw\ne 0).
	\label{eq:abwickelbar}
\end{align}

There is another property relating conjugacy of tangent vectors with
developable surfaces: Suppose we have a curve $\cw(t)$ contained in a
surface and a vector field $\rw(t)$ which is conjugate to the derivative
$\dot\cw(t)$. Then the ruled surface $\yw(t,s)=\cw(t)+s\rw(t)$ is
tangent to the original surface along the curve $\cw$, and is
itself developable \cite{liu+2006}. It is the envelope
of the tangent planes of the surface along the curve $\cw$, and it is
a geometric fact that the rulings of this envelope are {\em conjugate}
to the tangents of the curve --- see Fig.~\ref{fig:beruehrtorse}.

\subsection{Contact Element Nets}
\label{ss:contactelements}

It is our aim to express developability as a local property  of a quad
mesh whose edges and faces are allowed to be arbitrary. In
contrast to previous work
\cite{sauer:1970,liu+2006}
we do not require the faces to be planar, nor do we require the
edges to be aligned with special curves on the surface, as is done
by the previous references and by \cite{rabinovich+2018,Stein:2018:DSF}.
We also do not need an isometric mapping to a planar domain.

\subsubsection{Contact Element Nets From Weighted Vertices}
\begin{figure}[t]
\begin{overpic}[width=.49\columnwidth]{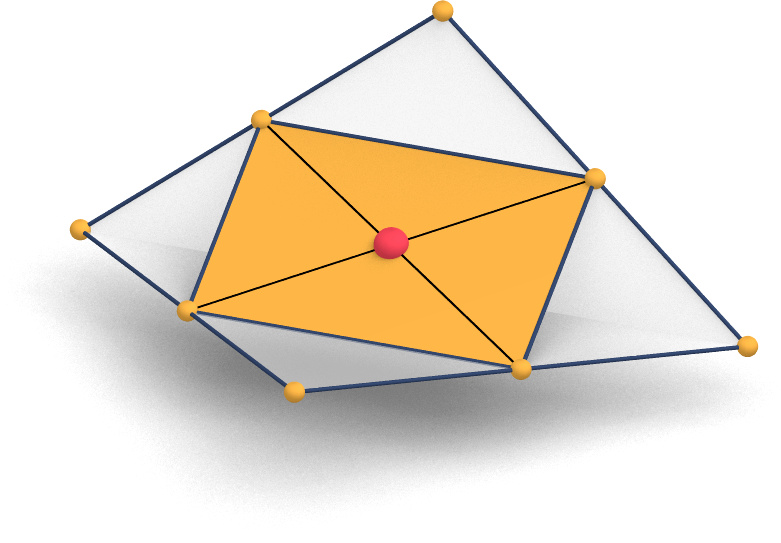}
    \put(0,60){$\R^4$}
	\lput(10,42){$ \wh v_0$}
	\ctput(38,14){$ \wh v_1$}
	\put(95,27){$\wh v_2$}
	\put(62,64){$\wh v_3$}
	\ccput(20,39){$\wh f$}%
	\ccput(35,37){$\wh\tau_f$}%
	\color{blau}%
	\color{rot}%
	\put(44,42.5){$\frac{\wh v_0+\!\cdots+\wh v_3}{4}$}
    \ltput(22,30){$\frac{\wh v_0+\wh v_1}{2}$}
    \ctput(65,17){$\frac{\wh v_1+\wh v_2}{2}$}
\end{overpic}\hfill
\begin{overpic}[width=.49\columnwidth]{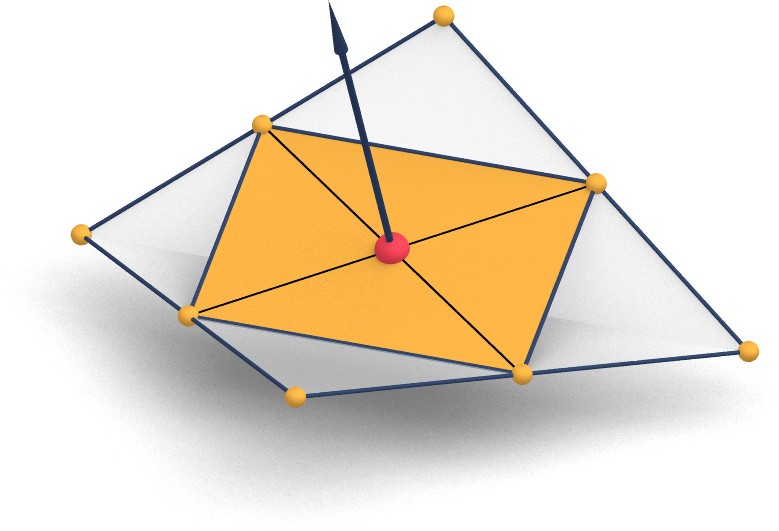}
    \put(0,60){$\R^3$}
	\lput(10,42){$v_0$}
	\ctput(38,14){$v_1$}
	\put(94,27){$v_2$}
	\put(62,64){$v_3$}
	\ccput(20,36){$f$}%
	\ccput(37,35){$\tau_f$}%
	\color{blau}%
	\lput(38,65){$\nw_f$}
	\color{rot}%
	\put(52,42){$b_f$}
    \lput(22,25){$m_{e_0}$}
    \ctput(65,17){$m_{e_1}$}
    \put(78,44){$m_{e_2}$}
    \lput(35,56){$m_{e_3}$}
\end{overpic}\vskip-1.5em
\caption{A contact element associated with the weighted
vertices of a face $f$ is defined by points $m_{e_i}$
on the edges $e_i = v_iv_{i+1}$, the plane $\tau_f$ and its normal vector $\nw_f$,
and the contact point $b_f$. The plane $\tau_f$ serves
as a tangent plane associated with face $f$.
\label{fig:parallelogram}}
\end{figure}

We propose a developability condition that uses a generalized
version of the {\em checkerboard
patterns} approach used  e.g.\ by \cite{jiang-2020}.
For each edge they consider the midpoint, and for
each face $f$, the
inscribed parallelogram formed by those edge midpoints ---
see Fig.~\ref{fig:parallelogram}.

The center of the parallelogram
together with a normal vector already form a contact element as it is.
However, we aim for greater generality; we wish to consider not
only parallelograms, but {\em any} planar quadrilateral inscribed in the
faces of meshes.

Consider a quadrilateral, which needs not be planar, with vertices $v_0,v_1,v_2,v_3$,
and \emph{edge points} $m_{e_i}$ on each edge $e_i = v_iv_{i+1}$ (indices modulo 4).
The edge points shall not coincide with the vertices.

\begin{lem}
    \label{lem:1}
    The edge points are co-planar if and only if the ratios of distances
    \begin{align*}
        (v_i,m_{e_i},v_{i+1}) := \frac{\| m_{e_i} - v_i \|}{\| v_{i+1} - m_{e_i} \|}
    \end{align*}
    fulfill the equation
    \begin{align}
        \label{eq:ratio}
        (v_0,m_{e_0},v_1) (v_1,m_{e_1},v_2)(v_2,m_{e_2},v_3)(v_3,m_{e_3},v_0)=1.
    \end{align}
\end{lem}

\begin{proof}
    Note that $(v_i, m_{e_i}, v_{i+1})$ equals
    the ratio of distances of points to any plane through $m_{e_i}$
    not containing $v_i, v_{i+1}$
    Taking this plane as the one which spans 3 edge points, the
    product of ratios equals 1 if and only if the 4th point lies on the same plane, so that all
    distances in the product cancel out.
\end{proof}

The edge points can be described by weights $w_i>0$ associated with the vertices $v_i$ as
\begin{align}
    \label{eq:weights}
    m_{e_i} & = \frac{1}{w_{i,i+1}} (w_i v_i + w_{i+1} v_{i+1} ), \quad w_{i,j}  = w_i+w_j.
\end{align}

\begin{lem}
    \label{lem:2}
    Any choice of weights $w_i>0$ leads to co-planar edge points.
    Conversely, any co-planar edge points can be described by vertex weights.
\end{lem}

\begin{proof}
    Given edge points as in Eq.~\eqref{eq:weights}, the ratios of distances
    become $(v_i,m_{e_i},v_{i+1}) = w_{i+1}/w_i$, so Eq.~\eqref{eq:ratio} holds.
    Conversely, to show the representation via weights, if Eq.~\eqref{eq:ratio} holds, we can obviously find weights $w_i>0$ to
    represent the points $m_{e_i}$.
\end{proof}

The setting of Lemma \ref{lem:2} is shown by Fig.~\ref{fig:parallelogram}.
For any given quad mesh we therefore attach a weight to each vertex and define edge points on the
edges by Eq.~\eqref{eq:weights}.
It is convenient to represent weighted points in $\R^3$ by their homogeneous coordinates,
letting
\begin{align*}
    \wh v_i = (w_iv_i,w_i)\in\R^4.
\end{align*}
Given a weight $\lambda$, provided that $\lambda\ne 0$, any point $(x_1,x_2,x_3,\lambda)\in\R^4$
corresponds to the point $(\frac{x_1}{\lambda},\frac{x_2}{\lambda},\frac{x_3}{\lambda})$ $\in$ $\R^3$.
It is elementary that the edge points
$m_{e_i}$ are represented by homogeneous coordinate vectors
$\frac{1}{2}(\wh v_i + \wh v_{i+1}) \equiv \wh v_i + \wh v_{i+1}$. This
correspondence between vectors of $\R^4$ and points of $\R^3$
is illustrated by Fig.~\ref{fig:parallelogram}.

We think of the given quad mesh as approximating a smooth surface. For
each face $f$, the plane containing the edge points represents a tangent plane $\tau_f$.
Every face is thus naturally equipped with a unit normal vector $\nw_f$. It is natural
to define the contact point $b_f$ as the weighted center of mass of vertices:
\begin{align*}
    \wh b_f &= (\wh v_0 + \cdots +\wh v_3)/4  \implies
    \\b_f &= (w_0 v_0 + \cdots + w_3 v_3)/w_f, \quad
        w_f =w_0+w_1+w_2+w_3.
\end{align*}
Here $w_f$ is a weight associated with the face $f$.
It is easy to show that this contact point is the intersection of diagonals
$m_{e_0}m_{e_2}$ and $m_{e_1}m_{e_3}$ --- see Fig.~\ref{fig:parallelogram}.
Note that we consider all
unit normal vectors $\nw_f$ to be  consistently oriented
and pointing to one side of the mesh.

Lifting the mesh to $\R^4$ yields a checkerboard pattern in the
original sense of \cite{jiang-2020}, where each face is equipped with
an inscribed parallelogram (Fig.~\ref{fig:parallelogram}, left), defining
a tangent plane $\wh \tau_f$.
The given quad mesh together with its edge points is a checkerboard
pattern in $\R^3$ only if all vertex weights are equal.

\begin{figure}[t]
\begin{overpic}[width=0.5\linewidth]{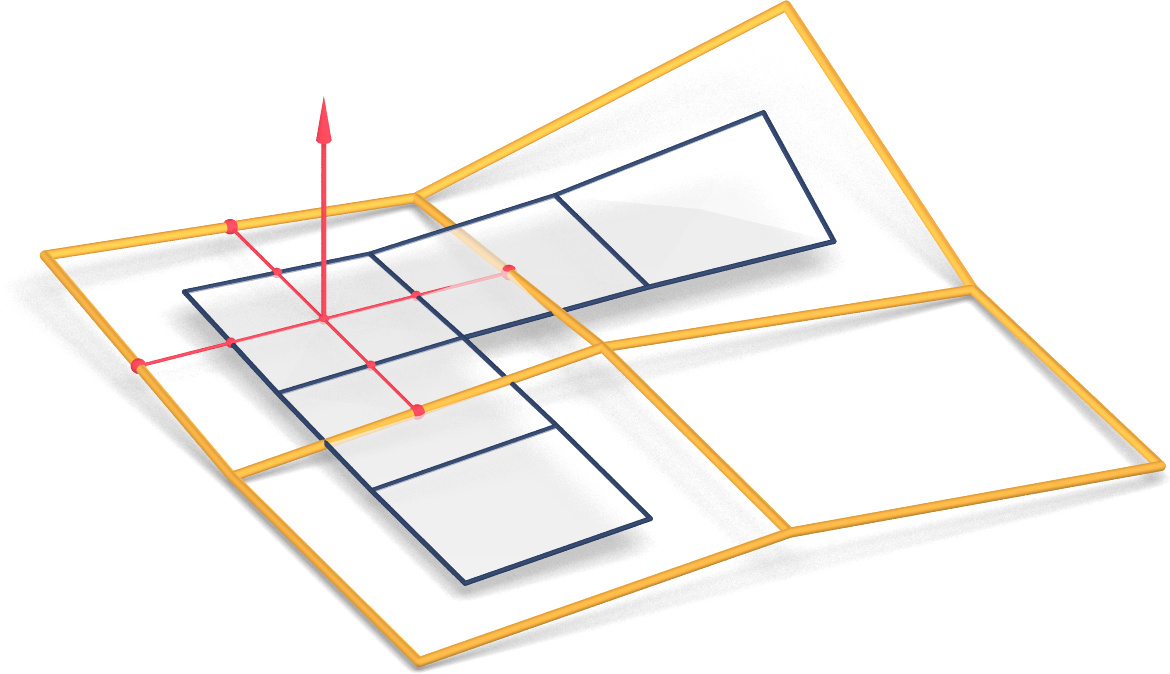}
  \put(0,4){$\R^3$}%
\color{gelb}%
  \lput(5,42){$v_0$}%
  \ltput(20,15){$v_1$}%
  \rtput(55,28){\contour{white}{$v_2$}}%
  \cput(33,45){$v_3$}%
  \color{blau}\footnotesize%
  \ltput(24,24){$v_{1,f}'$}%
  \lput(16,34){$v_{0,f}'$}%
  \color{rot}%
  \lput(31,54){$\nw_f$}
  \lput(32,35){\contour{white}{$b_f$}}
  \end{overpic}
\begin{overpic}[width=0.5\linewidth]{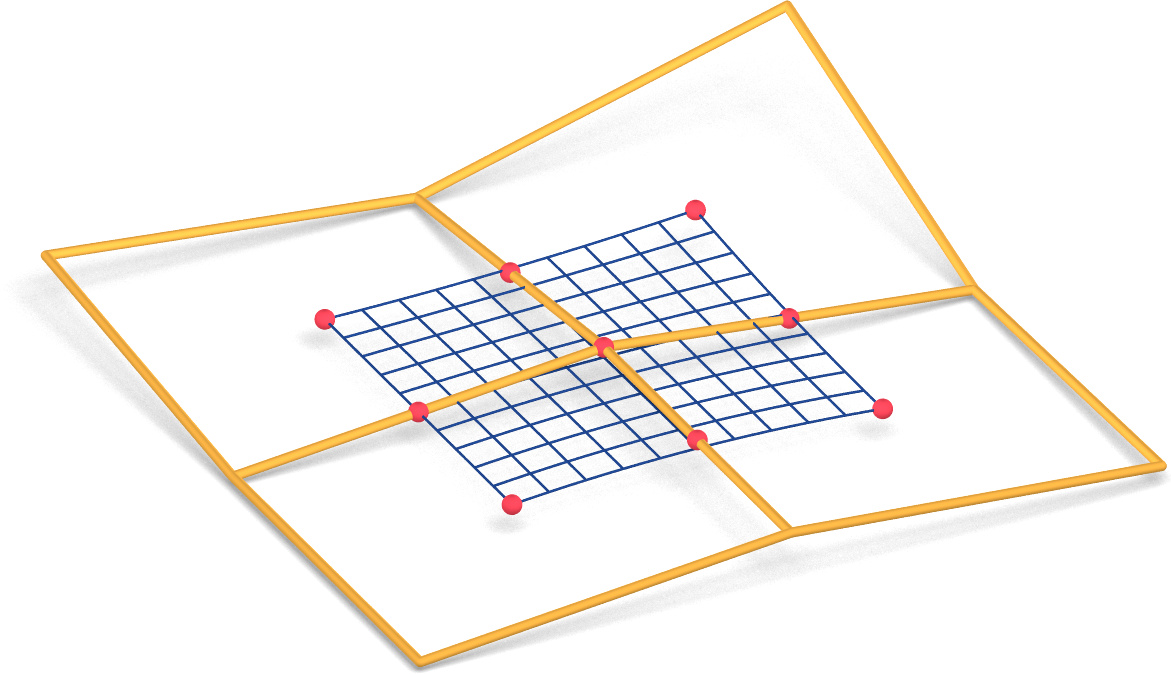}
  \put(0,4){$\R^4$}%
\color{gelb}%
  \lput(5,42){$\wh v_0$}%
  \ltput(20,15){$\wh v_1$}%
  \cput(33,45){$\wh v_3$}%
    \color{rot}%
      \rtput(55,28){\contour{white}{$\wh v_2$}}%
  \lput(32,35){\contour{white}{$\wh b_f$}}
  \end{overpic}\vskip-1ex
\caption{{\it Left:} Doo-Sabin refinement acting on weighted vertices interpolates
both the center $b_f$ of faces and the normal vectors $\nw_f$. {\it Right:} The limit surface
of the subdivision is a biquadratic rational spline surface composed of rational B\'ezier patches.
The control elements in $\R^4$ of the latter are derived from the homogeneous coordinates $\wh v_i$
of vertices, from their midpoints $(\wh v_i+\wh v_j)/2$, and from face midpoints, as shown in the figure.}
\label{fig:doosabin}
\end{figure}

\subsubsection{Subdivision of Contact Element Nets}
\label{sss:subdivision}
It is interesting that so-called {\em dual} quad-based subdivision rules
are able to refine contact elements in a natural way, yielding a
smooth limit surface. The well-known Doo-Sabin refinement rule even
interpolates contact elements, as follows. We are going to apply it
to homogeneous coordinate vectors $\wh v_i\in\R^4$.
For each vertex $\wh v_j$ of a quadrilateral face $f=(v_0v_1v_2v_3)$
we construct a new vertex
\begin{align*}
    \wh v_{j,f}'= \frac{1}{16}(9\wh v_j+3\wh v_{j+1}+3\wh v_{j-1}+\wh v_{j+2})
            \quad \text{(indices mod 4)}.
\end{align*}
In this way, each vertex, each edge, and also each face
of the original mesh is naturally associated with a cycle of new vertices,
yielding a new face each --- see Fig.~\ref{fig:doosabin}.
For the refinement rule for $n$-gons with $n\ne 4$ we refer to
\cite{peters:2008}.

For our purposes, the following observation is relevant: The center of mass
$\wh f=(\wh v_0\wh v_1\wh v_2\wh v_3)$ of a face is invariant
under subdivision:
\begin{align*}
    \wh b_f = \frac{1}{4} \sum \wh v_{i,f}' = \frac{1}{4} \sum \wh v_i.
\end{align*}
So are  diagonals in inscribed quads, which span the
tangent plane:
\begin{align*}
    \frac{1}{2} \Big( \frac{1}{2} (\wh v_0 + \wh v_1) - \frac{1}{2} (\wh v_2 + \wh v_3) \Big) =
    \frac{1}{2} (\wh v_{0,f}' + \wh v_{1,f}') - \frac{1}{2} (\wh v_{2,f}' + \wh v_{3,f}').
\end{align*}
An illustration is provided by Fig.~\ref{fig:doosabin}. Summing up, we
get:

\begin{prop} \label{prop:doosabin}
    The Doo-Sabin refinement scheme, acting on the
    homogeneous coordinate representation of a regular quad mesh,
    interpolates contact elements. In the limit, it produces a $C^1$-smooth
    spline surface of the rational biquadratic type,
    whose spline control points are the weighted
    vertices we started from. That spline surface   interpolates the contact
    elements of the original quad mesh.
\end{prop}

The nature of the limit referred to in
Prop.~\ref{prop:doosabin} is well known \cite{peters:2008}.
If the original mesh is not a regular quad mesh, the statement remains
true away from extraordinary points. Prop.~\ref{prop:doosabin}
is relevant because it shows how to construct a smooth surface
naturally connected to the data we work with. Note that
Doo-Sabin subdivision surfaces in their simple form as shown here do not
interpolate the boundary.

\begin{figure}[b]\centering
\begin{overpic}[width=.8\columnwidth]{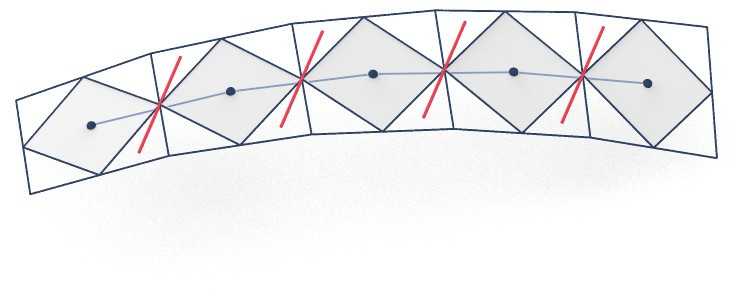}
    \cput(21,28){\contour{white}{$m_{e_0}$}}
    \cput(41,32){\contour{white}{$m_{e_1}$}}
	\ctput(24,17){$e_0$}%
	\ctput(42,19){$e_1$}%
	\ctput(62,20){$e_2$}%
	\ctput(80,18){$e_3$}%
	\color{blau}%
	\lput(14,26){\contour{white}{$b_{f_0}$}}%
	\ccput(31,31){\contour{white}{$b_{f_1}$}}%
	\color{rot}%
	\cput(25,36){\contour{white}{$r_{e_0}$}}%
	\cput(43,39){\contour{white}{$r_{e_1}$}}%
	\cput(63,40){\contour{white}{$r_{e_2}$}}%
	\cput(82,39){\contour{white}{$r_{e_3}$}}%
\end{overpic}\vskip-3em
\caption{{\it A discrete version of conjugacy}. This discrete version of
Fig.~\ref{fig:beruehrtorse} shows a strip of contact elements comprised of
faces $\{f_i\}$ with tangent planes $\{\tau_{f_i}\}$ and points of contact
$\{b_{f_i}\}$. It represents a
developable tangentially circumscribed to a surface
along the discrete curve $\{b_{f_i}\}$. In analogy to the smooth
case it is natural to define that discrete tangents $b_{f_{i+1}}-b_{f_i}$ and
intersections $r_{e_i}=\tau_{f_i}\cap\tau_{f_{i+1}}$ are conjugate.}
\label{fig:diskretetorse}
\end{figure}

\subsection{Developable Contact Element Nets}
\label{ss:discretedevelopables}
\subsubsection{Conjugacy and Fields of Rulings}

Consider a sequence of faces $f_0,f_1,\ldots$ which have
common edges $f_i\cap f_{i+1}$ --- see Fig.~\ref{fig:diskretetorse}. Each is
equipped with a tangent plane $\tau_{f_i}$.
We now perform a construction that is a discrete analog of the envelope of tangent planes shown by
Fig.~\ref{fig:beruehrtorse}: We construct the intersection lines of successive planes,
\begin{align*}
	r_{e_i}=\tau_{f_i} \cap \tau_{f_{i+1}}, \quad \text{where}\	e_i=f_i\cap f_{i+1}.
\end{align*}
The line $r_{e_i}$ passes through the edge point $m_{e_i}$ as defined by
Eq.~\eqref{eq:weights}.
The line $r_{e_i}$ it is a discrete ruling of the discrete envelope of
tangent planes along the discrete curve $b_{f_0},b_{f_1},\ldots$
\cite{sauer:1970}.
In analogy to the smooth case shown, we postulate:

\begin{defn} \label{defn:conjugacy}
  The discrete envelope of tangent planes $\tau_{f_0},\tau_{f_{1}},\ldots$
  along a discrete curve $b_{f_0},b_{f_1},\ldots$ is developable if
  discrete tangents $b_{f_{i+1}}-b_{f_i}$ and
  intersections $r_{e_i}=\tau_{f_i}\cap\tau_{f_{i+1}}$ are conjugate.
\end{defn}

The direction of the ruling $r_{e_i}$
is indicated by a {\em ruling vector} associated with an oriented edge
(\emph{half-edge}) $\vec e_i$. If $\vec e_i$ and $-\vec e_i$ are the two half-edges
corresponding to the edge $e_i=f_i\cap f_{i+1}$, we let
\begin{align}
\label{eq:rulingvector}
	\rw_{\vec e_i} = \nw_{f_i}\times \nw_{f_{i+1}}.
\end{align}
Here we assume that $f_i$ is to the left and $f_{i+1}$ to the right, and we
also  have $\rw_{-\vec e_i}=-\rw_{\vec e_i}$.

The vector $\rw_{\vec e_i}$ computed as a
cross product in Eq.~\eqref{eq:rulingvector} is zero if neighbouring
tangent planes $\tau_{f_i},\tau_{f_{i+1}}$ coincide. This happens
e.g. if the mesh is flat, or the strip under consideration is flat.

\begin{figure}[b!]
\hskip-1cm
\begin{overpic}[width=.8\columnwidth]{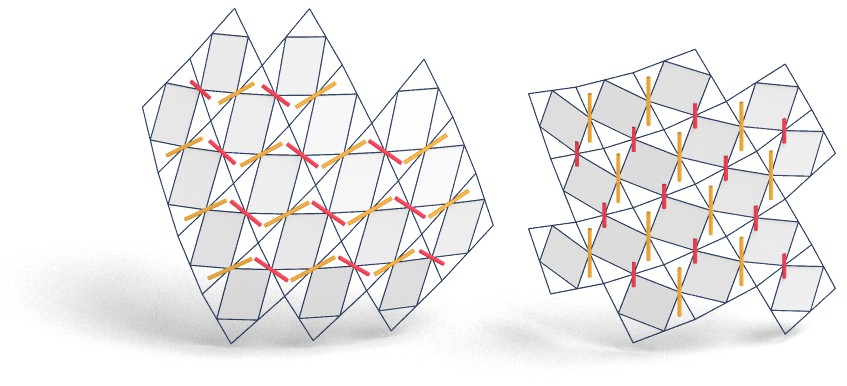}
\end{overpic}
\vskip-1em
\caption{{\it Left:}
On a general surface, the rulings associated with edges
correspond to two distinct
line fields (yellow and red).
{\it Right:} Developability is characterized by the
property that these line fields coincide.}
\label{fig:zweiflaechen}
\begin{minipage}[b]{.6\columnwidth}
    \caption{Characterization of developability involving
    the ruling vectors $\rw_{\vec e_0},\ldots,\rw_{\vec e_3}$ associated with
    the cycle of edges around a face $f$.}
    \label{fig:alternative}
\end{minipage}\hfill
\begin{minipage}[b]{.4\columnwidth}
    \begin{overpic}[width=\columnwidth]{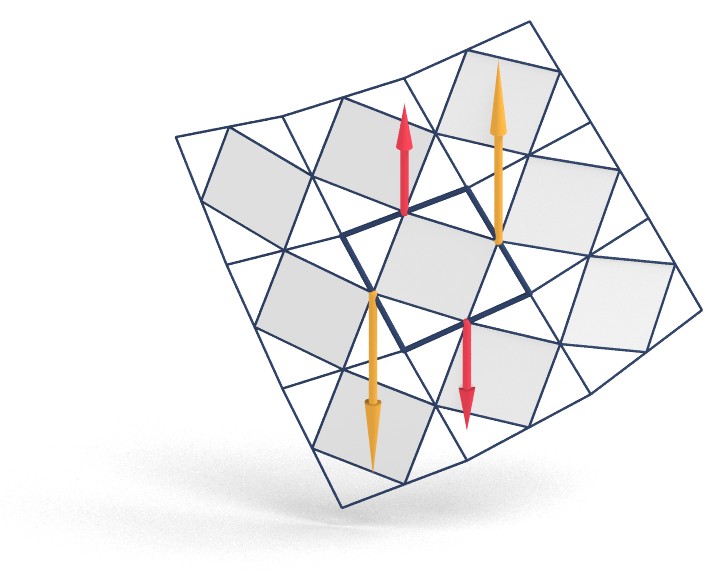}
    	\color{gelb}\ctput(51,12){\contour{white}{$\rw_{\vec e_0}$}}%
    	\color{gelb}\cput(68,74){\contour{white}{$\rw_{\vec e_2}$}}%
    	\color{rot}\cput(54,67){\contour{white}{$\rw_{\vec e_3}$}}%
    	\color{rot}\ctput(63,20){\contour{white}{$\rw_{\vec e_1}$}}%
    \end{overpic}
    \vspace*{-10mm}
\end{minipage}
\end{figure}

Figure \ref{fig:zweiflaechen} shows what happens when we assign a
ruling $r_e$ to {\em all} edges of a quad mesh. The rulings
$r_e$ associated with the edges of a mesh are usually not
samples of a continuous line field. However, in the case of a developable
mesh, they indicate the kernel of the 2nd fundamental form, so they
correspond to a single continuous line field.

This way of computing rulings amounts to numerical differentiation.
The definition of ruling vectors in Eq.~\eqref{eq:rulingvector} is
valid provided a fair mesh that discretizes a smooth surface.
We consider only such meshes where the edge polylines themselves are {\em fair}, so that they
could be interpreted locally as a discrete version of the parameter lines of
a $uv$ parametrization --- which is no restriction.

\subsubsection{Definition of Developable Quad Meshes}
\label{sss:developablemesh}

We express developability via the following property:
A surface is developable if and only if its
2nd fundamental forms do not have full rank \cite[p.~194]{docarmo:1976}. For a discrete
version of this statement, it is useful to associate part of
a 2nd fundamental form with each face of a mesh.

Consider a face $f=v_0v_1v_2v_3$ with its boundary cycle
$\vec e_0,\dots,\vec e_3$, assuming $e_i=f\cap f_i$.  We define
edge vectors $\ew_i= v_{i+1}-v_i$ (indices modulo $4$). We use
Eq.~\eqref{eq:rulingvector} to compute ruling
vectors $\rw_{\vec e_0},\ldots,\rw_{\vec e_3}$:
\begin{align*}
	\rw_{\vec e_i} = \nw_f\times \nw_{f_i}.
\end{align*}
Figure \ref{fig:alternative} shows this configuration. Note that
the ruling vectors associated with opposite edges point in the opposite
direction, like the edge vectors themselves.

We now partly define
the matrix $\II_f$ of a 2nd fundamental form
attached to the face center $b_f$. $\II_f$ governs
conjugacy, and in fact Def.~\ref{defn:conjugacy} already
states such a conjugacy relation {\em per edge}.
In order to formulate a
conjugacy condition {\em per face}, we define average edge vectors
\begin{align}
    \label{eq:edgevectors}
    \ew_{02} &= m_{e_2} - m_{e_0} = \frac{1}{w_{e_2}} (w_2v_2+w_3v_3) - \frac{1}{w_{e_0}} (w_1v_1+w_0v_0) \\
    \nonumber
    \ew_{13} &= m_{e_3} - m_{e_1} = \frac{1}{w_{e_3}} (w_0v_0+w_3v_3) - \frac{1}{w_{e_1}} (w_2v_2+w_1v_1),
\end{align}
and we postulate that these average edge vectors are conjugate
to average ruling vectors
\begin{align*}
    \rw_{\vec e_{13}} = \frac{1}{w_f} \Big(w_{e_1}\rw_{\vec e_1}+w_{e_3}(-\rw_{\vec e_3})\Big) \quad
    \rw_{\vec e_{02}} = \frac{1}{w_f} \Big(w_{e_0}\rw_{\vec e_0}+w_{e_2}(-\rw_{\vec e_2})\Big).
\end{align*}
The bilinear conjugacy relation per face then reads
\begin{align}
    \label{eq:II:implizit}
	\ew_{13}^T \cdot \II_f \cdot \rw_{\vec e_{13}}
	=\ew_{02}^T\cdot \II_f \cdot \rw_{\vec e_{02}}=0,
\end{align}
where $\II_f$ is the symmetric $2\times 2$ matrix of a discrete second fundamental
form associated with the face $f$.
Equations \eqref{eq:II:implizit} determine $\II_f$ uniquely up to a factor.

We now propose a definition of developability of quad meshes which uses the notions introduced above and discretizes several properties of
smooth developables simultaneously:

\begin{defn} \label{defn:developability}
A quad mesh is developable if for all faces the
average ruling vectors are parallel, i.e.,
\begin{align}
	\label{eq:cond:developability}
	\rw_{\vec e_{13}}\times \rw_{\vec e_{02}}=0.
\end{align}
\end{defn}

This expresses the fact that each face is equipped with a single ruling direction,
where rulings arise as intersections of neighboring tangent planes. Further,
Eq.~\eqref{eq:cond:developability} implies that the conjugacy relation
\eqref{eq:II:implizit} is degenerate, because now two different average edge
vectors are conjugate to the same ruling direction. It follows that
\begin{align*}
	\det(\II_f)=0.
\end{align*}
A further equivalent condition is
$\det(\rw_{\vec e_{13}},\rw_{\vec e_{02}},\nw_f)=0$.

\begin{figure}[b]
    \begin{minipage}[b]{.5\columnwidth}
        \Inc[width=\columnwidth]{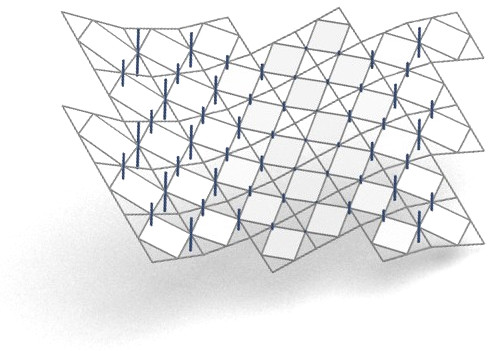}
        \vspace*{-10mm}
    \end{minipage}\hfill
    \begin{minipage}[b]{.45\columnwidth}
        \caption{This developable has an inflection ruling visible
        as the zero level set of the length of ruling vectors. Ruling
        vectors are here shown without arrows.}
        \label{fig:wendeerzeugende}
    \end{minipage}
\end{figure}

Ruling vectors might vanish for several reasons:
(i) In planar parts of a developable, rulings vectors $\rw_{\vec e_i}$
vanish and Eq.~\eqref{eq:cond:developability} is fulfilled automatically.
The same is true if the developable has an {\em inflection ruling}:
Fig.~\ref{fig:wendeerzeugende} shows that already close to the inflection
ruling, vectors $\rw_{\vec e_i}$ approach zero. (ii) An individual ruling
vector $\rw_{\vec e_i}$ vanishes if and only if the normal vectors
$\nw_f,\nw_{f_i}$ are parallel. This situation
happens if faces $f,f_i$ are arranged along a ruling
contained in the developable mesh; similar to the previous
special cases, it is consistent with
Eq.~\eqref{eq:cond:developability}.

\begin{remark}
In case vertex weights are equal,
Eq.~\eqref{eq:cond:developability} has interesting
consequences for the Gauss image. We illustrate this in the generic
case (away from an inflection ruling), where a developable locally is convex.
Definition \ref{defn:developability} is expressed by
the condition that
\begin{align*}
	\nw_f\times(\nw_{f_1} - \nw_{f_3}), \quad
	\nw_f\times(\nw_{f_0}-\nw_{f_2})\quad
	\text{are parallel}.
\end{align*}
\par\noindent\hangindent=-.25\columnwidth\hangafter=-4
All vectors involved here approximately lie in a plane orthogonal to
$\nw_f$ (namely the unit sphere's tangent plane in $\nw_f$). If the
quad $\nw_{f_0},\ldots,\nw_{f_3}$ were planar, then
Eq.~\eqref{eq:cond:developability} would express the fact that
diagonals are parallel (see the inset figure; diagonals are yellow).
This characterizes quads of zero oriented area,
and is nicely consistent with the fact that the Gauss image of
a developable is curve-like and has zero area.
\end{remark}

\AddToShipoutPictureFG*{
    \put(\LenToUnit{0.435\paperwidth},\LenToUnit{0.68\paperheight}){\vtop{{\null}\makebox[0pt][c]{\begin{overpic}[width=.18\columnwidth]{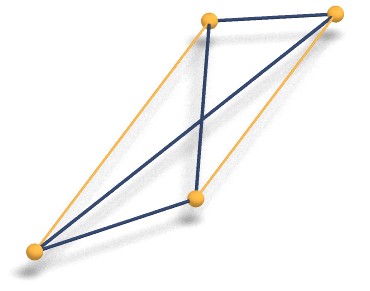}
	\color{gelb}\small
	\lput(10,20){$\nw_{f_0}$}
	\lput(110,83){$\nw_{f_1}$}
	\lput(52,75){$\nw_{f_2}$}
	\rcput(55,10){$\nw_{f_3}$}
	\end{overpic}}}}%
}%

\paragraph{Projective Transformations}
Developability is well known to be invariant under
projective transformations. This property is numerically
verified by Fig.~\ref{fig:remeshing}, and is in part reflected in
our approach using homogeneous coordinates (in this way
projective transformations are simply expressed as
linear mappings). Our constructions up to and including the computation
of the ruling line fields are projectively invariant.
Only the developability condition of Def.~\ref{defn:developability}
itself, which checks equality of ruling line fields, is not.
We employ such a projective transformation later --- see Fig.~\ref{fig:paraboliccylinder}.

\begin{figure}[b]
    \Inc[width=.40\columnwidth]{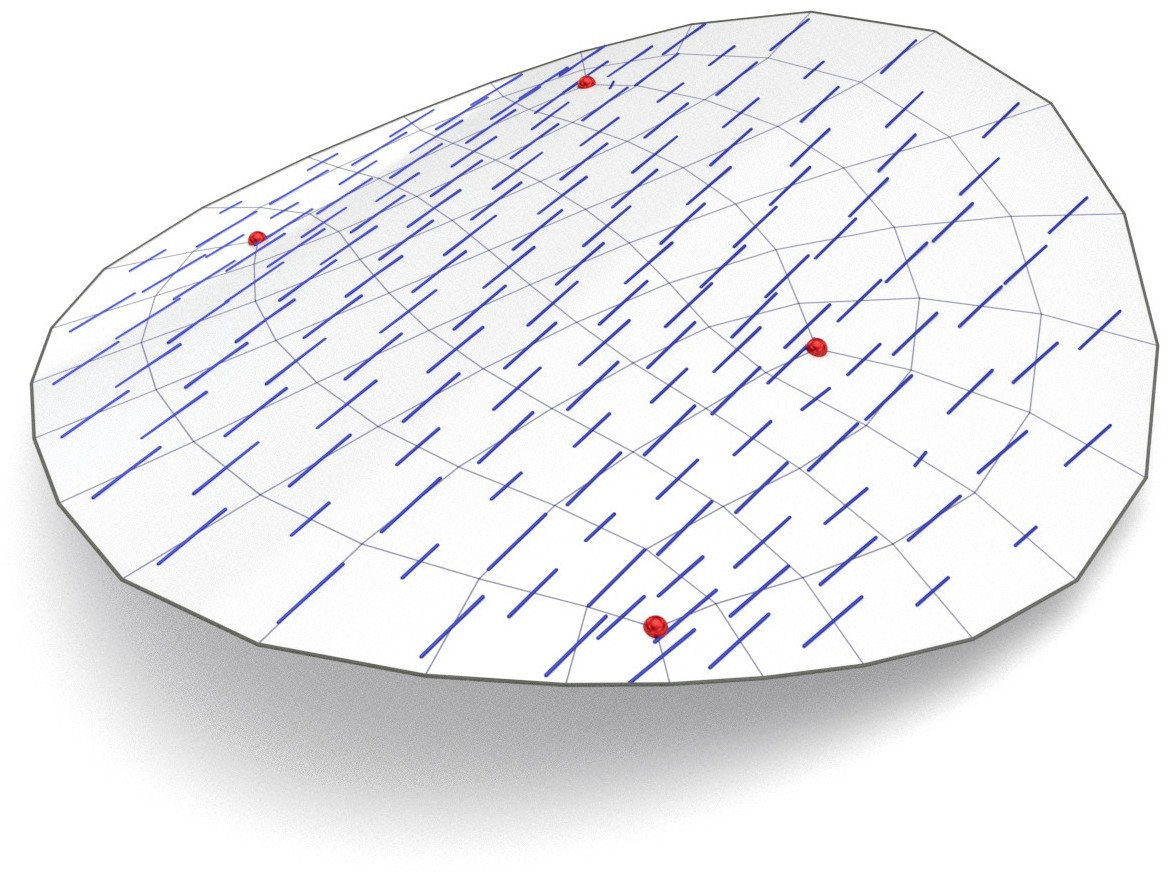}\hfill
    \raise.03\columnwidth\hbox{\Inc[width=.55\columnwidth]{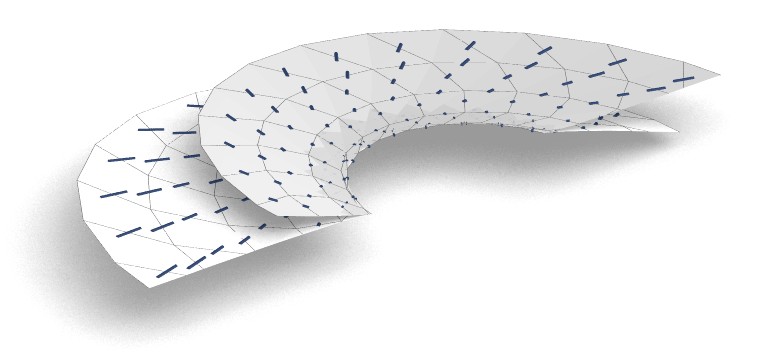}}
    \vskip-3ex
    \caption{{\em Singularities.} {\it Left:} We show a discrete-developable mesh
    and the good behaviour of ruling vectors $\rw_{\vec e}$ in
    the vicinity of combinatorial singularities (red). {\it Right:}
    This mesh exhibits a geometric singularity typical for developables, namely
    a sharp curve of regression. The singularity is not detectable via
    checking fairness of mesh polylines and is made visible by clipping
    by a plane. The mesh fulfills the developability
    condition of Def.~\ref{defn:developability}.}
    \label{fig:sing}
\end{figure}

\paragraph{Singularities of Developable Quad Meshes}
Developables can exhibit geometric singularities like a cone's
vertex or the line of regression in a torsal ruled
surface. The developability criterion of Def.~\ref{defn:developability}
does not prevent singularities from emerging, and in fact, extremely
singular meshes formed by the tangents of a curve may
well enjoy discrete developability. We prevent
such singularities by fairness imposed on the normal vector field
$\nw_f$. Combinatorial singularities do not pose a problem,
since also in this case the geometric interpretation
of Eq.~\eqref{eq:cond:developability} remains valid --- see Fig.~\ref{fig:sing}.

\begin{figure}[t]\fboxsep.2pt\fboxrule.2pt\flushleft\small
    \centerline{\begin{overpic}[width=.5\columnwidth]{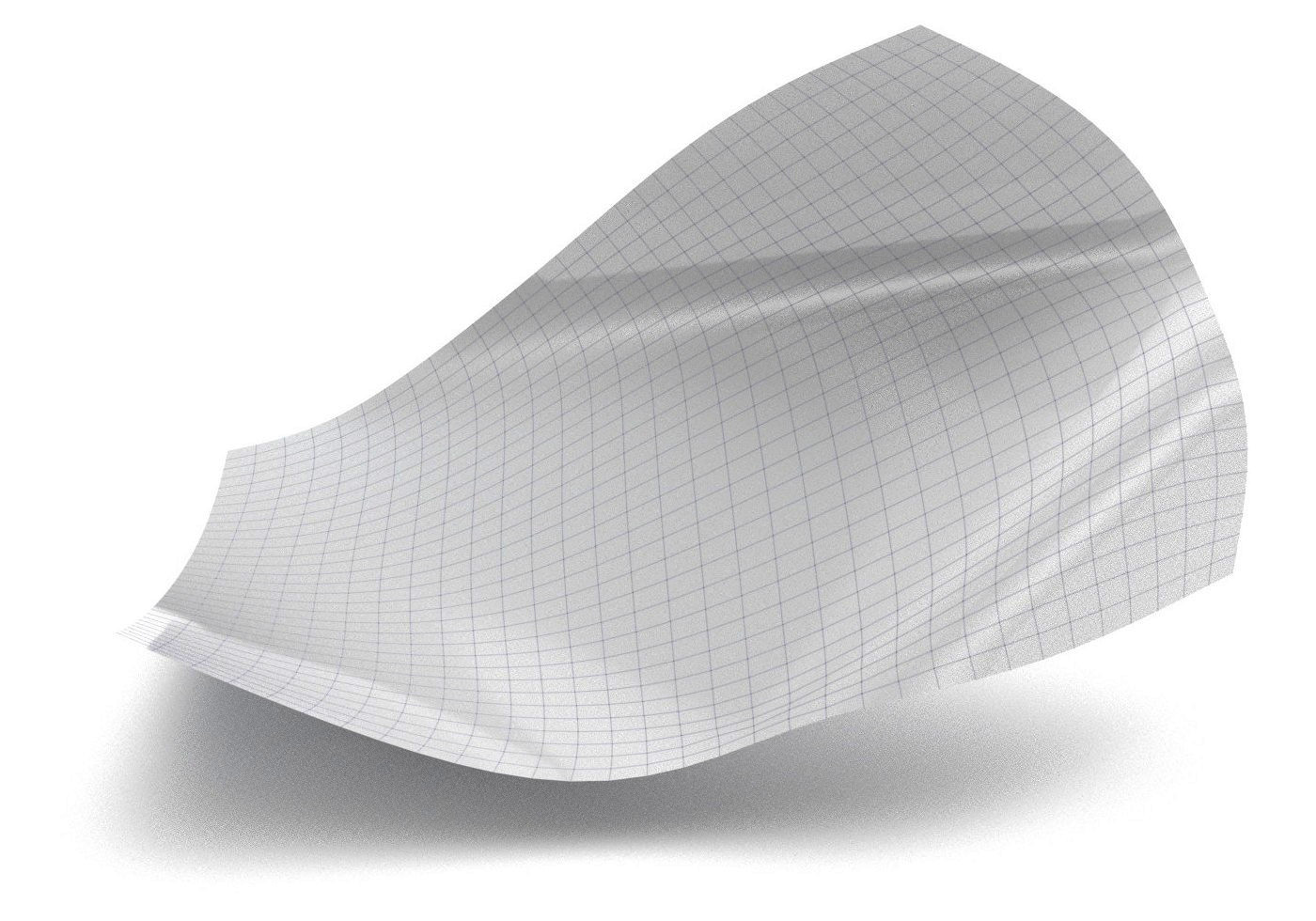}
    \put(0,10){(a)}
    \end{overpic}\hfill
    \begin{overpic}[width=.5\columnwidth]{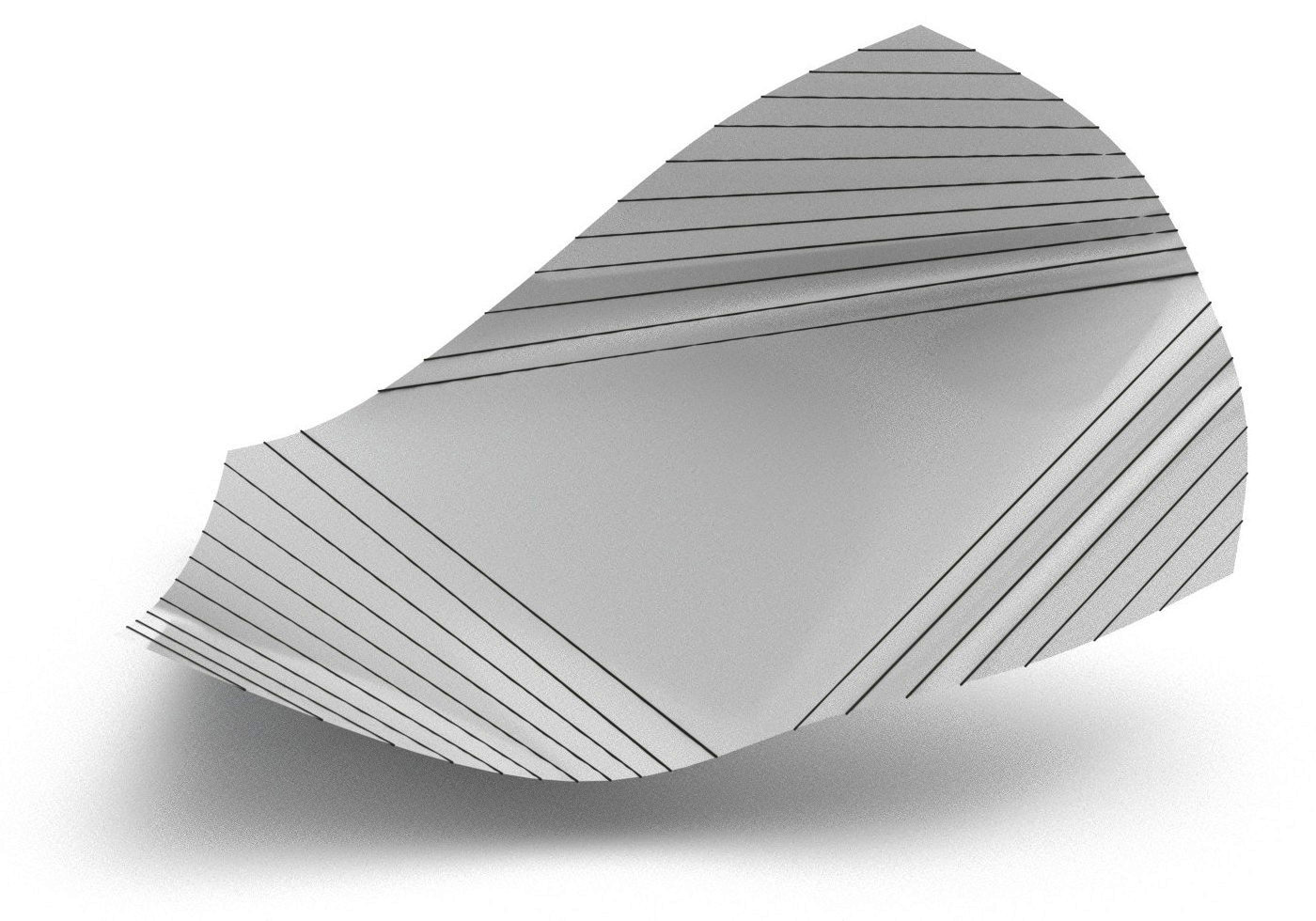}
    \put(0,10){(b)}
    \end{overpic}}
    ~\hfill\raise1ex\hbox{(c)}\Inc[width=.20\columnwidth]{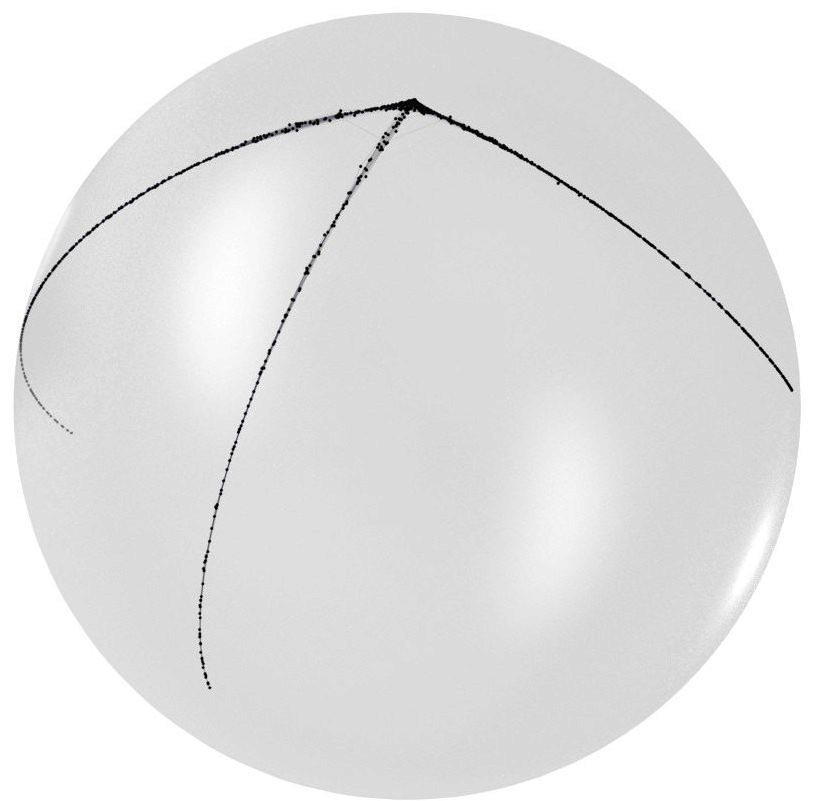}\hfill
    \raise1ex\hbox{(d)}\quad {\scalebox{0.9}{\fbox{\clipbox{0mm 7mm 0mm 10mm}{\Inc[height=3.5cm]{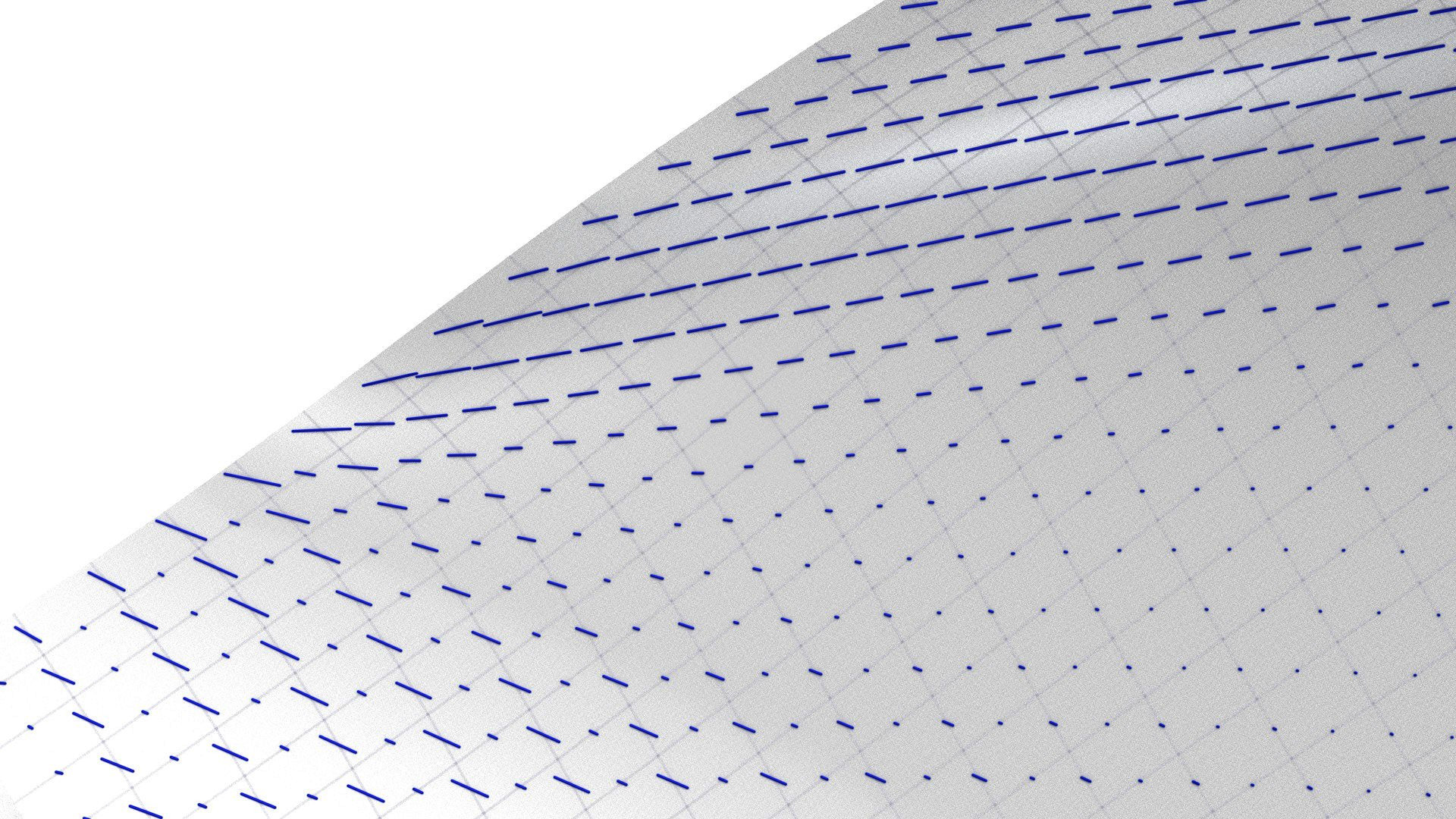}}}}}
    \caption{A simple developable computed with our method. (a)
    An optimized quad mesh exhibiting discrete developability. (b)
    Rulings found by integrating the ruling vector field. (c)
    The Gauss image consisting of face normal vectors $\nw_f$. (d)
    Detail of the  discrete ruling vector field $\rw_{\vec e}$.}
    \label{fig:simple}
\end{figure}

\subsection{Computation}
\label{ss:computation}

All our computations are based on an optimization procedure
which achieves constraints
and yields low values of fairness functionals. We operate with a
quad mesh $(V,E,F)$ which is thought to follow the parameter lines of a smooth
surface. We have
regular grid combinatorics except for isolated singularities, and we assume
that we have
a consistent orientation of faces.
Our variables are vertices $v_i$, vertex weights $w_i$, re-weighted
vertices $\wt v_i=w_iv_i$, a unit normal vector $\nw_f$ of each face $f\in F$,
appropriately re-weighted edge vectors $\wt\ew_{02,f},\wt\ew_{13,f}$ per face,
a ruling vector $\rw_{\vec e}$ for each oriented edge $\vec e\in E$, and re-weighted ruling vectors $\wt\rw_{02,f},\wt\rw_{13,f}$ per face.

Constraints involving vertices are
\begin{align*}
    c_{\vert,1}(v):=\wt v_i-w_iv_i=0, \ \ \
    c_{\vert,2}(v):=w_i-\omega_i^2-1.0=0.
\end{align*}
We use one dummy variable $\omega_i$ per weight to ensure that
weights remain above the threshold $1.0$. The precise value of the threshold is not relevant because all other constraints are
homogeneous.
Average edge vectors per face, in the notation
used by Eq.~\eqref{eq:edgevectors}, are defined by the constraints
\begin{align*}
  c_{\ev,1}(f) &:= \wt\ew_{02,f}
        - \big((w_0+w_1)(\wt v_2+\wt v_3) -
            (w_2+w_3)(\wt v_1+\wt v_0)\big) = 0,
            \\
   c_{\ev,2}(f) &:= \wt\ew_{13,f}
        - \big((w_1+w_2)(\wt v_0+\wt v_3)
            -(w_0+w_3)(\wt v_2+\wt v_1)\big)=0
\end{align*}
Vectors $\wt\ew_{02,f},\wt\ew_{13,f}$
are the previously defined average edge vectors $\ew_{02},\ew_{13}$,
multiplied by a combination of weights which makes denominators vanish.

The normal vector $\nw_f$
is initialized according to Fig.~\ref{fig:parallelogram},
while taking the orientation into account. Our average edge
vectors are precisely the diagonals in the inscribed quad depicted
in Fig.~\ref{fig:parallelogram}. We therefore use the following
constraints to handle normal vectors:
\begin{align*}
	c_{\norm ,1}(f)
		& := \<\nw_f, \wt\ew_{02,f}\>  = 0,\quad
	c_{\norm ,2}(f)
		 := \<\nw_f, \wt\ew_{13,f} \>  = 0,
	\\
	c_{\norm ,3}(f)
		& := \|\nw_f\|^2 - 1 = 0.
\end{align*}
These normal vectors are recomputed after each round of optimization,
since we cannot be sure that the implicit conditions above are sufficient to
keep a consistent orientation.
For every oriented half-edge $\vec e=v_iv_j$,
we have a ruling vector $\rw_{\vec e}$ defined by the constraint

\noindent
\begin{align*}
	c_{\rul}(\vec e) := \rw_{\vec e} - \nw_f\times\nw_{f'} = 0,
	\quad \text{where}\ \vec e = f\cap f',
\end{align*}
with $f$ to the left and $f'$ to the right of the half-edge $\vec e$.
We define re-weighted ruling vectors per face by the constraints
\begin{align*}
    c_{\rul,1}(f)&:=\wt\rw_{13,f}-\big((w_1+w_2)\rw_{v_1v_2}+(w_3+w_0)(-\rw_{v_3v_0})\big)=0,\\
    c_{\rul,2}(f)&:=\wt\rw_{02,f}-\big((w_0+w_1)\rw_{v_0v_1}+(w_2+w_3)(-\rw_{v_2v_3})\big)=0.
\end{align*}
Developability is expressed by
Eq.~\eqref{eq:cond:developability}:
	\begin{align*}
	c_{\dev}(f) :=
	\wt\rw_{13,f}\times \wt\rw_{02, f}=0.
	\end{align*}
The sums of squares of the constraints define energy functionals,
$E_{\vert} = \sum\nolimits_{v\in V,j} c_{\vert,j}(v)^2$,
$E_\norm   = \sum\nolimits_{f\in F,j} c_{\norm,j}(f)^2$,
and analogously for energies $E_\ev$, $E_\rul$, $E_\dev$.

To ensure that the mesh polylines approximate smooth parameter lines
of a surface, we employ fairness functionals: We define
\begin{align*}
	E_\fair^V = \sum\nolimits_{\text{triples}\,v_iv_jv_k}
		\|v_i-2v_j+v_k\|^2,
\end{align*}
where the sum is over all triples $v_iv_jv_k$
of successive vertices on a discrete parameter polyline. Likewise we
measure the fairness of the  normal field and ruling fields by energies
\begin{align*}
	E_\fair^n = \hspace*{-1.5em}\sum\limits_{\text{triples}\,f_if_jf_k}
    \hspace*{-1em}
		\|\nw_{f_i}-2\nw_{f_j}+\nw_{f_k}\|^2,
   E_\fair^r = \hspace*{-1.5em}\sum\limits_{\text{triples}\,\vec e_i,\vec e_j,\vec e_k}
   \hspace*{-1em}
		\|\rw_{\vec e_i}-2\rw_{\vec e_j}+\rw_{\vec e_k}\|^2.
\end{align*}
The sum in $E_\fair^n$ is over all triples of successive faces arranged in a strip
like shown in Fig.~\ref{fig:diskretetorse}. We found that imposing
fairness on the normal vector field prevents singularities like the
one in Fig.~\ref{fig:sing}, right. The sum in $E_\fair^r$ is over triples
of successive half-edges.

Summing up, in our optimization we minimize the functional
\begin{align}
    \label{eq:totalenergy}
    \nonumber
	E = E&_{\norm} + \lambda_{\vert} E_{\vert} + \lambda_{\ev} E_{\ev} + \lambda_{\rul} E_{\rul} + \lambda_{\dev} E_{\dev} \\
          &+ \lambda_{\fair}^V E_{\fair}^V + \lambda_{\fair}^n E_{\fair}^n + \lambda_{\fair}^r E_{\fair}^r.
\end{align}
The weights $\lambda_{\vert},\lambda_{\rul},\ldots$ have to be chosen according
to the particular application --- see Table~\ref{tab:stat}.
In the last steps of the iteration, weights of terms used for
regularization are set to $0$.  This enables $E_{\dev}$ to approach zero
itself. Table~\ref{tab:stat} refers to the status immediately before.

\begin{figure}[b]
    \Inc[width=\columnwidth]{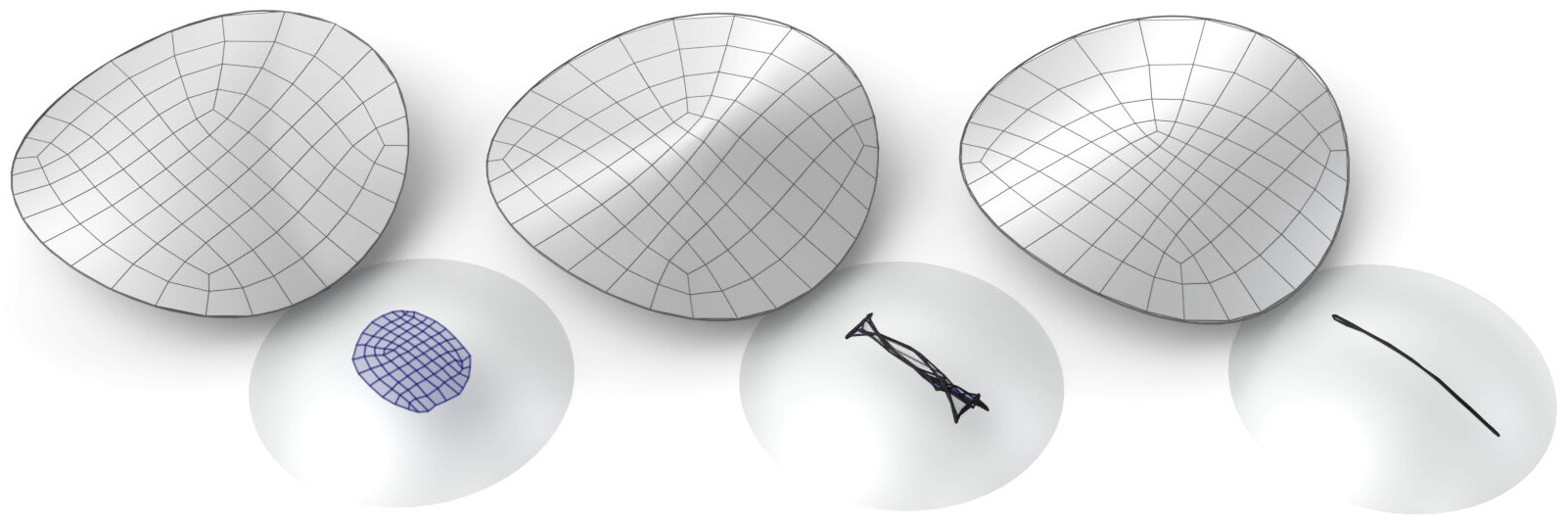}\vskip-1ex
    \caption{{\it The influence of vertex weights $w_j$}.  A coarse mesh (left) has been
    optimized to become developable, setting all point weights to $1$
    (center) and with  weights as variables (right). Gauss images show that
    the additional degrees of freedom provided by the weights have
    a beneficial influence.}
    \label{fig:weights:better}
\end{figure}

\begin{remark}
Both the definition of discrete developability and the optimization setup are
much simplified if the vertex weights are equal. In such case, we can, without
loss of generality, simply let $w_i=1$ for all $i$. Our choice of
variables for optimization is guided by the empirical rule that the polynomial
degree of constraints should not exceed 2. If all weights equal $1$, the
number of variables is reduced not only because the weights themselves are
constants now, but we also would not need edge vectors as variables --- they could
be replaced by a linear combination of vertices. Similarly, the average
face ruling vectors could be replaced by a difference of edge ruling vectors.
Figure \ref{fig:weights:better} shows that
the additional degrees of freedom provided by the weights have
a beneficial influence, which is particularly visible for coarse meshes.
\end{remark}

\begin{figure}[t]
    \Inc[width=\columnwidth]{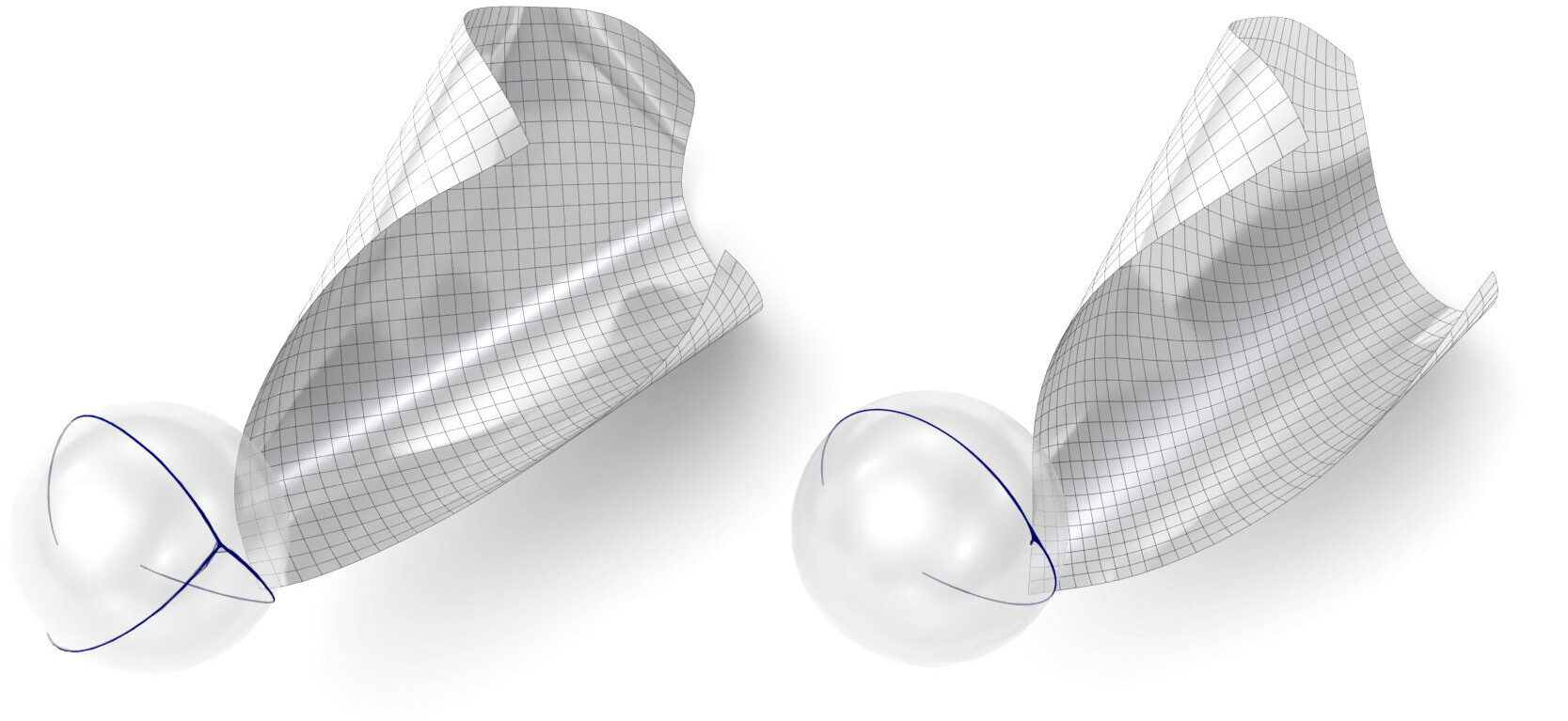}
    \vskip-1em
    \caption{{\em The influence of ruling fairness}. The left- and right-hand
    images show results of optimization with ruling fairness
    disabled ($\lambda_\fair^r=0$) resp., enabled.
    The effect is that the resulting developable does not as easily
    decompose into several ruled pieces.}
    \label{fig:rulingsfairness}
\end{figure}

\subsection{Approximation Power of Discrete Developables}

We argue that contact element nets as introduced in \S\,\ref{ss:discretedevelopables}
are a suitable discretization for developable surfaces, as they
are capable of representing developables up to 2nd order approximations.

Locally, a surface $\Psi$ is represented as a height field $z=f(x,y)$.
We approximate $f$ with a 2nd order Taylor polynomial, which yields
an approximating surface $\Phi$ defined by
\begin{align*}
    z = g(x,y) = f(0,0) + \nabla f(0,0)^T \binom{x}{y} + \frac{1}{2} (x \;\; y) \nabla^2 f\binom{x}{y}.
\end{align*}
If $\Psi$ is developable, then
$\det\nabla^2 f=0$ \cite[p.~163]{docarmo:1976}. It follows that $\Phi$ is a parabolic cylinder.

Interestingly, there are many contact element nets that are
developable in the sense of Def.~\ref{defn:developability}, and
whose associated B-spline surface reproduces the above-mentioned
cylinder $\Phi$. The construction is the following:

\begin{figure}[b]
    \flushleft\fboxsep0pt\fboxrule.1pt
    \scalebox{0.71}
    {{\clipbox{4mm 0 5mm 2mm}{\begin{overpic}
            [width=50mm]{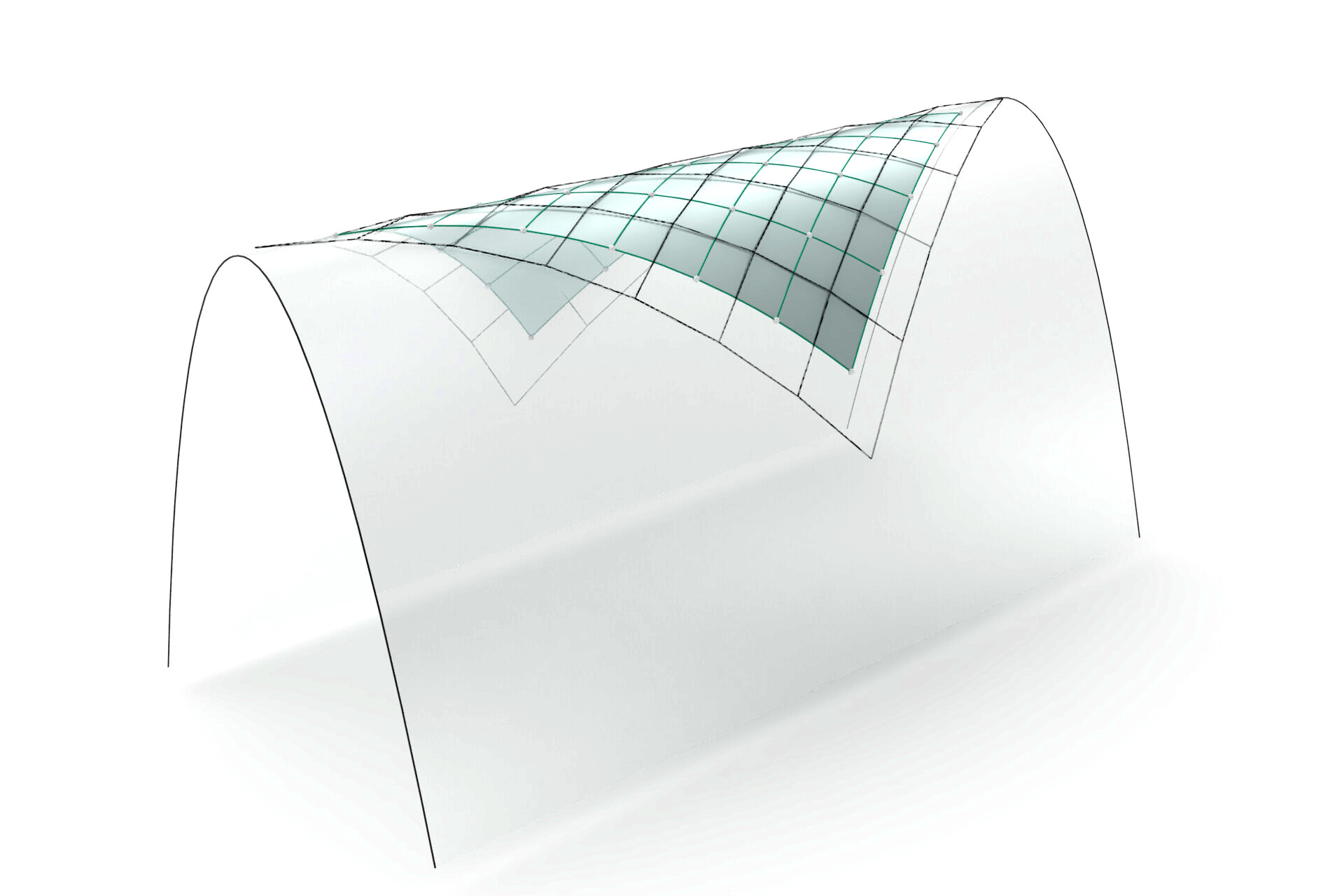}
            \put(40,20){$\Phi_1$}
            \end{overpic}}}\relax
    {\clipbox{7mm 0 7mm 2mm}{\begin{overpic}
            [width=50mm]{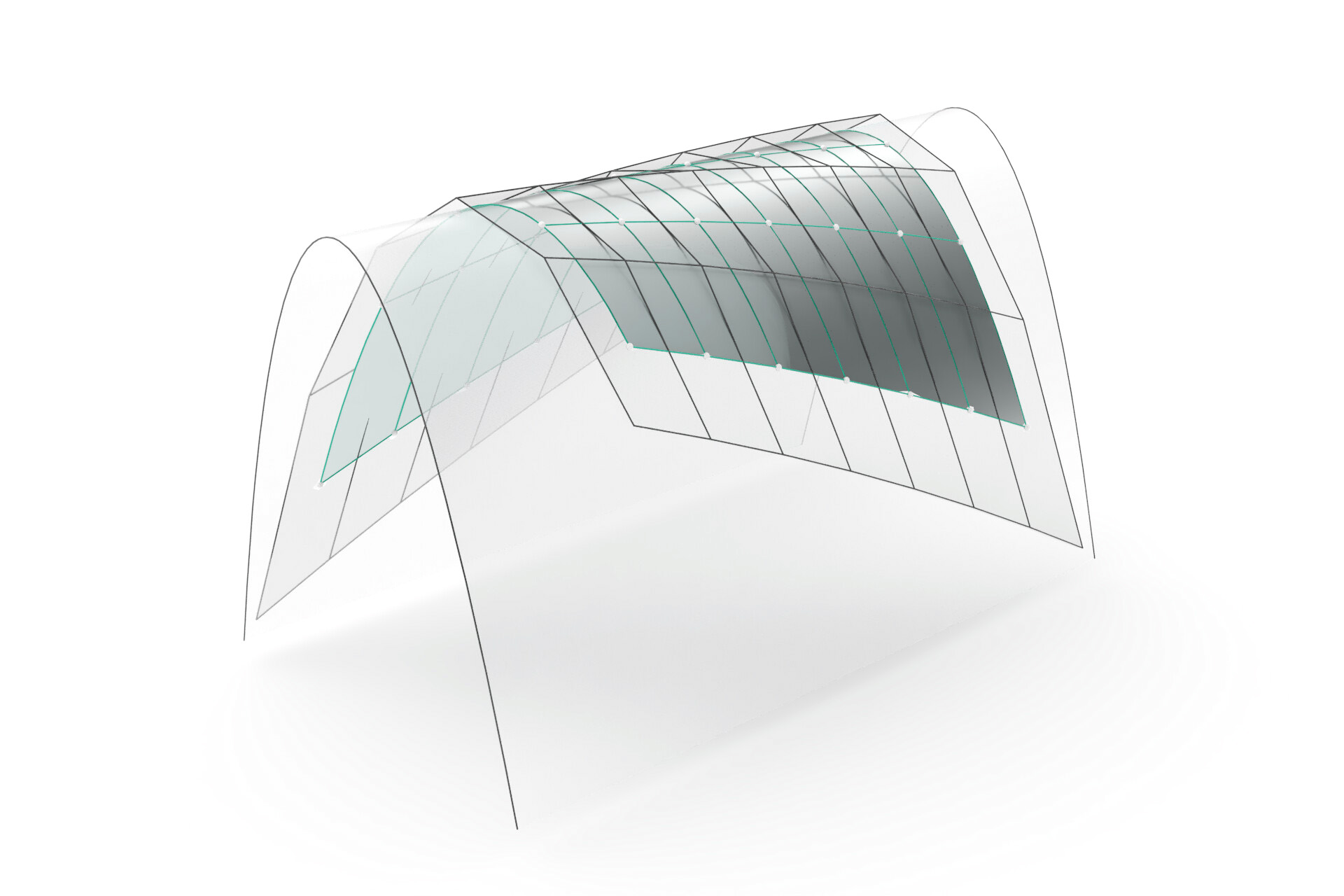}
            \put(40,20){$\Phi_1$}
            \end{overpic}}}\relax
    {\clipbox{2mm 0 6mm 2mm}{\begin{overpic}
            [width=50mm]{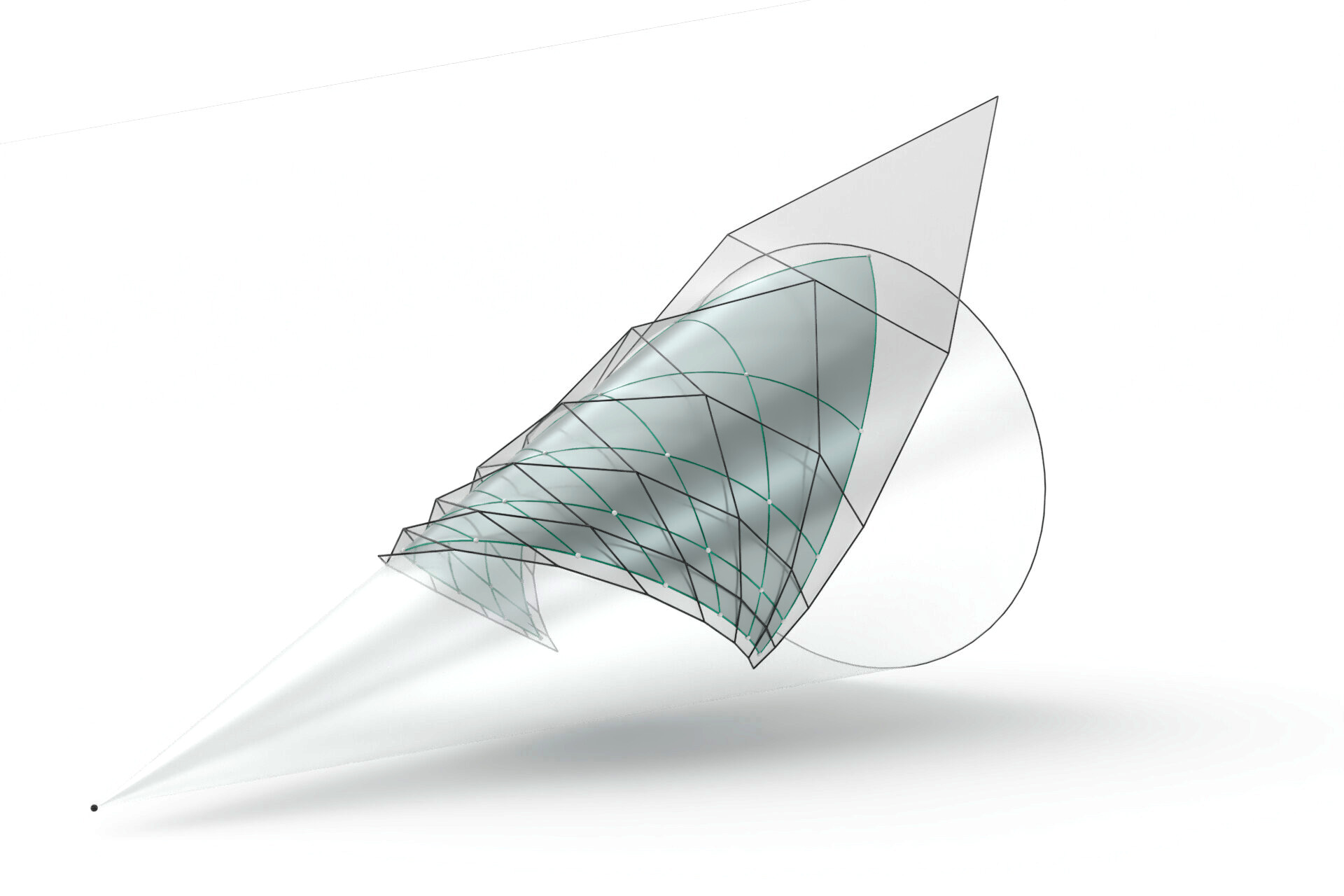}
            \put(25,12){$\Phi_2$}
            \end{overpic}}}}
    \vspace*{-0.5em}
    \caption{A projective transformation maps the parabolic cylinder $\Phi_1$
    to a quadratic cone $\Phi_2$. $\Phi_1$ has many exact representations
    by a discrete-developable quad mesh and associated
    biquadratic spline surface (left and center). The projective
    transformation maps spline control points of
    $\Phi_1$ to weighted spline control points of $\Phi_2$, leading
    to a quadratic NURBS representation. This low degree is possible
    with polynomial (un-weighted) splines only in special cases, namely,
    if parameter lines are rulings.}
    \label{fig:paraboliccylinder}
\end{figure}


\begin{figure*}[t!]
    \Inc[width=\textwidth]{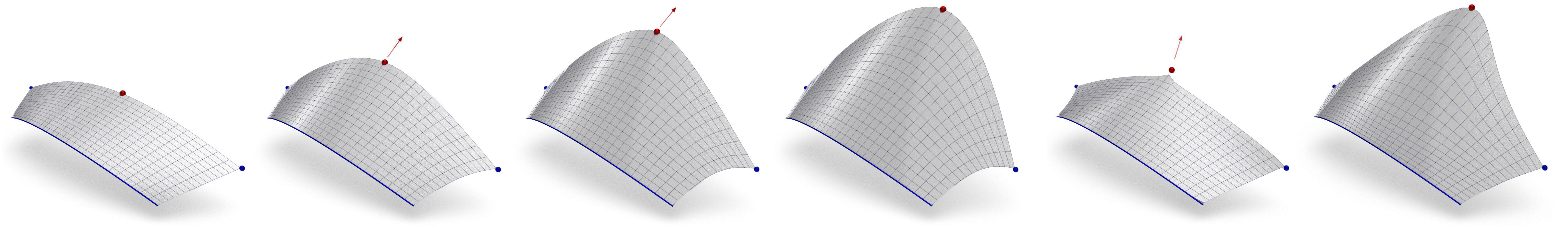}\vskip-6ex
    \leftline{\hspace*{.01\textwidth}\small
    	(a)\hskip.15\textwidth
    	(b)\hskip.15\textwidth
    	(c)\hskip.16\textwidth
    	(d)\hskip.15\textwidth
    	(e)\hskip.15\textwidth
    	(f)}
    \caption{{\it Interactive editing}. The sequence of images (a)--(d) shows
    a developable patch being interactively modified by a user who keeps
    one boundary segment and two corner vertices (blue) fixed, and is dragging
    on another boundary vertex (red). Images (e) and (f) show the influence
    of the weight we give to $E_\iso$ in our optimization. While
    in subfigures (a)--(d) we use isometry to the previous step for
    its regularizing effect, in (e) we employ isometry to the original
    patch. This is a constraint that is not compatible with the user's
    desire to move the red vertex. Subfigure (f) is similar to (a)--(d),
    but with a lower weight of $E_\iso$.}
    \label{fig:material-behaviour}
\end{figure*}

\begin{figure*}[t!]\fboxsep0pt\fboxrule.1pt
    \scalebox{0.73}{\relax
    {\clipbox{5mm 2mm 2mm 7mm}{\Inc[height=4.5cm]{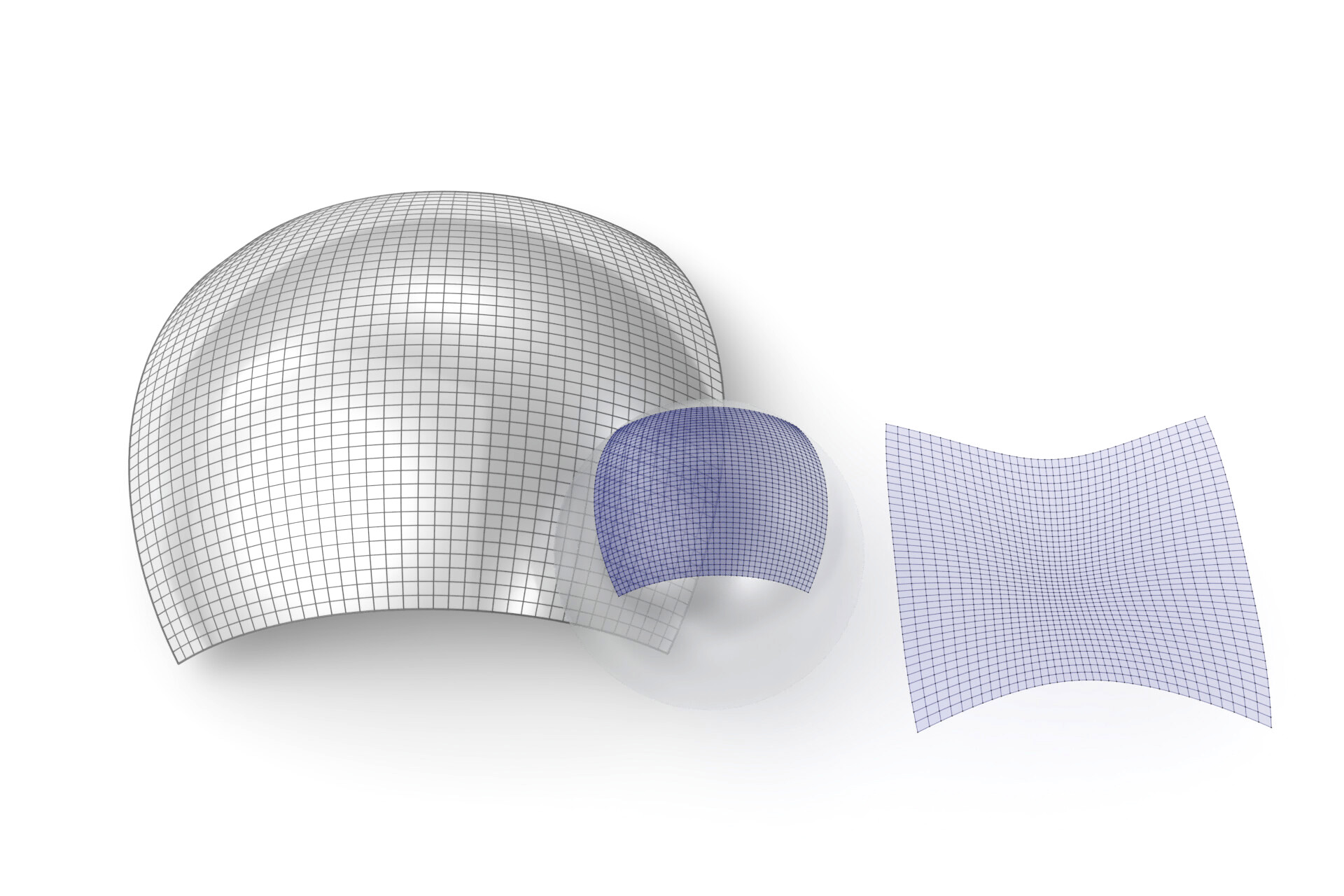}}}%
    {\clipbox{4mm 2mm 1mm 7mm}{\Inc[height=4.5cm]{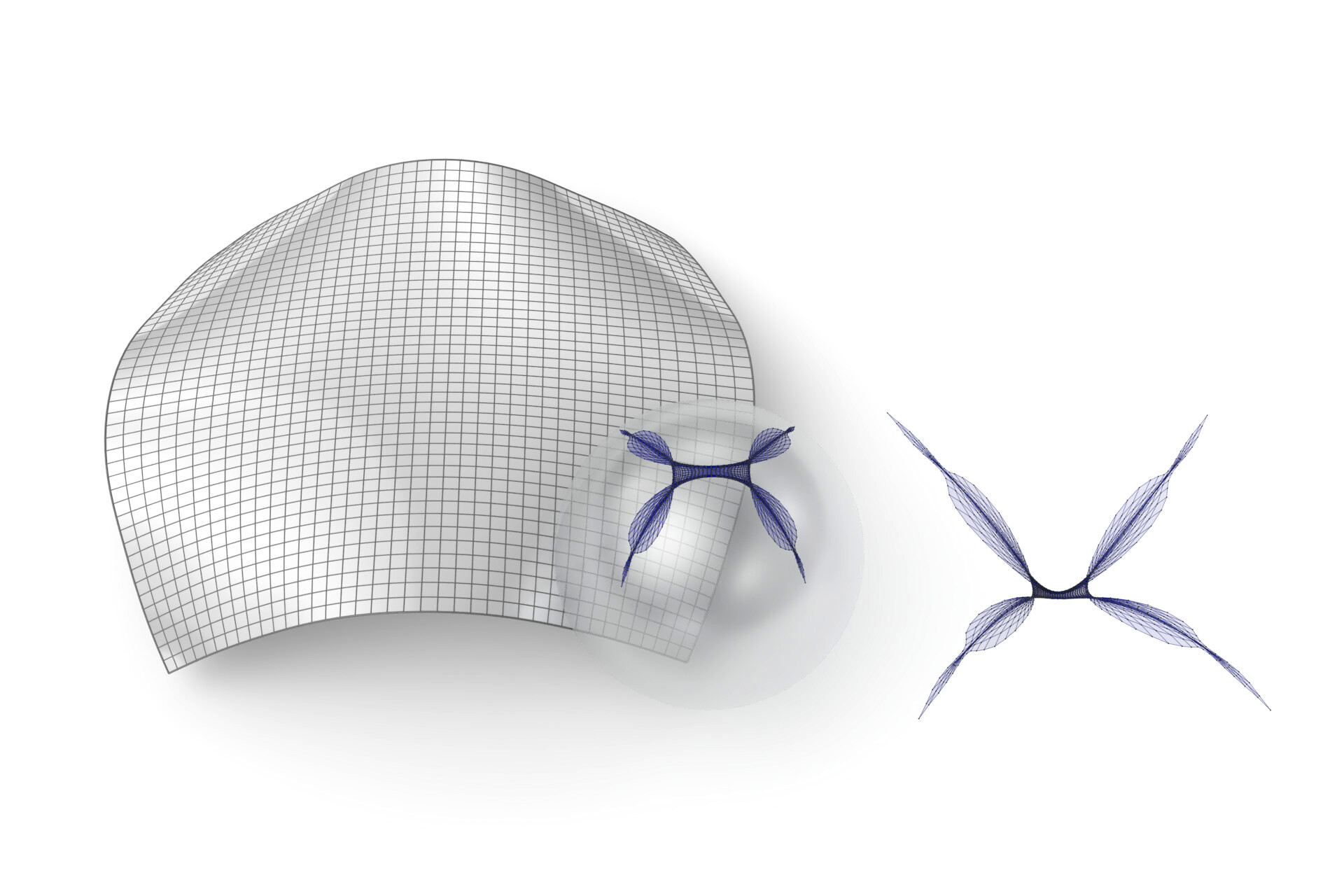}}}%
    {\clipbox{5mm 2mm 1mm 7mm}{\Inc[height=4.5cm]{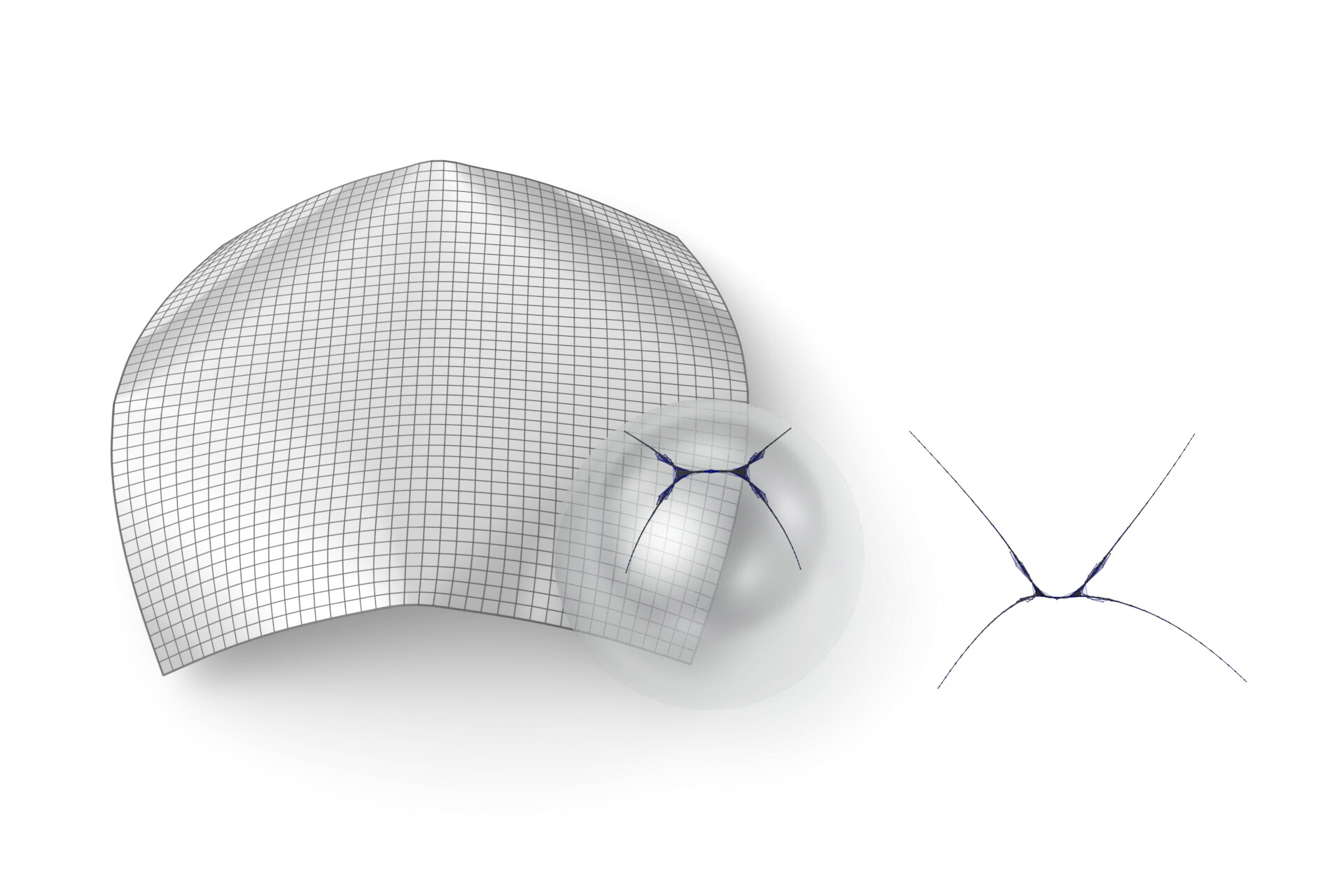}}}%
    {\clipbox{5mm 2mm 1mm 7mm}{\Inc[height=4.5cm]{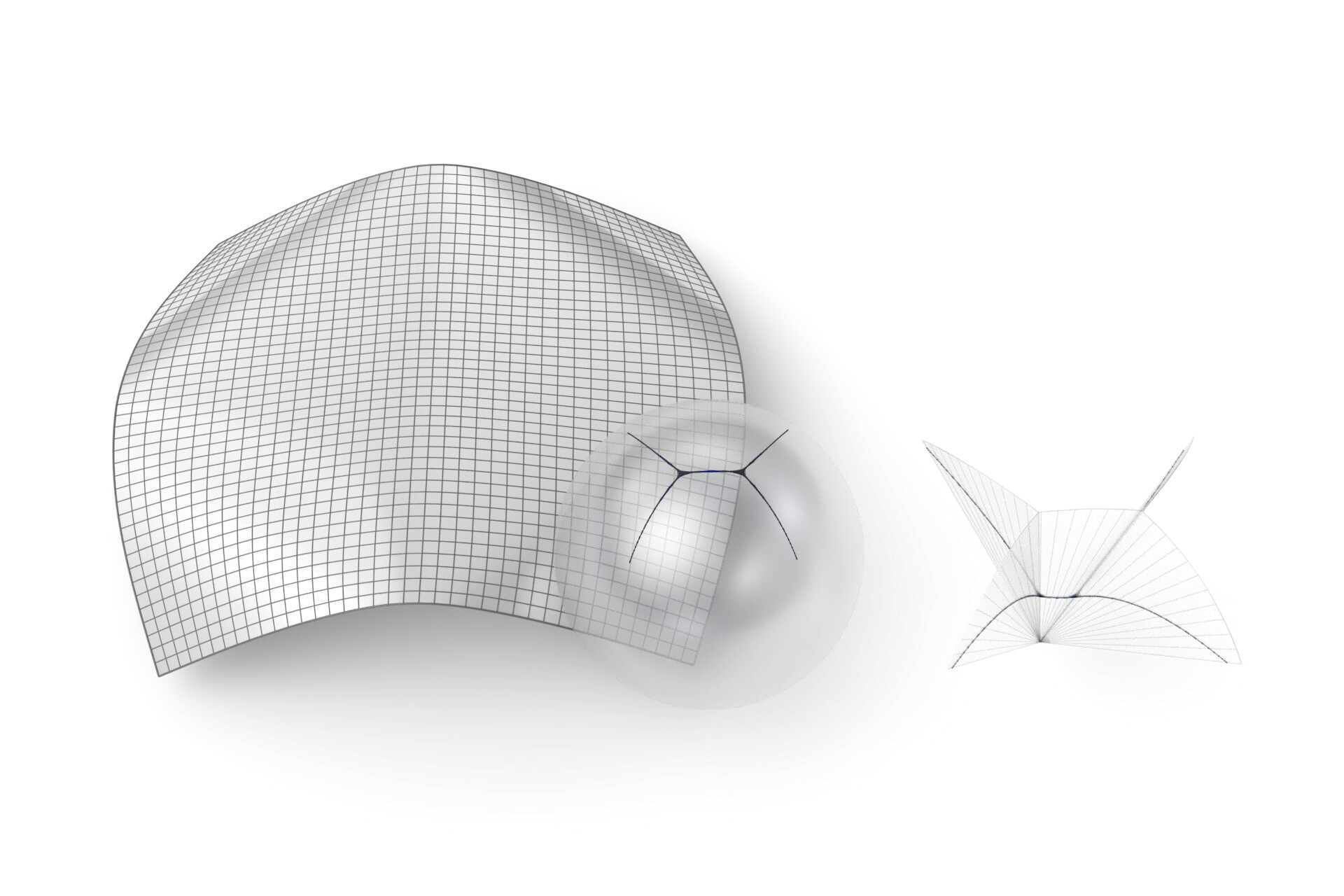}}}%
    }
    \vskip-2.0em
    \caption{The action of our optimization procedure on a non-developable
    initial mesh. From left to right, we show the initial mesh
    and the result of 2, 5, and 10 rounds of optimization. After 10
    rounds, both
    Gauss image and orthotomic surface (\S\,\ref{ss:validation}) are curve-like and
    no pockets of non-developability remain.}
    \label{fig:sequence}
\end{figure*}


Consider the special case $g(x,y)=x^2$ first.
Define a parametrization of the $xy$ plane and $\Phi$ by
$x(u,v)=a_1u+b_1v$,
$y(u,v)=a_2u+b_2v$, and
$z(u,v)=x(u,v)^2=(a_1u+b_1v)^2$.
Its polar form \cite{prautzsch:2002} reads
\begin{align*}
    G(u_1,u_2;\ v_1,v_2) =
    \big( & a_1 \frac{u_1+u_2}{2} + b_1 \frac{v_1+v_2}{2}, a_2 \frac{u_1+u_2}{2}+ b_2 \frac{v_1+v_2}{2}, \\
    & a_1^2 u_1 u_2 + 2 a_1 b_1 \frac{u_1+u_2}{2} \frac{v_1+v_2}{2} + b_1^2 v_1 v_2 \big).
\end{align*}
The function $G$ defines spline control points $v_{ij}  = G(i,i+1;\ j,j+1)$, where $i,j$
run in the integers. The bi-quadratic B-spline surface with
control points $v_{ij}$ exactly reproduces $\Phi$.
By \S\,\ref{sss:subdivision}, it is at the same time
the limit surface when the net of control
points undergoes Doo-Sabin subdivision.

A general parabolic cylinder $z=g(x,y)$ is either generated from
the special case $z=x^2$ by applying an affine transformation, or
directly by computing control points via the polar form of
$g(x(u,v),y(u,v))$ \cite{prautzsch:2002}. Figure \ref{fig:paraboliccylinder}
shows examples.

The contact element net with vertices $v_{ij}$ is exactly developable
in the discrete sense. This is because discrete tangent
planes $\tau_f$ by \S\,\ref{sss:subdivision} are tangent to $\Phi$,
therefore intersect in lines parallel to the rulings of $\Phi$. It follows
that all discrete rulings are parallel, implying developability.

Developables have the same tangent plane in all points of a single ruling. The parabolic
cylinders mentioned above have 2nd order contact only in a single point. However, applying
projective transformations yields the class of quadratic cones, which are capable of
2nd order approximation along an entire ruling \cite[\S\,6.1]{pottmann:2001}. This means
that contact element nets with appropriately weighted vertices can reproduce developables
up to 2nd order along an entire ruling.

Summing up, contact element nets are capable of approximating developables in the sense of
a 2nd order Taylor approximation; they are even capable of exactly reproducing said
Taylor approximation.

For an exact reproduction, it is not necessary that the edges of the contact net
are aligned with rulings.

\section{Design Tools for Developable Surfaces}
\label{sec:desig}

We here discuss several tools for modeling developables. All are based on
minimizing a target functional composed of individual energies expressing
either constraints or fairness.

\subsection{Interactive Editing}
\label{ss:editing}

A basic way of design is the interactive manipulation
of a surface by pulling on handles, and by requiring that certain
vertices stay close to prescribed
positions. Positional constraints are handled by
adding an energy of the form $E_\pos=\sum \| v_i - v^*_i \|^2$ to the
total energy of Eq.~\eqref{eq:totalenergy}.
Here the sum is over all vertices $v_i$ for which target positions
$v^*_i$ are available.
Figure \ref{fig:material-behaviour} shows how different positional constraints influence editing.

\begin{figure}[b]\centering
\begin{overpic}[width=.7\columnwidth]{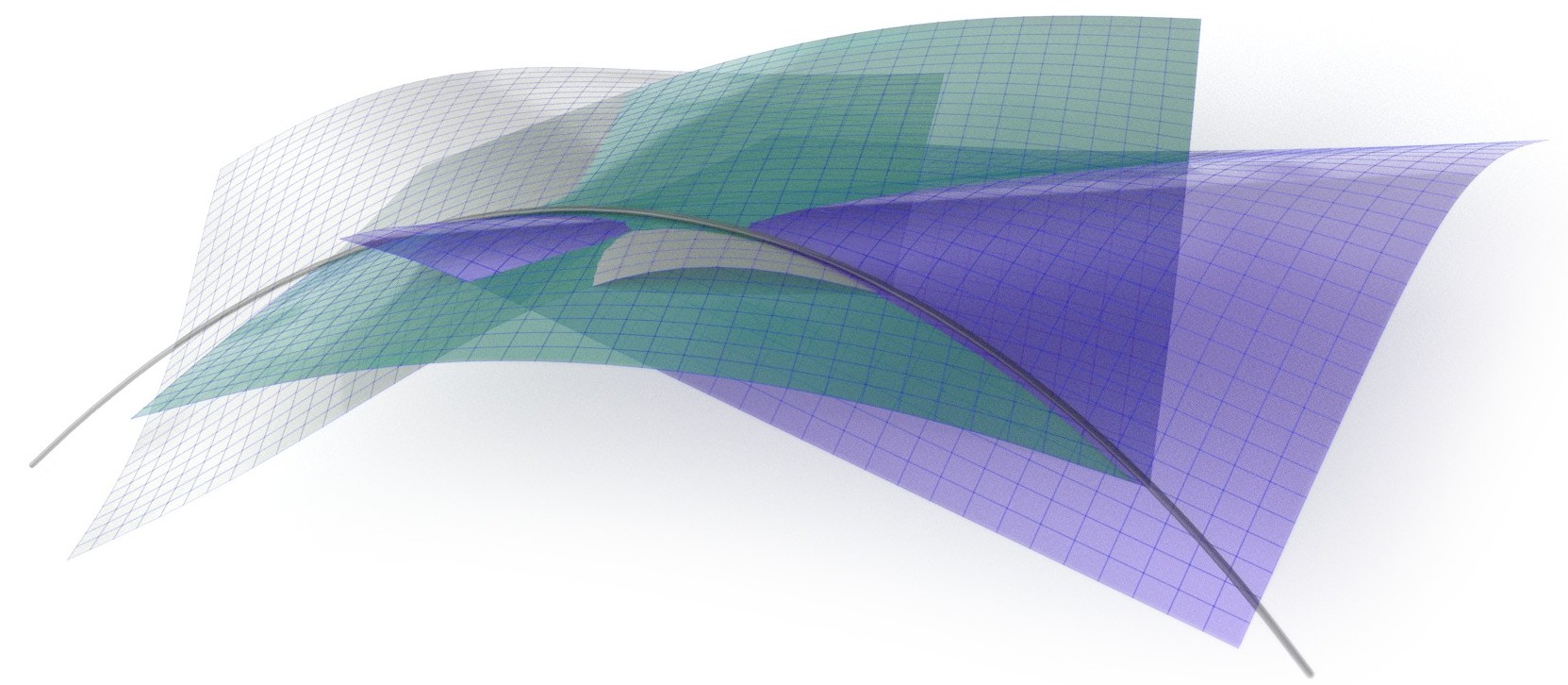}
	\put(4,19){$\Phi$}
\end{overpic}\hskip-5cm\hfill\begin{minipage}[b]{.28\columnwidth}
\caption{{\it Gliding Constraint}. 3 positions of
a developable gliding along a curve $\Phi$.}
\label{fig:gliding}
\end{minipage}
\end{figure}

\paragraph{Initializing Variables For Editing}
In all our examples concerning editing, vertex weights are
set to $1$ and are never modified. The vertices of the mesh
to be edited are assumed to be given. The remaining variables
are initialized directly via their respective constraints.

The initial mesh can
be arbitrary; it evolves toward developability in a way
which is defined by the positional constraints imposed
by the user. Figure \ref{fig:sequence} shows an example of how
a non-developable initial mesh quickly becomes developable.

\begin{figure*}[t]
	\hskip-0.15cm
	\Inc[height=.19\textwidth]{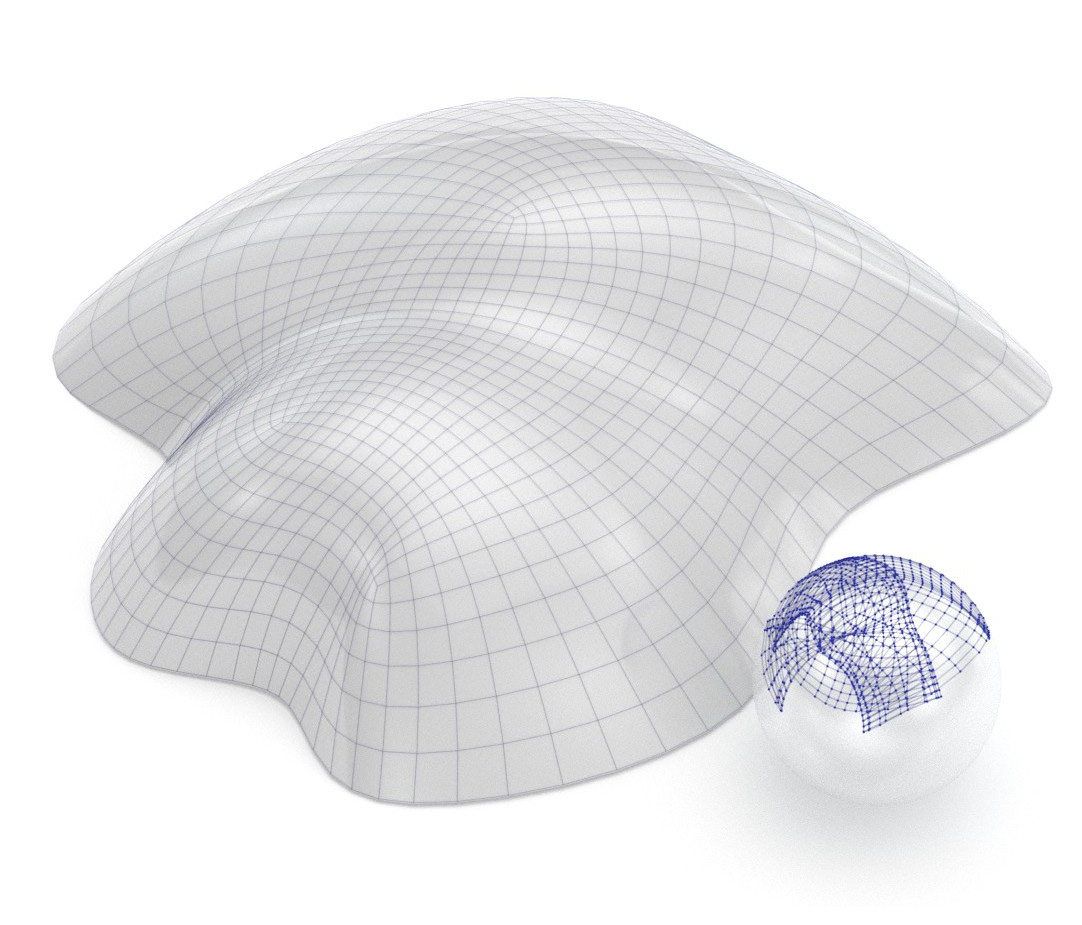}
	\hskip0.1cm
    \Inc[height=.19\textwidth]{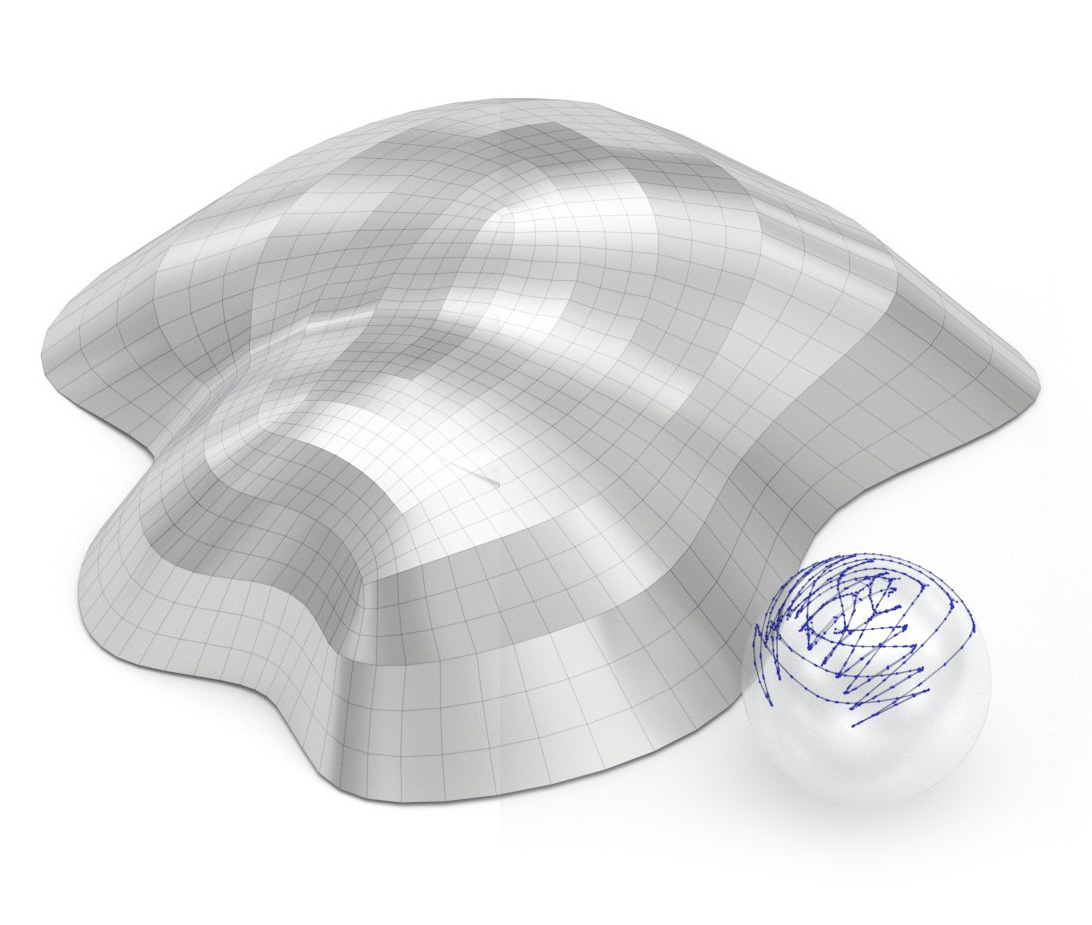}
    \hskip0.2cm
    \scalebox{0.79}{\clipbox{1.0cm 2mm 0.5cm 3mm}{\Inc[width=.4\textwidth]{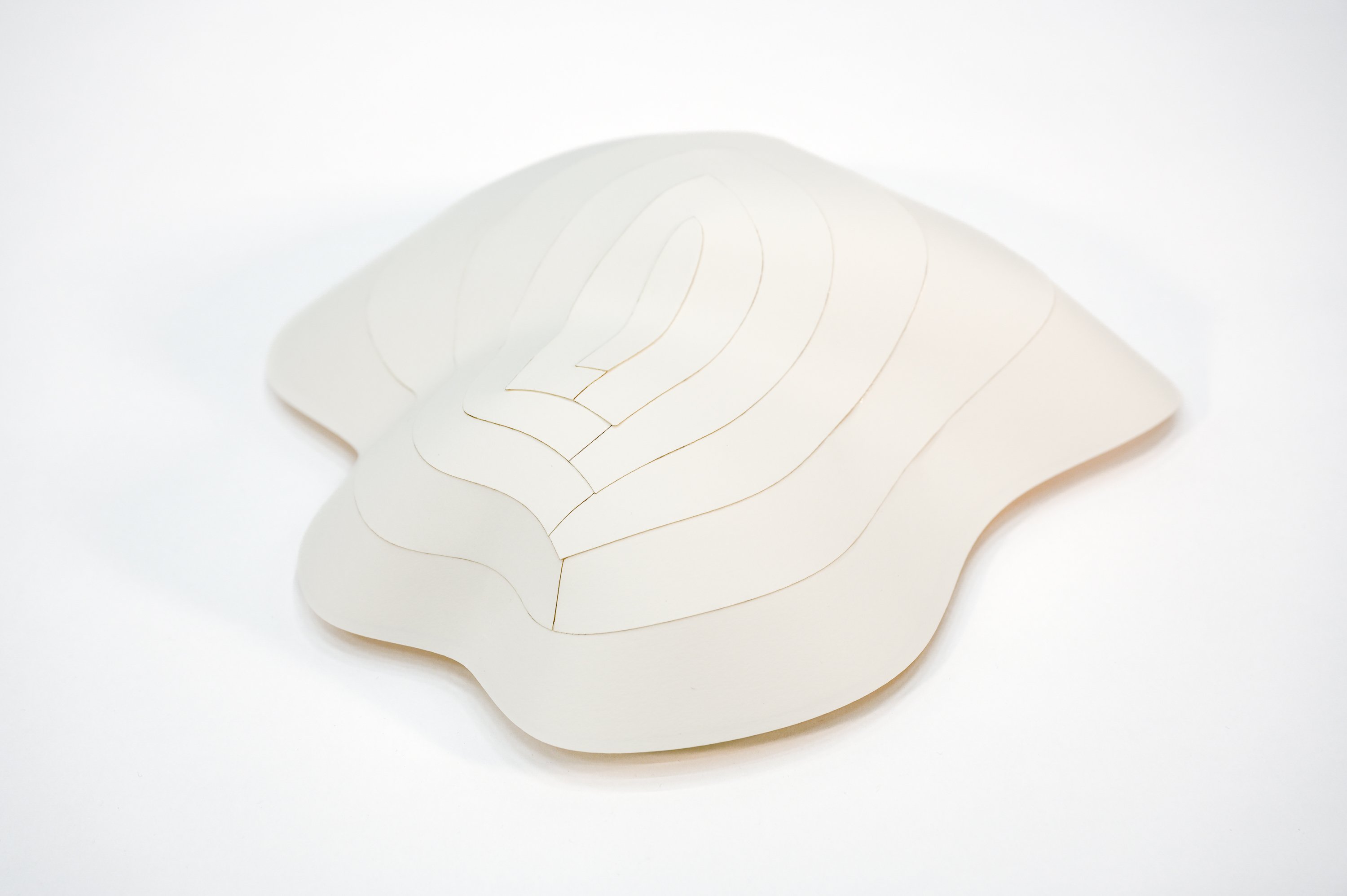}}}
    \hfill
    \begin{minipage}[b]{.26\textwidth}
    \caption{The surface on the left
    is made piecewise developable by partitioning it into
    strips along mesh polylines, and making
    those strips developable while boundaries remain fixed.
    We also show the Gauss image, whose curve components correspond to the
    developable parts of the surface. The figure on the right shows a paper model.}
    \label{fig:strips}
    \end{minipage}
\end{figure*}

\paragraph{Gliding Constraint.}
Another basic design requirement would be  that our
surface $M$ is to glide through a reference shape represented
by a point cloud $\Phi$ (e.g.\ a curve). For all $p\in\Phi$ we
compute the closest point projection $p^*\in M$ which is contained
in some face $f(p)$. We now require that $p$ does not deviate
from the tangent plane associated with this face, which passes through
the face midpoint $b_{f(p)}$ and has normal vector
$\nw_{f(p)}$.  This is expressed by a low value of the
energy
\begin{align*}
	E_\prox = \sum\nolimits_{p\ \text{active}} \<p-b_{f(p)},\
			\nw_{f(p)} \>^2
\end{align*}
The sum is over all \emph{active} points $p$ in the cloud $\Phi$ (which
are not variables), where active means they
are not too far from the variable mesh $M$. The faces $f(p)$ are recomputed in each round
of the optimization. The approximation of the distance field of $M$
by distances to tangent planes is done on the basis of
\cite{registration2006}. It is known to be accurate to 2nd order in the
case of zero distance, and it prevents unwanted effects if the pool
of active points in $\Phi$ is not entirely correct. Figure \ref{fig:gliding}
shows an example where a developable is gliding along a curve.

\paragraph{Soft Isometry Constraints as Regularizers.}
In interactive modelling applications, we offer to the designer
different kinds of material behaviour as
illustrated by Fig.~\ref{fig:material-behaviour}.

For the soft isometry constraints,
we follow \cite{jiang-2020}. We express isometry between faces
$f=v_0v_1v_2v_3$ and
$f'=v_0'v_1'v_2'v_3'$ by
	\begin{align*}
	 c_{\iso,1} (f, f')
	&:= \| v_2 - v_0 \|^2 - \| v'_2 - v'_0 \|^2 = 0,\\
	 c_{\iso,2} (f, f')
	&:=  \| v_3 - v_1 \|^2 - \| v'_3 - v'_1 \|^2 = 0,\\
	 c_{\iso,3} (f, f')
	&:= \<v_2-v_0,v_3-v_1\>-\<v'_2-v'_0,v'_3-v'_1\>= 0.
	\end{align*}
These constraints yield a contribution to the target
functional in optimization, namely
\begin{align*}
	E_\iso=\sum\nolimits_{(f,f')}\sum\nolimits_{k=1,2,3}
	c_{\iso,k}(f,f')^2.
\end{align*}

\Bullet If the design surface $M$ is to behave like an
inextensible material, geometric design can be done
by including the property of being
isometric to the reference mesh, using the term $E_\iso$
with a large associated weight $\lambda_\iso$.

\Bullet The design surface may behave in an elastic way.
In our implementation, we can choose to include the
property of being isometric either to the reference
mesh (with a smaller weight $\lambda_\iso$), or to
the previous state/iteration.
Both provide a regularizing effect to varying extents,
without pulling the surface back to its initial position.
We do not claim to accurately model any exact
material behavior.

\subsection{Developable Lofting}
\label{ss:lofting}

\begin{figure}[b]
    \leftline{\Inc[width=1.05\columnwidth]{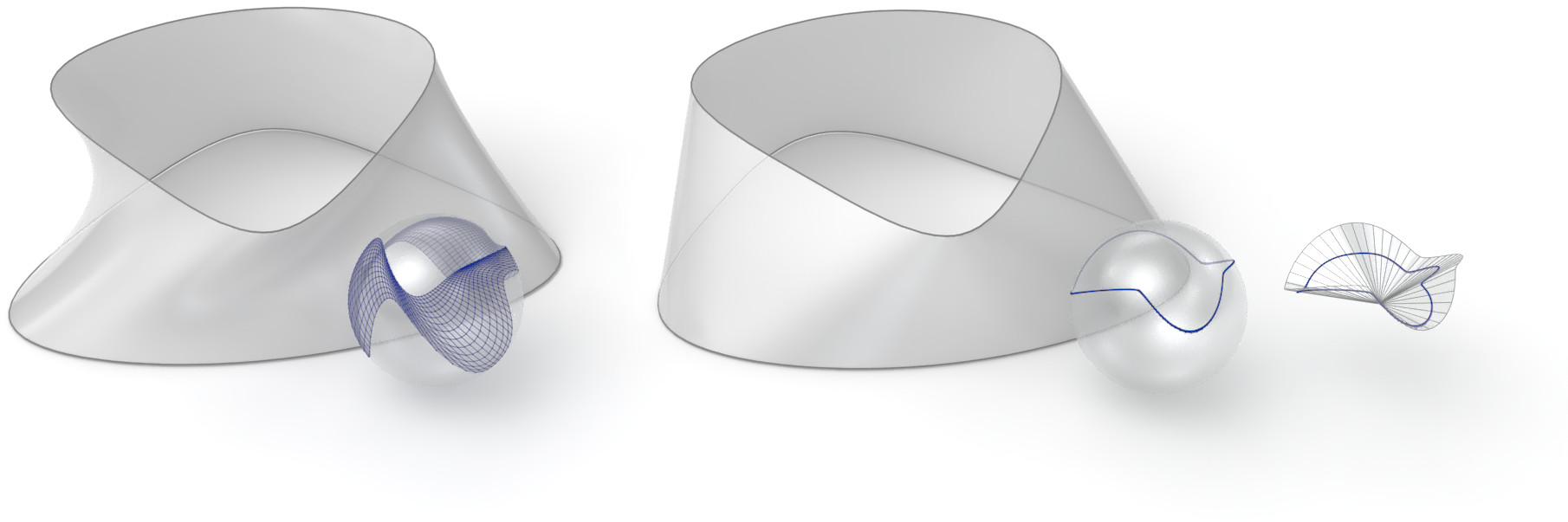}\hss}
    \vskip-2em
    \caption{{\it Developable lofting}. The surface with cylinder topology on the
    left is optimized for developability so that the two boundary
    curves remain unchanged. We visualize developability not only via the Gauss
    image, but also with the orthotomic curve mentioned in \S\,\ref{ss:validation}.}
    \label{fig:cylinder}
\end{figure}

A basic way how a designer may specify a developable is to prescribe
two boundary curves --- see Fig.~\ref{fig:cylinder}.
This procedure is referred to as {\em lofting}. Developable lofting
is a problem with a long history, and early solutions for special cases.
Within the framework of our optimization, we set up lofting as follows:
We connect the two given curves by
an arbitrary quad mesh (e.g.\ by a ruled surface) which we subsequently
optimize.

Previous approaches
to discrete developables cannot take this road easily:

\Bullet The semidiscrete representation
of piecewise-developables by Pottmann
et al.\ \shortcite{pottmann-2008-strip} uses
quad meshes with {\em planar faces}.
Edges correspond either to boundary curves, or rulings. Apart from the
the difficulty of having planar faces, such a mesh also cannot describe
strips that are cut off in an arbitrary way, not exactly along a ruling.
Since in our approach rulings are not aligned with edges, such problems
do not occur.

\Bullet The ruling-based method of
Tang et al.\ \shortcite{tang-dev-2015} suffers the same deficiencies.

\Bullet The methods based on orthogonal geodesic nets, such as
\cite{rabinovich+2018a} and follow-up contributions, cannot
describe a collection of developable strips without trimming, since the
boundary curves are, in general, not geodesics.

\Bullet The method of Jiang et al.\ \shortcite{jiang-2020}, which is based
on isometries, cannot easily
solve examples like that of Fig.~\ref{fig:cylinder}. This is because
the development, which does not even exist globally, has to be initialized
and then optimized simultaneously with the surface.

\smallskip

While some of these drawbacks might be solvable through trivial engineering
solutions (such as trimming, or matching local developments), our method
does not require any computational overhead, making it more efficient,
and hence suitable, for interactive design.

\begin{figure}[t]
    \fboxsep0pt\fboxrule.2pt\flushleft
    \newdimen\tmplen\tmplen0.008\columnwidth
    {\begin{overpic}[width=77\tmplen]{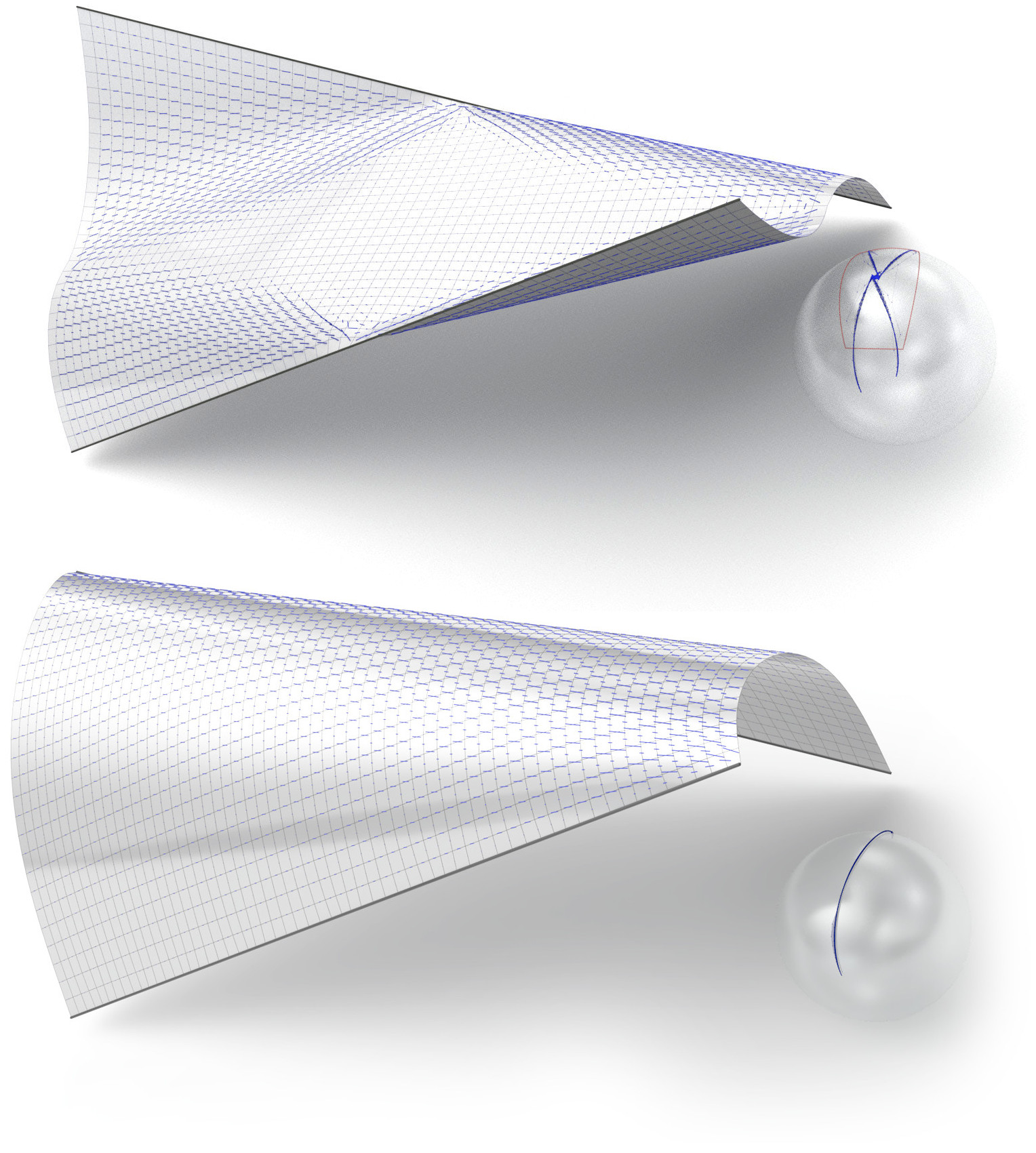}
    \end{overpic}}\hfill
    \raise30\tmplen\hbox{\fboxsep2pt\fbox{\scalebox{2.4}
    	{\clipbox{3mm 5mm 25mm 5mm}
    		{\begin{overpic}[width=4cm]{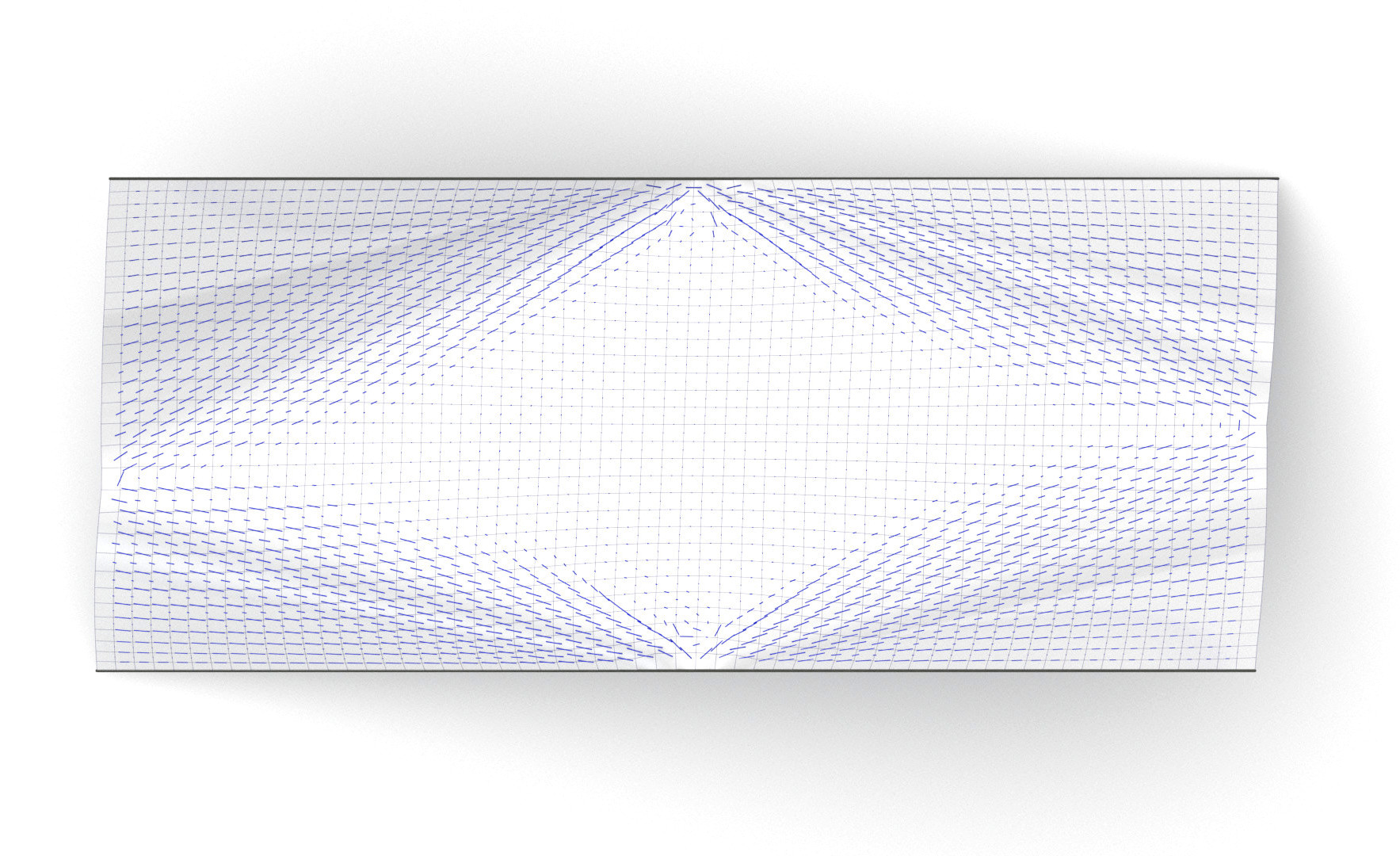}
    		\end{overpic}}}}}\\
    \vskip-1em
    \caption{{\it Developable lofting of skew straight lines.} A lofting
    solution usually findings a developable with rulings transverse to the
    prescribed boundaries. However, this is not possible if those boundaries
    are straight lines. Our method allows us to find a solution with
    singularities and a planar piece (top) and another solution where the
    boundaries are actually rulings (bottom). The latter is found automatically
    if the {\em ruling fairness} term is given a higher weight.}
    \label{fig:failwithcrease}
\end{figure}

\paragraph{Solvability of the Lofting Problem}
Lofting is actually a difficult problem,
which does not need to have a smooth solution: It
has many continuous solutions such as a union of cones whose
vertices lie on the given boundary curves in an alternating way.
Figure \ref{fig:failwithcrease} shows the behavior of our algorithm
in such a failure case. A developable containing two skew lines
exists and can be found (Fig.~\ref{fig:failwithcrease}, bottom).
However, when disabling ruling fairness, our optimization gets
stuck in a local minimum and tries to find a developable whose
rulings are transverse to the given boundary. In this special
case, it is known that no solution exists, so we use this example
to simulate a failure case. Our algorithm produces a developable
with singularities at the boundary (Fig.~\ref{fig:failwithcrease}, top).

Generally, developable lofting is known to
be challenging. An example demonstrating the capabilities of our method
is illustrated by the architectural design shown in Fig.~\ref{fig:bridge}.

\subsection{Multiresolution Modeling}
\label{ss:multiresolution}

In \S\,\ref{sss:subdivision} we described the watertight spline surface
associated with a quad mesh, and how it occurs as the limit of weighted Doo-Sabin
subdivision. We use these tools for a multiresolution approach to modeling
developables.

We start with a coarse quad mesh $M$ equipped with vertex weights.
Vertices and weights are optimized such that $S^k M$, the result of $k$ rounds of
subdivision, is a discrete developable. Here typically $k=1$ or $k=2$. The idea
of this procedure is to define a near-developable spline surface $S^\infty M$
by a small optimized control mesh $M$. In case the result of optimization does
not yield the desired quality, we subdivide, let $M:=S^1 M$, and
repeat the procedure.

\begin{figure}[t]
    \Inc[width=\columnwidth]{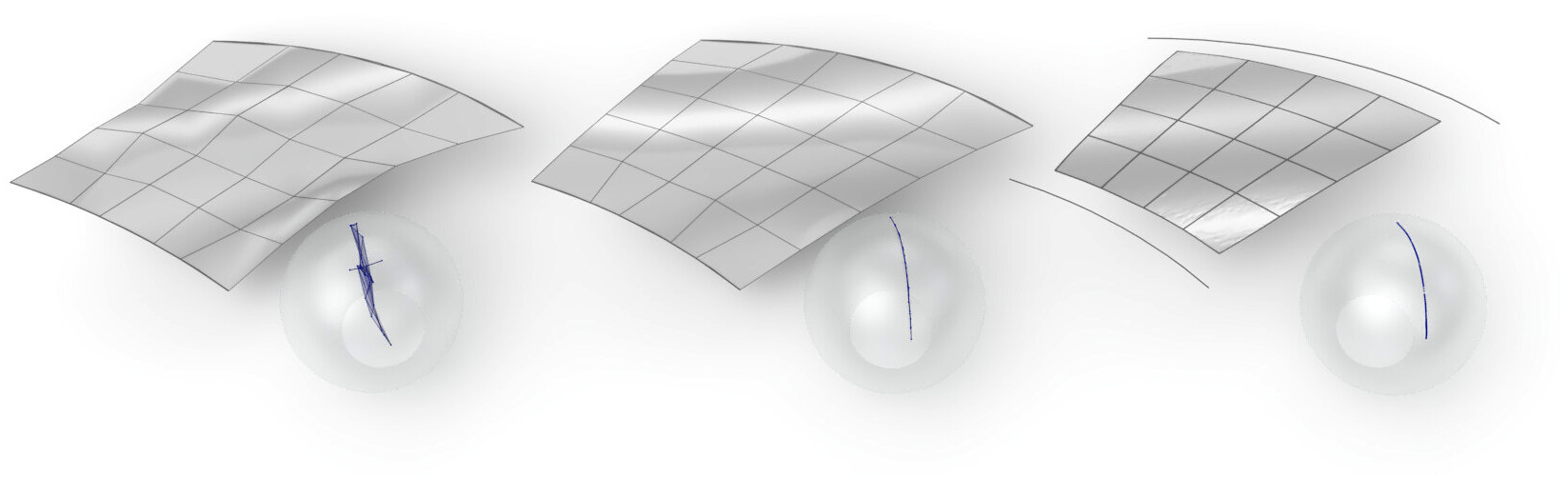}
    \vskip-1em
    \caption{{\it Combining lofting with multiresolution.} {\it Left:} A coarse mesh $M$ which
    is optimized towards developability. The corresponding Gauss image shows that the goal
    has not been achieved, owing to the low resolution. {\it Center:}  A coarse mesh
    $M'$, constrained to the same two boundary curves as $M$, is optimized such that
    a subdivided mesh $S^2 M'$ is discrete-developable. The Gauss image of
    $M'$ demonstrates a high degree of developability. {\it Right:} The biquadratic spline
    surface defined by $M'$, consisting of 16 B\'ezier patches.  It does not interpolate
    the boundary (for that, we would have to impose that certain face centers  are constrained
    to the boundary, see Fig.~\ref{fig:doosabin}, right).}
    \label{fig:multiresolution}
\end{figure}

The resulting spline surface consists of as many biquadratic
rational B\'ezier patches as the mesh has faces. Our aim is to achieve a small
number of patches. Figure \ref{fig:multiresolution} shows an example of
multiresolution modeling.

\section{Discussion}
\label{sec:discussion}

\subsection{Implementation}
\label{ss:implementation}

All interactive design tools described in \S\,\ref{sec:desig} were implemented as part of
{\it Rhinoceros3D}. Our plugin is a C++ dynamic-link Windows\Reg\ library that
can directly interact with the Rhino application. The benefit of
our implementation choice is two-fold: first, it seamlessly combines with
all the existing geometry processing tools already in Rhino. This
results in a powerful design environment. Secondly, Rhino is
widely used within the architectural community, and thus provides a
natural, broad user base for the new algorithms. Our plugin will be made
available as open-source software.

The optimization of \S\,\ref{ss:computation} and
\S\,\ref{ss:editing} was solved using a Levenberg-Marquardt method
according to \cite[\S 3.2]{madsen:2004}, with a damping parameter of
$10^{-6}$. In the interactive application, the optimization is restarted
on every user input. We terminate the optimization loop when the energy
value falls below a certain threshold, or when there is no more
improvement.

Our implementation uses McNeel's \textsc{openNURBS} toolkit
for elementary geometric manipulations, the
Intel\Reg\ \textsc{oneAPI} Math Kernel library
for
efficient sparse matrix manipulations,
and Intel\Reg's\ \textsc{oneMKL} \textsc{pardiso} for solving large sparse
symmetric linear systems.

\begin{figure}[b]
\hspace*{1.5em}
\begin{overpic}[height=.4\columnwidth]{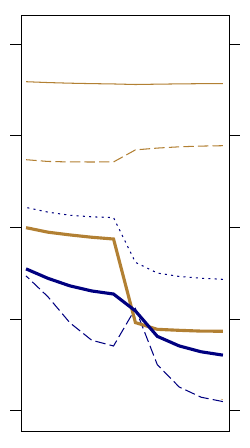}%
	\color{gelb}\small%
	\color{black}\scriptsize%
  \lcput(-2,10){$10^{-6}$}
  \lcput(-2,32){$10^{-4}$}
  \lcput(-2,52){$10^{-2}$}
  \lcput(-2,72){$10^{0}$}
	\ctput(22,0){\scriptsize 7 iterations}
	\small\ctput(22,95){Fig.~\ref{fig:simple}}
\end{overpic}\hskip-0.6ex
\begin{overpic}[height=.4\columnwidth]{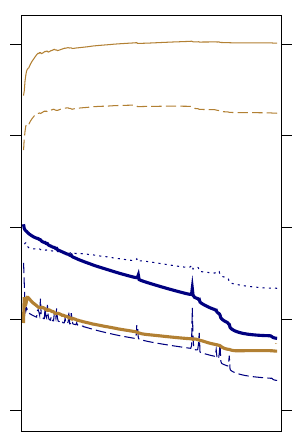}%
	\small\ctput(35,95){\contour{white}{Fig.~\ref{fig:failwithcrease}, top}}
	\scriptsize%
	\ctput(35,0){\scriptsize 400 iterations}%
	\color{blau}\small
\end{overpic}\hfill
\begin{minipage}[b]{.44\columnwidth}
\begin{pp}
\put(0,-5){\Inc[width=.6\linewidth]{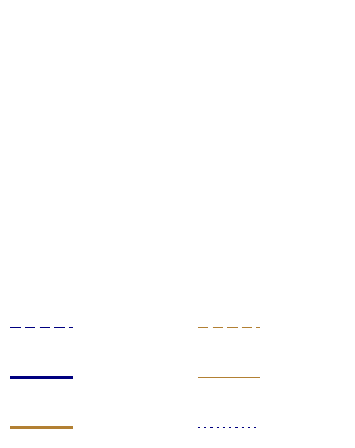}}
	\footnotesize
	\rcput(18,-2){\color{gelb}$E_\pos$}
	\rcput(18,7){\color{blau}$E_\dev$}
	\rcput(18,16){\color{blau}$E_\rul$}
	\rcput(48,-2){\color{blau}$E_\norm$}
	\rcput(48,7){\color{gelb}$E_\fair^V$}
	\rcput(48,16){\color{gelb}$E_\fair^n$}
\end{pp}
\caption{Energies during optimization
for Fig.~\ref{fig:simple}
(weights change after 4 iterations),
and  for the  singular case of Fig.~\ref{fig:failwithcrease}, top.
Energy spikes correspond to buckling-like local shape changes.}
\label{fig:energies}
\end{minipage}
\end{figure}

The behaviour of energies during the course of optimization
is exemplarily shown by Fig.~\ref{fig:energies}.
Table~\ref{tab:stat} provides statistics on the size of optimization
problems, the choice of weights, and the time needed. Times refer to
an Intel\Reg\ Xeon\Reg\
CPU E5-2695 v3 $@$ 2.30GHz x64-based processor, running 64-bit Windows\Reg\
10. Our plugin is configured to use 64 parallel \textsc{openMP}
threads. This was chosen to achieve interactivity for the models shown
throughout the figures, but could be tuned for larger designs.

\begin{figure}[b]
    \Inc[width=.68\columnwidth]{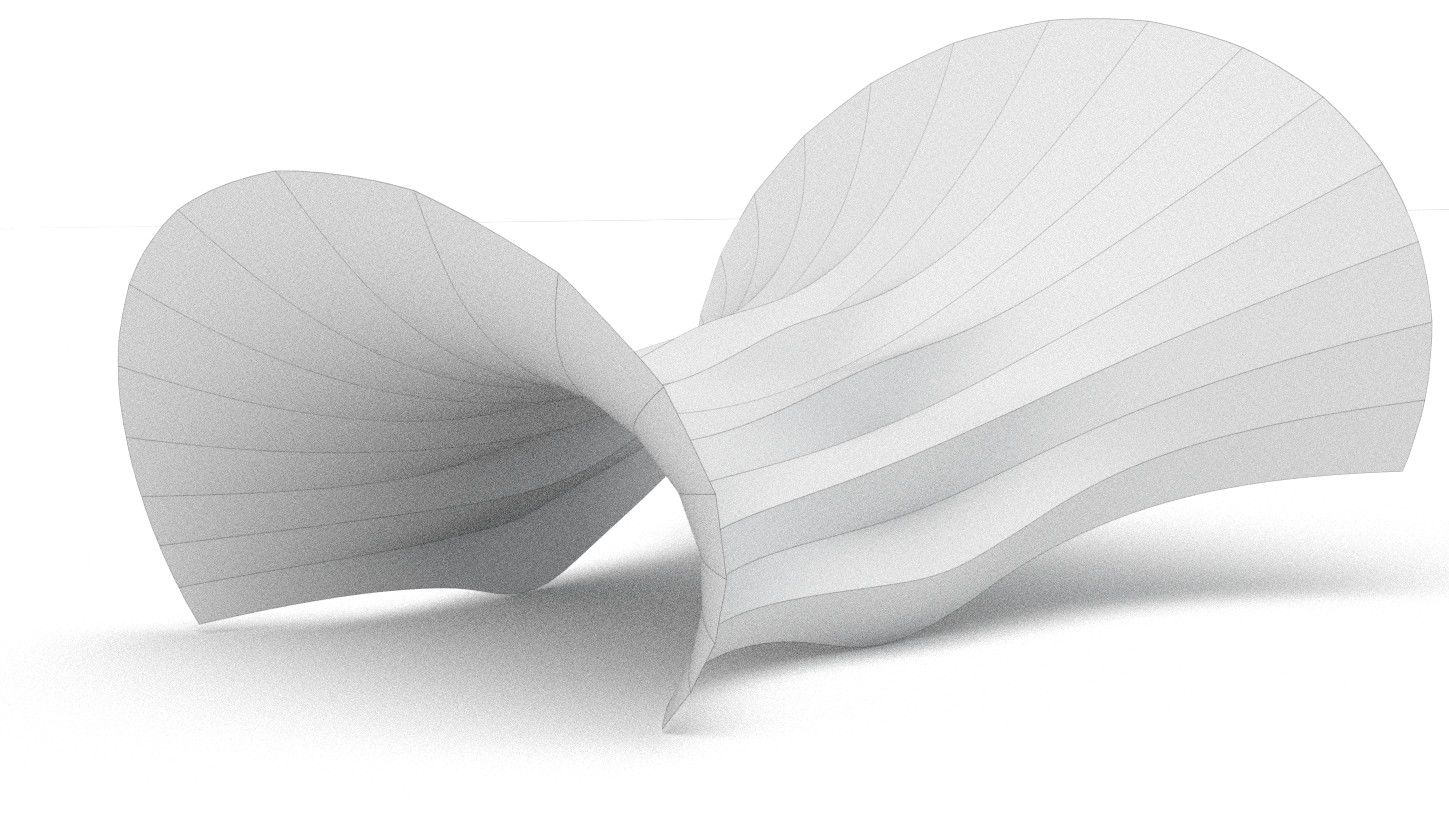}
    \caption{Creases as a design element of piecewise developables.}
    \label{fig:creases2}
\end{figure}

\begin{figure}[t]
	\hskip-0.4cm
    \Inc[width=1.0\columnwidth]{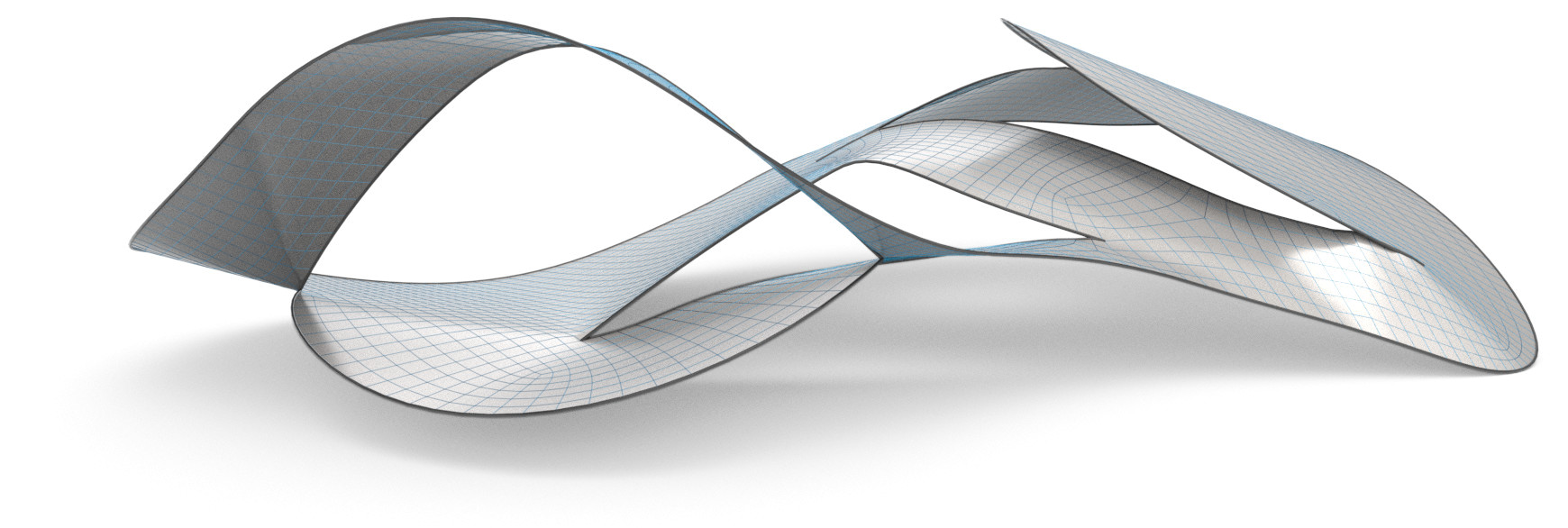} \\[-2mm]
    \hskip0.4cm
    \Inc[angle=90,width=.45\columnwidth]{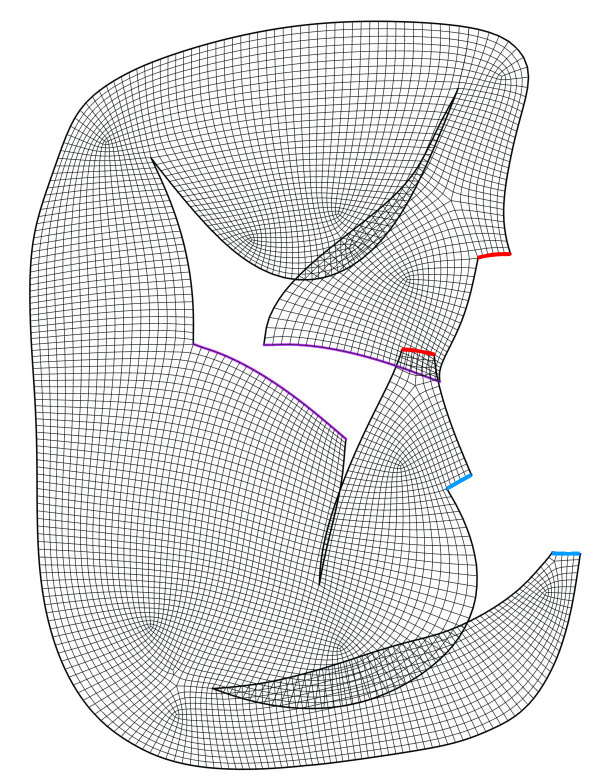}
    \hfill
    \begin{minipage}[b]{.44\columnwidth}
        \caption{A developable surface made with the design tool
        described in \S\,\ref{ss:editing}, starting from a planar mesh with
        3 incisions. After cutting free every hole, a development can be computed, using the method of \cite{jiang-2020}. \newline}
        \label{fig:cuts}
    \end{minipage}
\end{figure}

We do not give statistics for Fig.~\ref{fig:bridge}, Fig.~\ref{fig:creases2} or Fig.~\ref{fig:cuts} since
these examples were interactively designed, with optimization running
in the background continuously. The architectural design in Fig.~\ref{fig:bridge}
consists of 6 surfaces with a total of 15k faces.

\begin{table}[t]
    \caption{{\em Optimization details.} For selected examples,
    we give the number of faces, weights used in optimization, the
    number of individual surfaces this example consists of,
    the number of iterations (resp.\ an average number of iterations,
    if marked by ``$\sim$''), the average time in seconds needed for a single
    iteration and a single surface, and the total time used for optimization.
    \label{tab:stat}}
    \def\E#1{\cdot 10^{#1}}
\small\begin{tabular}{|@{\hskip0.1ex}c@{\hskip0.1ex}r
	@{\,\,\,}|
		*{7}{@{\,\,\,}l}
	@{}|@{\,}
		*{4}{r@{\,\,\,}}
	|}
\hline
	Fig.\vphantom{$F_{)_{)_)}}$}
	& $|F|$\hspace*{1ex}
	& $w_\fair^V$
	& $w_\fair^n$
    & $w_\fair^r$
	& $w_\rul$
	& $w_\dev$\!\!
	& $w_\pos$\!\!
	& $w_\iso$
	& \#surf\!\!\!\!\!\!
	& \,\#it\!\!
	& $T_{\text{single}}^{\text{per it}}$\!\!\!
	& $T_{\text{total}}$
\\ \hline
\ref{fig:simple}
    & 1024
    & 0.1
    & 1.0
    & 
    & 10.0
    & 10.0
    & 
    & 0.1
    & 1
    & 4
    & .201
    &
\\
    & 
    & 0.01
    & 0.1
    & 
    & 10.0
    & 10.0
    &
    & 0.1
    & 
    & 3
    & .207
    & 1.4
\\
\ref{fig:weights:better}$^{\text{cent}}$
    & 99
    & 0.1
    & 
    & 
    & 10.0
    & 100.0
    & 1.0
    & 
    & 1
    & 162
    & .009
    & 1.5
\\
\ref{fig:weights:better}$^{\text{right}}$
    & 99
    & 0.1
    & 
    & 
    & 10.0
    & 100.0
    & 1.0
    & 
    & 1
    & 133
    & .018
    & 2.4
\\
\ref{fig:sequence} 
    & 1638
    & 0.1
    & 
    & 
    & 10.0
    & 100.0
    & 
    &
    & 1
    & 10
    & .227
    & 2.3
\\
\ref{fig:strips}
    & 1076
    & 0.1
    & 0.01
    & 
    & 1.0
    & 10.0
    & 100.0
    &
    & 6
    & $\sim$21
    & .018
    & 2.7
\\
\ref{fig:cylinder}
    & 1500
    & 1.0
    & 0.01
    & 
    & 10.0
    & 100.0
    & 10.0
    & 
    & 1
    & 52
    & .211
    & 11.0
\\
\ref{fig:failwithcrease}$^{\text{top}}$
    & 1800
    & 0.01
    & 0.1
    & 
    & 10.0 
    & 10.0
    & 10.0
    & 0.1
    & 1 
    & 400
    & .395
    & 158 
\\
\ref{fig:failwithcrease}$^{\text{bot}}$
    & 1800
    & 0.1
    & 0.1
    & 2.0
    & 10.0
    & 100.0
    & 10.0
    & 0.1
    & 1
    & 380
    & .703
    & 267 
\\ 
\ref{fig:multiresolution}$^{\text{left}}$
    & 25
    & 0.05
    & 
    & 
    & 10.0
    & 100.0
    & 1.0
    & 
    & 1
    & 23
    & .004
    & .092
\\
\ref{fig:multiresolution}$^{\text{cent}}$
    & 25
    & 0.05
    & 
    & 
    & 10.0
    & 100.0
    & 1.0
    & 
    & 1
    & 106
    & .037
    & 3.9
\\
\hline
\end{tabular}

    \vskip1.5em
    \caption{{\em Measuring developability.}  This table gives
    the energy $E_{\text{dev}}$ for those examples
    where remeshing and projective transformations
    take place.}
    \label{table:2}
    \vskip-1ex
    \begin{minipage}{.3\columnwidth}\def\E#1#2{$#1\!\cdot\!10^{-#2}$}
\small\begin{tabular}{|*3{l@{\hskip2ex}}l|}   \hline
    Fig.
	& $E_{\text{dev}}^{\text{total}}$
	& $E_{\text{dev}}^{\text{per face}}$
	& $|F|$ 
	\\ \hline
	\ref{fig:remeshing}a
	& \E{5.5}{5}
	& \E{7.4}{8}
	& 734 	
	\\
	\ref{fig:remeshing}b
	& \E{2.0}{4}
	& \E{1.1}{6}
	& 175 
	\\
	\ref{fig:remeshing}c
	& \E{1.9}{5}
	& \E{2.1}{8}
	& 900
	\\
	\ref{fig:remeshing}d
 	& \E{2.5}{5}
 	& \E{3.4}{8}
	& 734
 \\ \hline
\end{tabular}\hss
\end{minipage}\hfill
\begin{minipage}{.6\columnwidth}\flushright\def\E#1#2{$#1\!\cdot\!10^{-#2}$}
\small\begin{tabular}{|*3{l@{\hskip2ex}}l|}   \hline
    Fig.
	& $E_{\text{dev}}^{\text{total}}$
	& $E_{\text{dev}}^{\text{per face}}$
	& $|F|$ 
	\\ \hline
	\ref{fig:paraboliccylinder}$^{\text{left}}$
	& \E{9.6}{30}
	& --
	& 77 	
	\\
	\ref{fig:paraboliccylinder}$^{\text{center}}$
	& \E{1.6}{28}
	& --
	& 77
	\\
    \ref{fig:paraboliccylinder}$^{\text{right}}$
	& \E{6.1}{5}
	& \E{7.9}{7}
	& 77
	\\
    &&&
    \\ \hline
\end{tabular}
\end{minipage}
\end{table}

\subsection{Validation} \label{ss:validation}

Our developability condition of Def.~\ref{defn:developability}
imposed on meshes is comparable to the requirement that the Gauss curvature
of a smooth developable vanishes.
The latter has a list of local and global implications, including
a curve-like Gauss image and existence of
rulings. These properties could be verified up to tolerances.

\paragraph{Visualization of Gauss Curvature}
The property of the Gauss image being curve-like is
a very sensitive indicator of developability. In contrast to
this, visualizing the numerical values of Gauss curvature is
not suitable for properly identifying developable surfaces, because the
numerical errors inherent in the approximate computation of Gauss
curvature are bigger than the value of Gauss curvature itself. This
was confirmed in
experiments on fair quad meshes sampled from
mathematically correct developables, using the
{\it jet fit} method of \cite{cazals03}. For this reason, we validate developability
via the Gauss image throughout this paper.

\paragraph{Visualization of Developability via Orthotomics}
For any surface $\Phi$, reflecting a source point in all
tangent planes yields the orthotomic surface $\Phi^*$
\cite{hoschek1984}. It degenerates
into a curve if and only if the original surface was
developable and thus is a good visual indicator of developability ---
see Fig.~\ref{fig:sequence} and Fig.~\ref{fig:cylinder}.

\subsection{Conclusion}

The developability criterion for quad meshes presented in this
paper has successfully been used to solve design problems with
developable surfaces, which is a well-known and difficult topic
with a long list of individual contributions. The fact that the edges
of our developable quad meshes do not have to be aligned with special curves,
represents a great practical advantage. Another advantage is the fact
that we do not have to consider the actual development at the same time
as the developable surface. These advantages are evident in
comparison with prior work.

\begin{figure}[t]
    {\scalebox{0.95}{\fboxsep0pt\fboxrule.1pt{\clipbox{8mm 0 8mm 0}{\Inc[height=4cm]{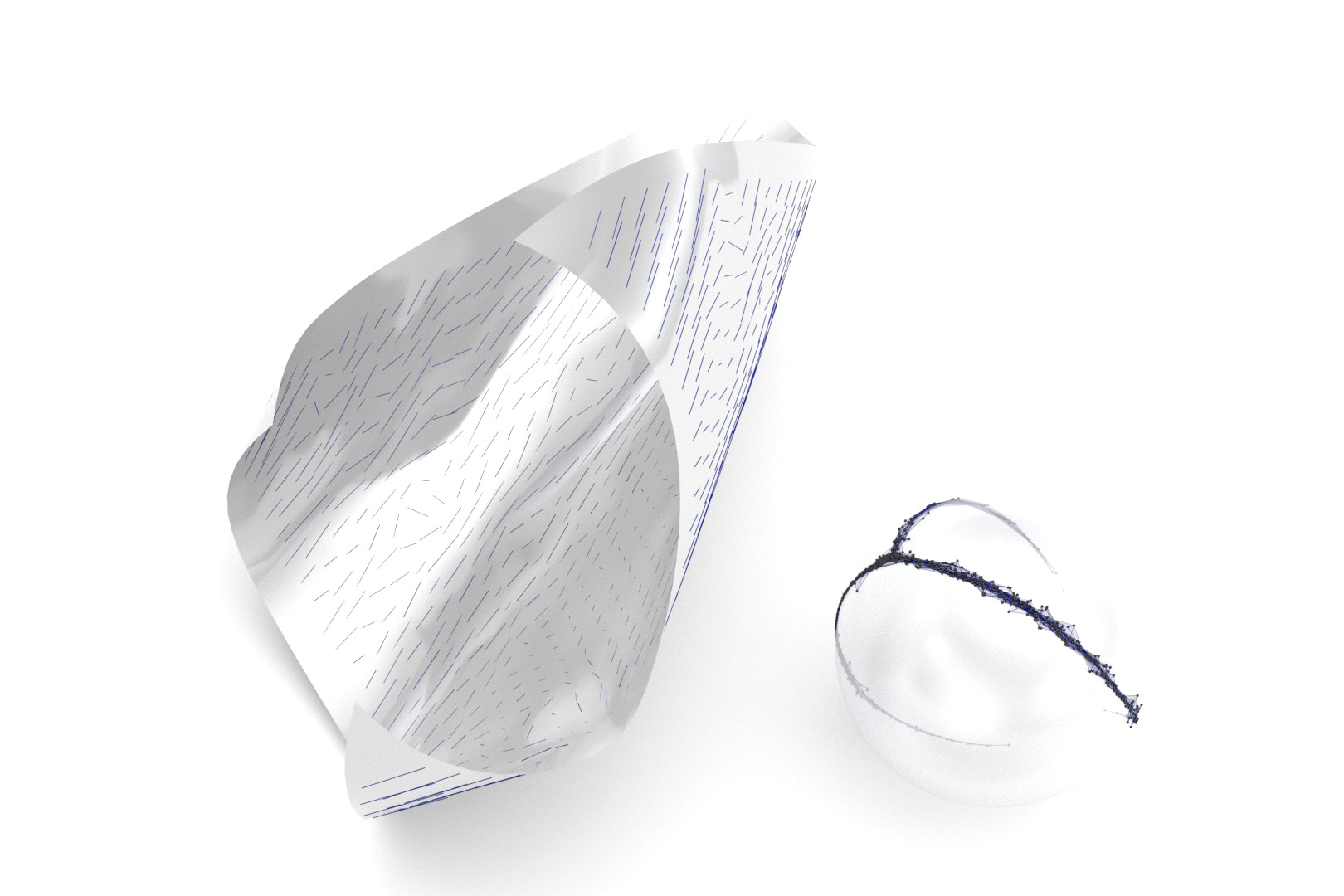}}}}}\hfill
    {\scalebox{0.95}{\fboxsep0pt\fboxrule.1pt{\clipbox{8mm 0 8mm 0}{\Inc[height=4cm]{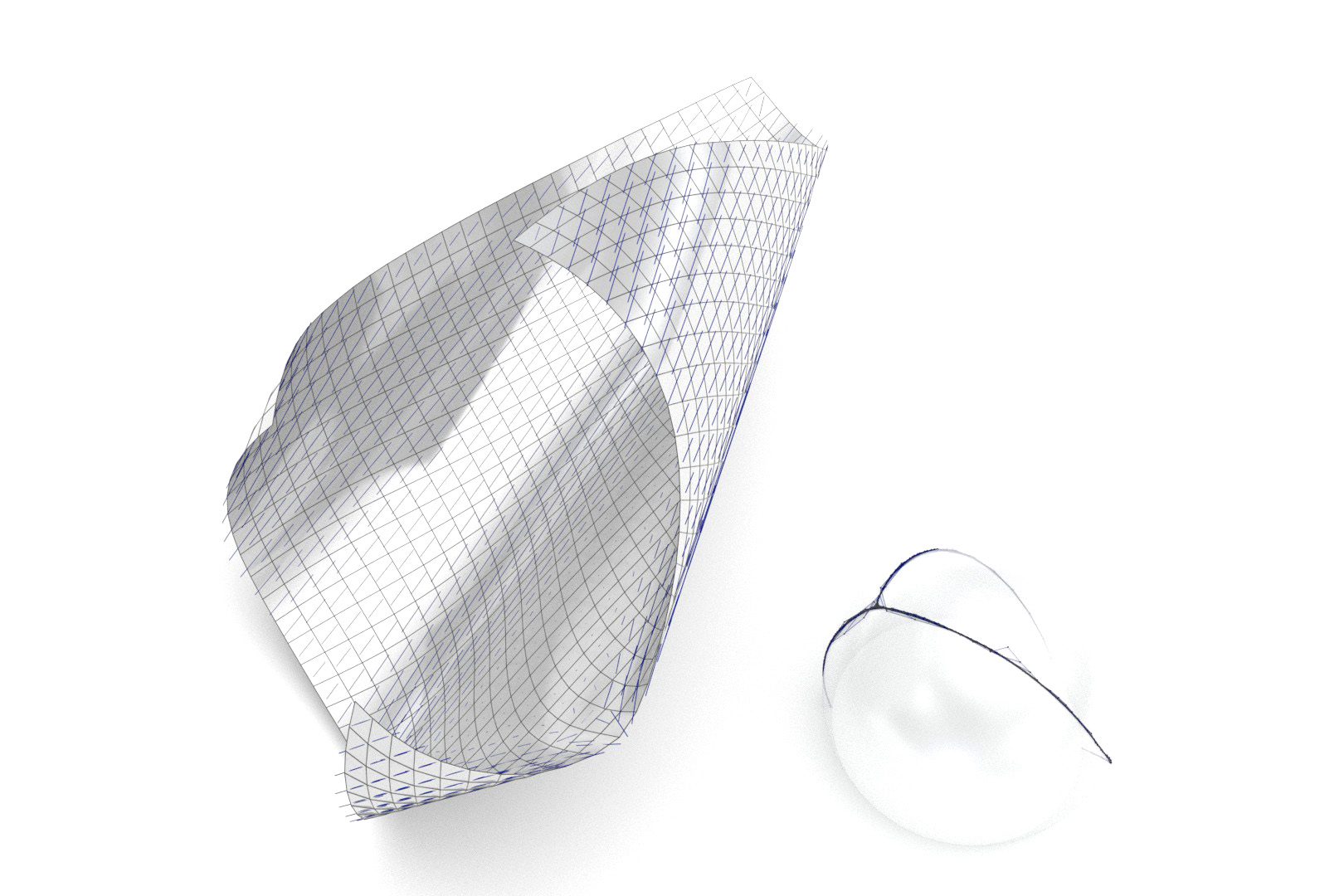}}}}}\vskip-1em
    \caption{The mesh on the left has been created using the method of
    \cite{rabinovich+2018a}. We illustrate rulings and Gauss image to show
    that this surface is developable. Further optimization by
    our method improves the quality (right). The
    main advantage of our method compared to that of Rabinovich
    et al.~\shortcite{rabinovich+2018a} is that theirs works with a special
    parametrization, limiting its capabilities e.g.\ for modeling.}
    \label{fig:dog}
\end{figure}

The method presented in this paper has a focus on the modeling of continuous
deformations of discrete developable surfaces for applications in \emph{design},
in line with the works of \cite{rabinovich+2018a,jiang-2020}.
We observed that typically
we could improve the quality of results obtained by prior work by
subjecting these results to optimization by our method. Figure \ref{fig:dog}
shows an example where this has been tested on a result obtained
by the method of \cite{rabinovich+2018a}.
In our experience, the Gauss image test
reveals that the results of \cite{rabinovich+2018a} have the highest quality
of developability among related work \cite{jiang-2020,sellan-2020,Stein:2018:DSF}.
Yet our method can improve the quality
of developability even further.

Our method can be used in other applications involving developables,
such as \emph{guided} surface approximation by piecewise
developables --- see Fig.~\ref{fig:strips}. We do
not compare approximation capabilities with previous contributions
in that area \cite{binninger-2021,ion-2020,sellan-2020,Stein:2018:DSF},
as currently our method would require a prior decomposition into piecewise
developables. This limitation is intentional, as our focus is to incorporate
design aesthetics that cannot be achieved without user-guided input.

\paragraph{Interactive Modeling}
Our method is interactive in two ways. Firstly it can be used to interactively
model developables --- see Fig.~\ref{fig:material-behaviour}.
Secondly, we provide immediate feedback to
the user: The Gauss image  directly
shows if developability has been achieved. The user can react and change
constraints, or the weights given to constraints.

\begin{figure}[b]
    \Inc[width=.95\columnwidth]{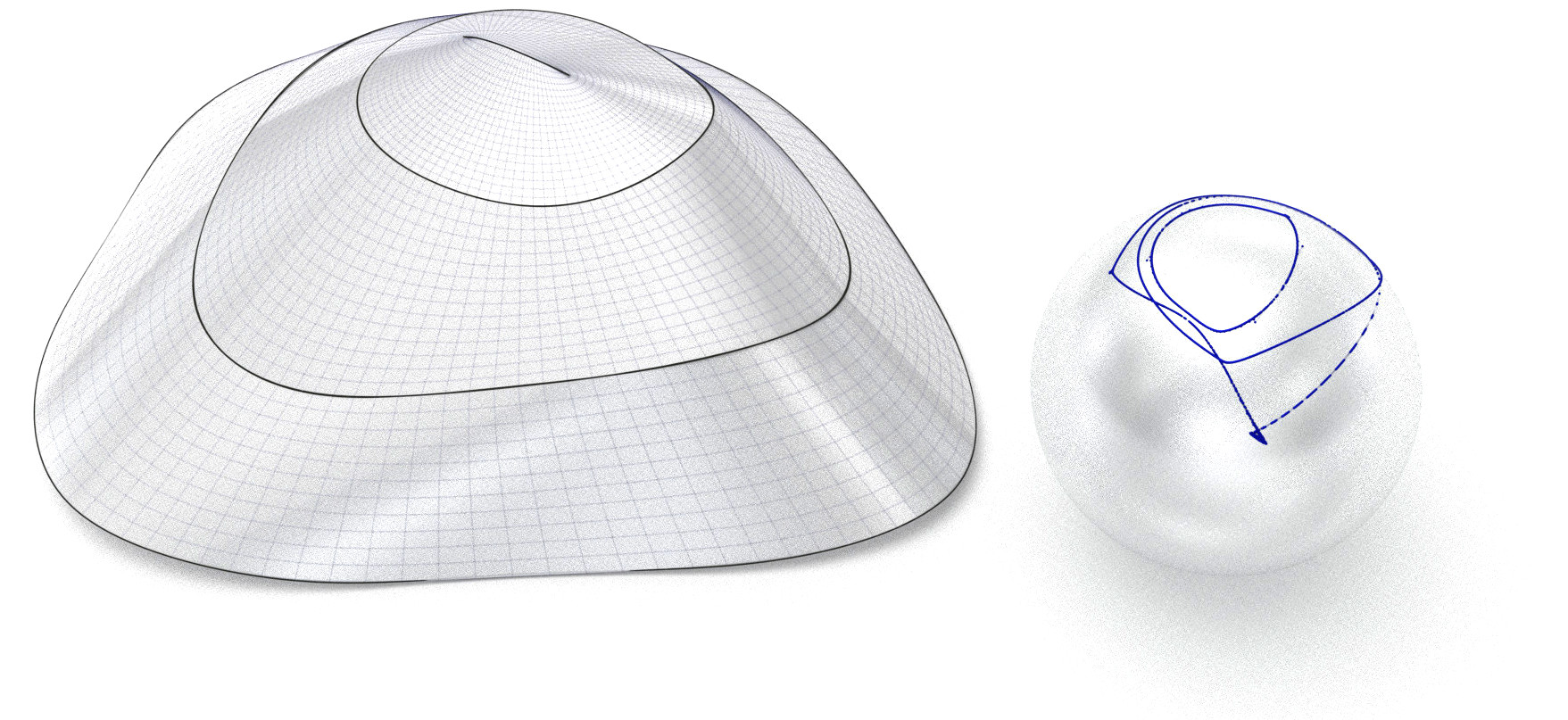}\vskip-1em
    \caption{{\it Lofting singularities.} By choosing boundaries
    and optimizing the surfaces between them we achieve piecewise developability.
    The singularity at the top is a strip boundary that doubles back
    onto itself.}
    \label{fig:conesing}
\end{figure}

Since developables can often be defined by their boundaries, lofting is
actually a very good method of design. We show several examples in
previous sections; here we only point out that we can
model creases as a design element, as shown in Fig.~\ref{fig:creases2},
as well as singularities, illustrated by Fig.~\ref{fig:conesing}.

\begin{figure}[t]
    \centerline{\relax
    {\clipbox{1mm 0 0mm 0}
    {\begin{overpic}[width=.45\columnwidth]{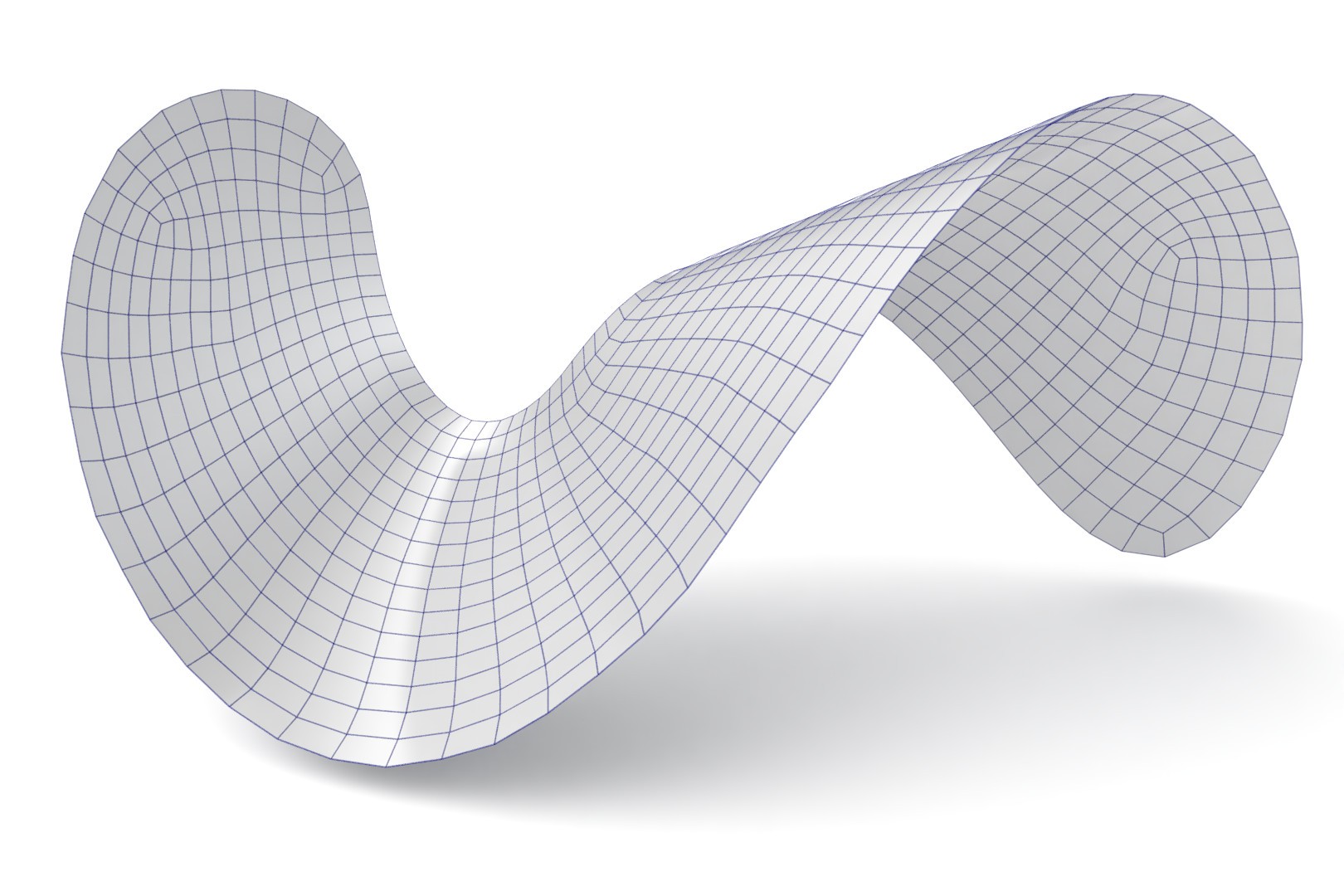}\put(2,10){(a)}\end{overpic}}}\hskip5mm
    {\clipbox{1mm 0 0mm 0}
    {\begin{overpic}[width=.45\columnwidth]{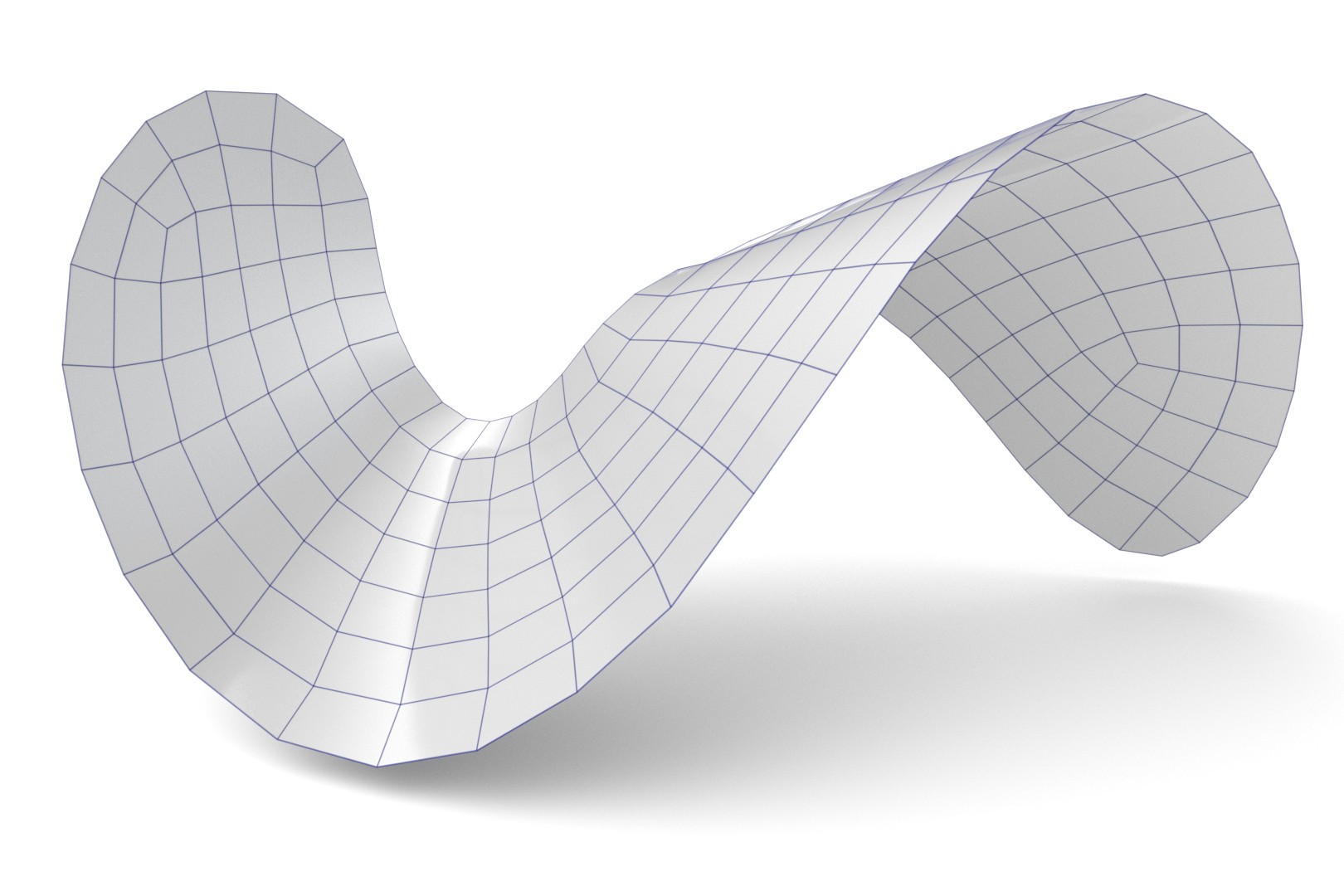}\put(2,10){(b)}\end{overpic}}}
    }
    \centerline{\relax
    {\clipbox{1mm 0 0mm 0}
    {\begin{overpic}[width=.45\columnwidth]{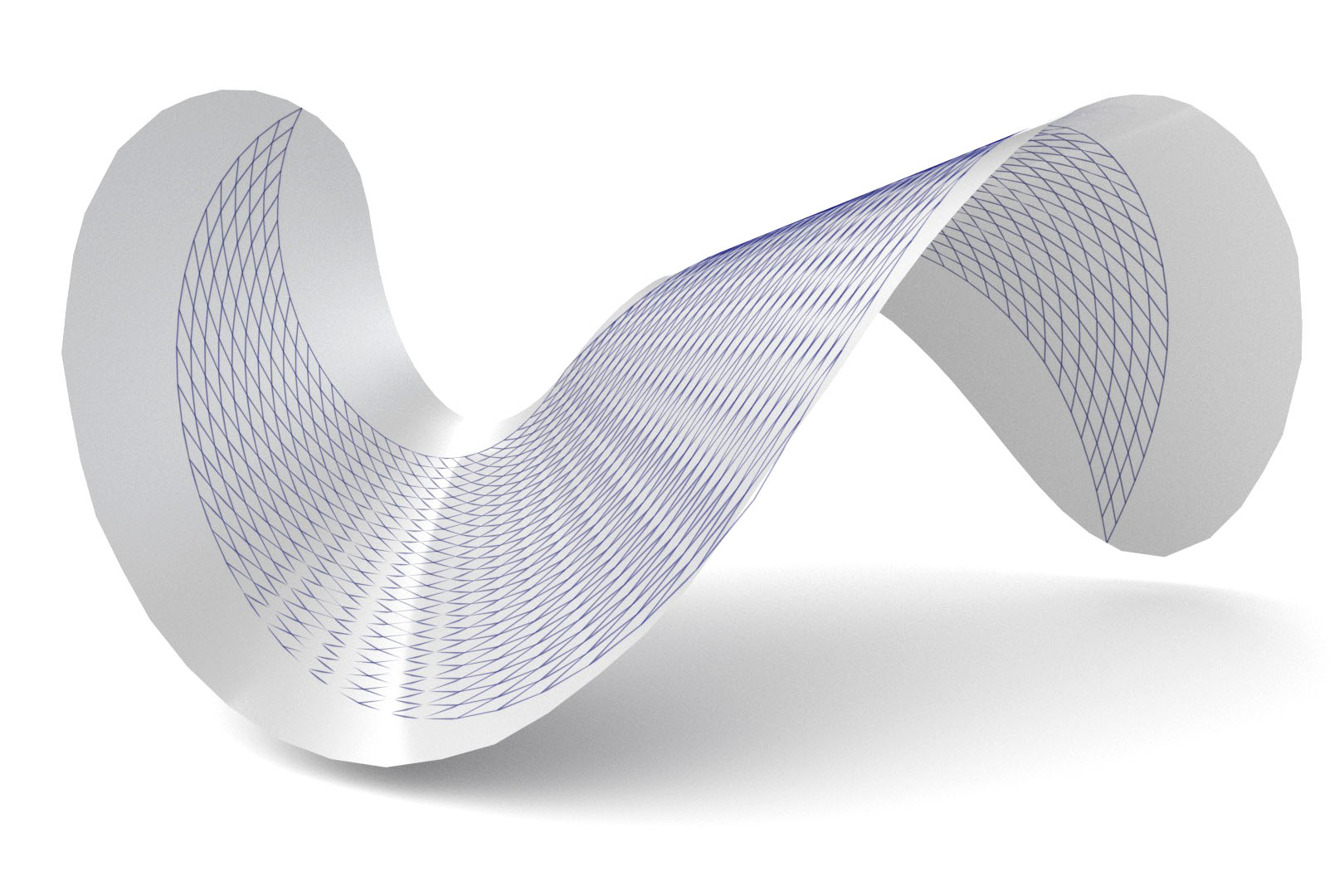}\put(2,10){(c)}\end{overpic}}}\hskip5mm
    {\clipbox{1mm 0 0mm 0}
    {\begin{overpic}[width=.45\columnwidth]{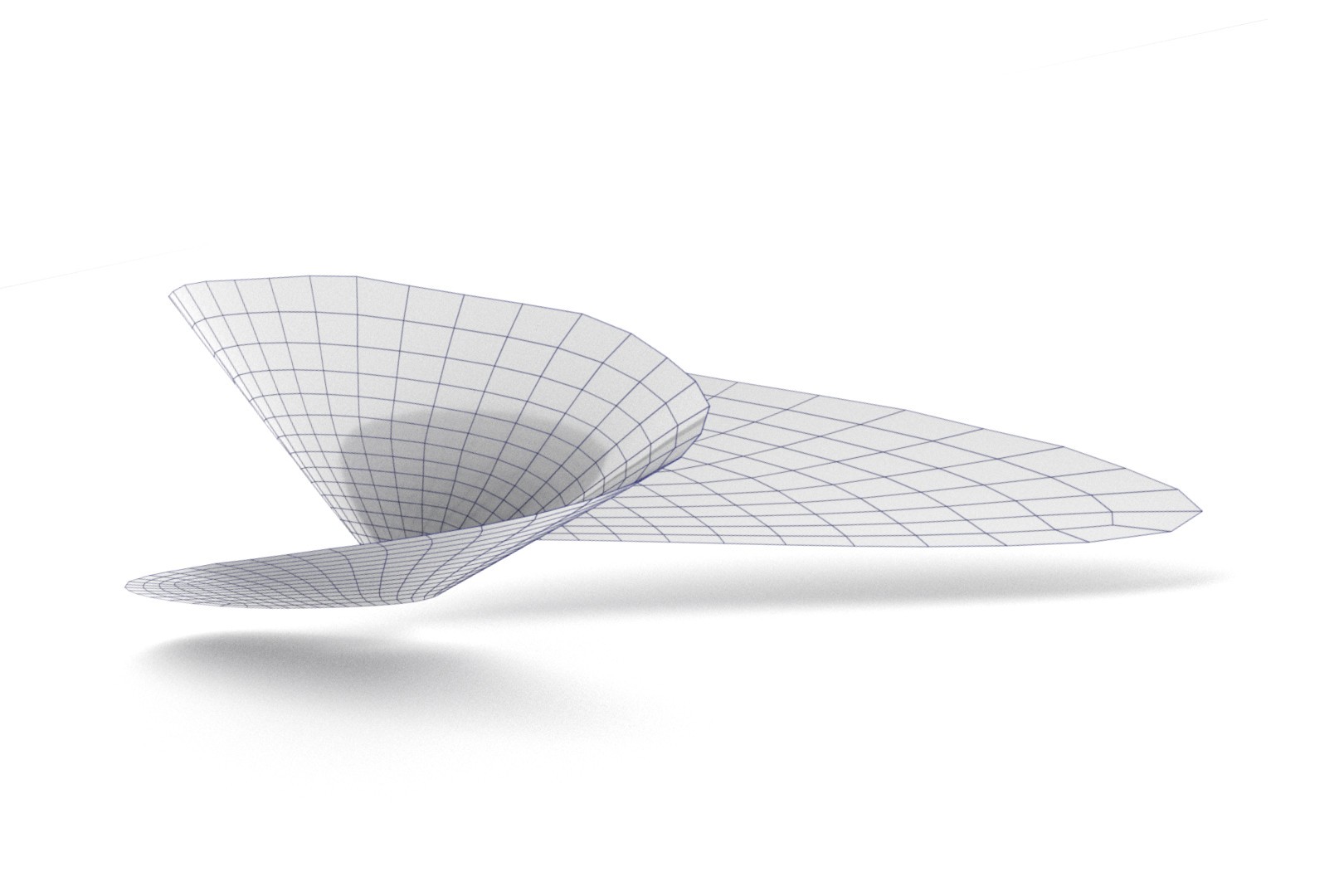}\put(2,10){(d)}\end{overpic}}}
    }
    \vskip-1em
    \caption{{\em Invariance of developability under remeshing and projective maps.}
    Remeshing converts mesh (a) to meshes (b), (c), and a
    projective transformation yields (d). $E_{\dev}$
    remains small --- see Table~\ref{table:2}.}
    \label{fig:remeshing}
\end{figure}

The developability condition proposed in this paper is almost-invariant
under remeshing --- see Fig.~\ref{fig:remeshing}. This means representing a given developable mesh
by another quad mesh with fair mesh polylines yields a mesh that
almost fulfills our criterion of Def.~\ref{defn:developability} again.
This is due to the underlying geometric property not being changed.
This property has been used to create the examples of
Fig.~\ref{fig:cuts}, and is extremely useful
e.g.\ for trimming and for joining patches. We emphasize that we can
work with most methods for user-guided quad remeshing, e.g.\
\cite{ebke-2016}.

Recall that in our setup we can directly leverage the
projective invariance of developability, cf. \S\,\ref{sss:developablemesh}.
Affine transformations were used in the modeling process that led to Fig.~\ref{fig:cuts}.

\paragraph{Limitations}

One major limitation of our methods is geometric in nature.
While an experienced user can generate developables
easily, this is more difficult for a user without prior knowledge of
the quirks of developable surfaces.
Our implementation currently requires the user to choose weights
meaningfully.

\paragraph{Future Research}

Our aim is to publish a plugin for the software {\em Rhino}, which
benefits from the possibility of conversion to NURBS format.
For practical applications, material properties and tolerances
need to be considered.
Regarding our contribution to geometry, we are confident that it can lead to
a more comprehensive theory of contact element nets.

\begin{acks}
This work was supported by the Austrian Science Fund via grants
I2978 (SFB-Transregio programme Discretization in geometry and
dynamics), and F77 (SFB grant Advanced Computational Design).
V. Ceballos Inza and F. Rist were supported by KAUST baseline funding.
\end{acks}

\bibliographystyle{references/ACM-Reference-Format}
\bibliography{references/references}

\clearpage

\end{document}